\documentclass[Afour,sageh,times]{sagej}

\usepackage{moreverb,url}

\usepackage{hhline}
\usepackage{colortbl}
\usepackage{arydshln}
\usepackage{amssymb}
\usepackage{algorithm}
\usepackage{algpseudocode}
\usepackage{algorithmicx}
\usepackage{MnSymbol}
\algrenewcommand\algorithmicrequire{\textbf{Input:}}
\algrenewcommand\algorithmicensure{\textbf{Output:}}
\usepackage{tikz}
\usetikzlibrary{decorations.pathreplacing,calc,fit}
\usetikzlibrary{calc}
\tikzset{every picture/.append style={remember picture},
	na/.style={baseline=-0.6ex}}
\definecolor{lblue}{RGB}{174,199,232}
\definecolor{lorange}{RGB}{255,187,120}
\definecolor{lgreen}{RGB}{152,223,138}
\definecolor{lred}{RGB}{255,152,150}
\definecolor{lpurple}{RGB}{197,176,213}
\definecolor{lgray}{RGB}{199,199,199}
\definecolor{lbrown}{RGB}{196,156,148}
\definecolor{lpink}{RGB}{247,182,210}
\definecolor{lmush}{RGB}{219,219,141}
\definecolor{lcyan}{RGB}{158,218,229}

\usepackage{bm}

\usepackage{subfig}
\usepackage{xcolor}
\newcommand{\bq}{\begin{equation}}
\newcommand{\eq}{\end{equation}}

\newcommand{\byte}{\mbox{B}}
\newcommand{\second}{\mbox{s}}

\newcommand{\flop}{\mbox{flop}}

\newcommand{\FS}{\mbox{\flop/\second}}

\newcommand{\MFS}{\mbox{M\flop/\second}}

\newcommand{\MiB}{\mbox{MiB}}

\newcommand{\rlm}{roof{}line model}

\newcommand{\likwidbench}{\texttt{likwid-bench}}

\definecolor{tumbleweed}{rgb}{0.87, 0.67, 0.53}

\newcommand*{\rom}[1]{\expandafter\@slowromancap\romannumeral #1@}

\def\addlegendimage{\csname pgfplots@addlegendimage\endcsname}

\definecolor{applegreen}{rgb}{0.55, 0.71, 0.0}
\definecolor{amethyst}{rgb}{0.6, 0.4, 0.8}
\definecolor{amber}{rgb}{1.0, 0.75, 0.0}

\definecolor{col0}{rgb}{  0.5508,    0.8242,    0.7773 }
\definecolor{col1}{rgb}{  0.98,    0.93,    0.36 }
\definecolor{col2}{rgb}{  0.7422,    0.7266,    0.8516 }
\definecolor{col3}{rgb}{  0.9805,    0.5000,    0.4453 }
\definecolor{col4}{rgb}{  0.5000,    0.6914,    0.8242 }
\definecolor{col5}{rgb}{  0.9883,    0.7031,    0.3828 }
\definecolor{col6}{rgb}{  0.6992,    0.8672,    0.4102 }
\definecolor{col7}{rgb}{  0.9844,    0.8008,    0.8945 }
\definecolor{col8}{rgb}{  0.6484,    0.8047,    0.8867 }
\definecolor{col9}{rgb}{  0.1211,    0.4688,    0.7031 }
\definecolor{col10}{rgb}{  0.6953,    0.8711,    0.5391 }
\definecolor{col11}{rgb}{  0.1992,    0.6250,    0.1719 }
\definecolor{col12}{rgb}{  0.9805,    0.6016,    0.5977 }
\definecolor{col13}{rgb}{  0.8867,    0.1016,    0.1094 }
\definecolor{col14}{rgb}{  0.9883,    0.7461,    0.4336 }

\usepackage[most]{tcolorbox}

\pgfdeclarepatternformonly{mystrikeout}{\pgfqpoint{-1pt}{-1pt}}{\pgfqpoint{11pt}{11pt}}{\pgfqpoint{10pt}{10pt}}%
{
	\pgfsetlinewidth{0.4pt}
	\pgfpathmoveto{\pgfqpoint{0pt}{0pt}}
	\pgfpathlineto{\pgfqpoint{10.1pt}{10.1pt}}
	\pgfusepath{stroke}
}
\newtcolorbox{tcbstrikeout}{breakable,
	enhanced jigsaw,
	opacityback=0,
	parbox=false,
	boxrule=0mm,
	top=0mm,bottom=0pt,left=0pt,right=0pt,
	boxsep=0pt,
	frame hidden,
	finish={\fill[pattern=mystrikeout, pattern color=red] (frame.north west) rectangle (frame.south east);}
}

\usepackage[colorlinks,bookmarksopen,bookmarksnumbered,citecolor=red,urlcolor=red]{hyperref}

\newcommand\BibTeX{{\rmfamily B\kern-.05em \textsc{i\kern-.025em b}\kern-.08em
T\kern-.1667em\lower.7ex\hbox{E}\kern-.125emX}}

\def\HiLi{\leavevmode\rlap{\hbox to \hsize{\color{yellow!50}\leaders\hrule height .8\baselineskip depth .5ex\hfill}}}

\def\volumeyear{2016}

\begin{document}

\runninghead{
	Lacey and 
	Alappat \textit{et al.}}

\title{%
Cache Blocking of Distributed-Memory Parallel Matrix Power Kernels}

\author{%
	Dane Lacey\affilnum{1}, 
	Christie Alappat\affilnum{1}, 
	Florian Lange\affilnum{1}, 
	Georg Hager\affilnum{1},
	Holger Fehske\affilnum{1,2}\\and
	Gerhard Wellein\affilnum{1,3,4}
}

\affiliation{%
	\affilnum{1}Erlangen National High Performance Computing Center (NHR@FAU), Friedrich-Alexander-Universit\"at Erlangen-N\"urnberg\\
	\affilnum{2}Institute of Physics, University of Greifswald\\
	\affilnum{3}Department of Computer Science, Friedrich-Alexander-Universit\"at Erlangen-N\"urnberg\\
	\affilnum{4}Delft Institute of Applied Mathematics, Delft University of Technology
}

\corrauth{Dane Lacey,
	Martensstraße 1, 91058 Erlangen, Germany
}

\email{dane.c.lacey@fau.de}

\begin{abstract}
Sparse matrix-vector products (SpMVs) are a bottleneck in many
scientific codes. Due to the heavy strain on the main memory interface
from loading the sparse matrix and the possibly irregular memory
access pattern, SpMV typically exhibits low arithmetic
intensity. Repeating these products multiple times with the same
matrix is required in many algorithms. This
so-called matrix power kernel (MPK) provides an opportunity for data
reuse since the same matrix data is loaded from main memory multiple
times, an opportunity that has only recently been exploited
successfully with the Recursive Algebraic Coloring Engine (RACE). Using RACE,
one considers a graph based formulation of the SpMV and employs s level-based implementation of SpMV for reuse of
relevant matrix data. However, the underlying data dependencies have restricted the use of this concept to shared memory parallelization and thus to single compute nodes.
Enabling cache blocking for distributed-memory parallelization of MPK is challenging due to the need for explicit communication and synchronization of data in neighboring levels.

In this work, we propose and implement a flexible method that interleaves the
cache-blocking capabilities of RACE with an MPI communication scheme
that fulfills all data dependencies among
processes. 
Compared to a ``traditional" distributed memory parallel MPK, our new Distributed Level-Blocked MPK yields substantial speed-ups on modern Intel and AMD architectures across a wide range of sparse matrices from various scientific applications. 
Finally, we address a modern quantum physics problem to demonstrate the applicability of our method, achieving a speed-up of up to $4\times$ on 832 cores of an Intel Sapphire Rapids cluster.
\end{abstract}


\maketitle

\section{Introduction and Related Work}
\label{sec:Introduction and Related Work}
Parallel solvers for linear systems or eigenvalue problems involving
large sparse matrices have been in wide use for decades in traditional
research fields using high-performance computing (HPC) such as quantum
physics, quantum chemistry, and engineering. In recent years, new
applications relying on powerful and efficient sparse matrix solvers
have been developed, ranging from social graph analysis as shown by
\cite{10.1145/3218176.3218232} to spectral clustering in the context
of learning algorithms, shown by
\cite{spectral_clustering,JMLR:v17:16-109}. Typically these solvers
use iterative subspace methods, which may include advanced
preconditioning techniques and rely on an efficient parallel
implementation of the sparse-matrix vector (SpMV) kernel $y \gets Ax$,
where $A$ is a sparse matrix and $x$, $y$
are dense vectors. Scalable and efficient SpMV
implementations have thus been an active field of investigation for a long
time, where an overview of optimization efforts is given by \cite{10.5555/1023242}.
Its low computational intensity makes SpMV strongly memory bound
on all modern compute devices,  and much research % over the years
focuses on efficient sparse matrix data layouts or matrix
bandwidth reduction to improve access locality in the dense vectors
involved in the SpMV. \cite{doi:10.1137/130930352} showed that this was
particularly relevant on GPGPUs and wide-SIMD many-core CPUs.
All of these efforts
targeted only a single SpMV operation, ignoring the potential matrix
data reuse in successive invocations of SpMV with the same matrix.

Certain algorithms can be reformulated to group SpMV invocations with
the same matrix together, as shown by \cite{4536305} for
CA-Krylov and by \cite{doi:10.1137/1.9781611976137.4} for
preconditioners based on matrix polynomials. These back-to-back SpMVs
constitute what we call the traditional \emph{Matrix Power Kernel} (MPK)
implementation. This kernel computes all vectors $y_p \gets A^px$ for
each power $p=1,\dots,p_m$, where the sparse (and necessarily square)
matrix $A$ is loaded from main memory each time SpMV is called. This
scenario presents an immense opportunity for raising the computational
intensity through \emph{cache blocking}, keeping relevant matrix data
in cache across successive SpMV invocations.

In recent years, the top CPU and GPGPU manufacturers have been rapidly increasing
the cache sizes on their server-grade chips. Shown in
Table~\ref{tab:Processor cache size trends} is a selection of
top-of-the-line CPU and GPGPU models from Intel, AMD, and Nvidia, and
their respective aggregate cache sizes (sum of all cache levels,
rounded to the nearest \MiB\endnote{We conform to the standard of
  describing quantities as powers of two, and performance metrics as
  powers of ten, e.g., $1\,\MiB = 2^{20}\,\byte$, $1\,\MFS = 10^{6}\,\FS$. })
over the last several years. These advancements in hardware
capabilities have only broadened the opportunities for cache blocking.

The Recursive Algebraic Coloring Engine (RACE), as introduced by
\cite{RACE}, can be used to construct an efficient, cache-blocked
shared-memory MPK by taking advantage of the level-based
formulation of SpMV. \cite{RACEMPK} describes the resulting
\emph{Level-Blocked Matrix Power Kernel} (LB-MPK), with
applications of LB-MPK to contemporary sparse iterative
solvers shown by \cite{alappat2023algebraic}. While
successful, this work is restricted to shared-memory
compute nodes. No concept or implementation to parallelize RACE for distributed-memory parallel systems using the Message Passing Interface (MPI) has been proposed until now.  
Satisfying the data dependencies of the level-based formulation among
parallel processes by message passing is a non-trivial task.

The main contribution of this work is an MPI adaptation of
LB-MPK. Other works on distributed MPK, such as those developed by
\cite{6877272,7013063}, are focused on reducing the MPI communication
overhead. At the time of writing, there is surprisingly little work
found in the direction of cache-blocking techniques for the
distributed MPK. There exists an analysis of a similar diamond tiling
strategy by \cite{10.1145/3368474.3368494}, but it is purely
theoretical. The closest work is likely from
\cite{10.1145/1654059.1654096}, but there are clear differences
between this approach and ours. Besides being MPI-only whereas ours is
a hybrid (MPI+OpenMP) approach, their MPK requires redundant
computations and/or indirect accesses to matrix elements with
bookkeeping to fulfill data dependencies. We will revisit this
comparison in Section~\ref{sec:DLB-MPK Methodology}.

When compared to the traditional ``back-to-back" SpMV implementation
of MPK, our novel Distributed Level-Blocked MPK algorithm shows
speed-ups of up to $2.7\times$ across various architectures for a wide
variety of matrices from the SuiteSparse Matrix Collection by
\cite{10.1145/2049662.2049663}.

\begin{table}
	\small\sf\centering
	\begin{center}
		\caption{Cache size trends for Intel, AMD, and Nvidia devices.\protect\endnote{\url{www.techpowerup.com/cpu-specs/}}$^,$\protect\endnote{\url{www.techpowerup.com/gpu-specs/}}}
		\label{tab:Processor cache size trends}
		\resizebox{\columnwidth}{!}{%
			\begin{tabular}{llllr} 
				\toprule
				Company & Year   & Model & Type & Aggregate Cache\\ \midrule
				Intel  & 2019 Q1 & Cascade Lake - 8280 & CPU & $68$ MiB\\
				& 2021 Q4 & Ice Lake - 8380 & CPU& $102$ MiB\\
				& 2023 Q1 & Saphire Rapids - 8480 & CPU& $221$ MiB\\
				& 2023 Q1 & Ponte Vecchio - MAX 1550 & GPGPU & $472$ MiB\\
				%AMD    & 2017 Q2 & EYPC 1 - 7601 & CPU& $83$ MiB\\
				AMD   & 2019 Q2 & EYPC 2 - 7742 & CPU & $294$ MiB\\
				& 2022 Q1 & EYPC 3 - 7773X & CPU& $804$ MiB\\
				& 2023 Q2 & EYPC 4 - 9684X & CPU& $1254$ MiB\\ 
				& 2023 Q4 & Aqua Vanjaram - MI300X & GPGPU & $277$ MiB\\
				Nvidia    & 2018 Q1 & Volta - V100 SXM3 & GPGPU &$16$ MiB\\
				& 2020 Q1 & Ampere - A100 SXM4 & GPGPU& $60$ MiB\\
				& 2023 Q1 & Hopper - H100 SXM5 & GPGPU&$83$ MiB\\
				& 2023 Q2 & Grace& CPU& $333$ MiB\\
				\bottomrule
			\end{tabular}
		}
	\end{center}
\end{table}

\section{Overview and Contributions}
\label{sec:Overview and Contributions}
The \emph{Distributed Level-Blocked Matrix Power Kernel} (DLB-MPK) algorithm
extends the LB-MPK algorithm to the distributed setting with MPI. Our
implementation is efficient in that it does not increase the MPI
overhead when compared to the traditional MPK implementation, and it does
not require any redundant computations.

This paper is organized as follows.
In Section \ref{sec:MPK RACE}, we begin with a brief summary of
shared-memory SpMV and MPK. Then, by exploring the
graph-matrix correspondence, we are able to understand the broad
strokes of how RACE performs LB-MPK. In order to generalize LB-MPK, we
must first understand distributed-memory parallel SpMV and MPK without
cache blocking, which we explore in Section~\ref{sec:Distributed_Challenges}. We close this section with a motivation of our method by comparing it against the distributed MPK implemented by
\cite{10.1145/1654059.1654096}. Section~\ref{sec:DLB-MPK Methodology}
details our DLB-MPK method and implementation.
In Section~\ref{sec:Results} we investigate the relevant
hardware characteristics of three modern multicore CPU
systems and their influence on performance of DLB-MPK. Performance predictions based on the \rlm\ by
\cite{10.1145/1498765.1498785} are derived, and we investigate
the influence of various parameters with RACE in the
distributed setting. We close the section with a strong scaling
analysis of DLB-MPK. In Section \ref{sec:Application} we examine the
weak scaling characteristics of DLB-MPK used in Chebyshev time propagation,
which has applications in quantum physics.

In this work, we make the following contributions:
\begin{itemize}
	\item We extend the level-based concepts in RACE to the distributed setting. 
	\item We detail the trapezoidal-like tiling strategy which enables our DLB-MPK to fulfill the data dependencies inherent in repeated SpMV invocations, and present an efficient implementation of the DLB-MPK.
	\item For a wide array of sparse matrices, we present a performance and scaling benefit summary of DLB-MPK on three modern CPUs from Intel and AMD.
	\item We investigate the weak scaling behavior of DLB-MPK when applied to the Chebyshev method for the time evolution of quantum states for the Anderson model of localization, and show the favorable scaling qualities as compared to the traditional MPK.
\end{itemize}

\section{RACE Applied to the Matrix Power Kernel}
\label{sec:MPK RACE}

\begin{figure*}[!htbp]
	\centering
	\captionsetup[subfigure]{justification=centering}
	\subfloat[Graph]{\raisebox{6.5mm}{%
	%  \tikzsetnextfilename{#2}%
	\edef\scaleFac{0.75}
%\begin{document}
	\begin{tikzpicture}[darkstyle/.style={circle,draw,fill=yellow!30!white,minimum size=12.5, inner sep=0.5pt}, stencilstyle/.style={circle,draw,red,fill=red!30!white,minimum size=13.2, inner sep=0.5pt}, 
	yscale=-1]
				
	\foreach \x in {0,...,4}
	\foreach \y in {0,...,3} 
	{\pgfmathtruncatemacro{\label}{\x+5*\y}
			\node [darkstyle]  (\x\y) at (\scaleFac*\x,\scaleFac*\y) {\fontsize{7.4}{9}\selectfont{\label}};} 
	  \draw[-] (10) to [out=180+45, in=45, looseness=0](01);
	  \draw[-] (20) to [out=180+45, in=45, looseness=0](11);
	  \draw[-] (30) to [out=180+45, in=45, looseness=0](21);
	  \draw[-] (40) to [out=180+45, in=45, looseness=0](31);
	  \draw[-] (20) to [out=0, in=180, looseness=0](30);
	  \draw[-] (30) to [out=0, in=180, looseness=0](40);
	  
	  \draw[-] (11) to [out=180+45, in=45, looseness=0](02);
	  \draw[-] (21) to [out=180+45, in=45, looseness=0](12);
	  \draw[-] (31) to [out=180+45, in=45, looseness=0](22);
	  \draw[-] (41) to [out=180+45, in=45, looseness=0](32);
	  \draw[-] (21) to [out=0, in=180, looseness=0](31);
	  
	  \draw[-] (12) to [out=180+45, in=45, looseness=0](03);
	  \draw[-] (22) to [out=180+45, in=45, looseness=0](13);
	  \draw[-] (32) to [out=180+45, in=45, looseness=0](23);
	  \draw[-] (42) to [out=180+45, in=45, looseness=0](33);
	  \draw[-] (12) to [out=0, in=180, looseness=0](22);	
	  
	  \draw[-] (03) to [out=0, in=180, looseness=0](13);
	  \draw[-] (13) to [out=0, in=180, looseness=0](23);
	  
	  \draw[-] (40) to [out=-90, in=-90, looseness=.6](00);
	  \draw[-] (41) to [out=-90, in=-90, looseness=.6](01);
	  \draw[-] (42) to [out=-90, in=-90, looseness=.6](02);
	  \draw[-] (43) to [out=-90, in=-90, looseness=.6](03);

	\end{tikzpicture}
%\end{document}  
%
%\includegraphics{tikz_cache/paper-figure\theplotCtr.pdf}
%\stepcounter{plotCtr}
\label{fig:stencil_graph}}}
	\hfill
	\subfloat[Sparsity pattern of matrix]{\includegraphics[width=.5\columnwidth]{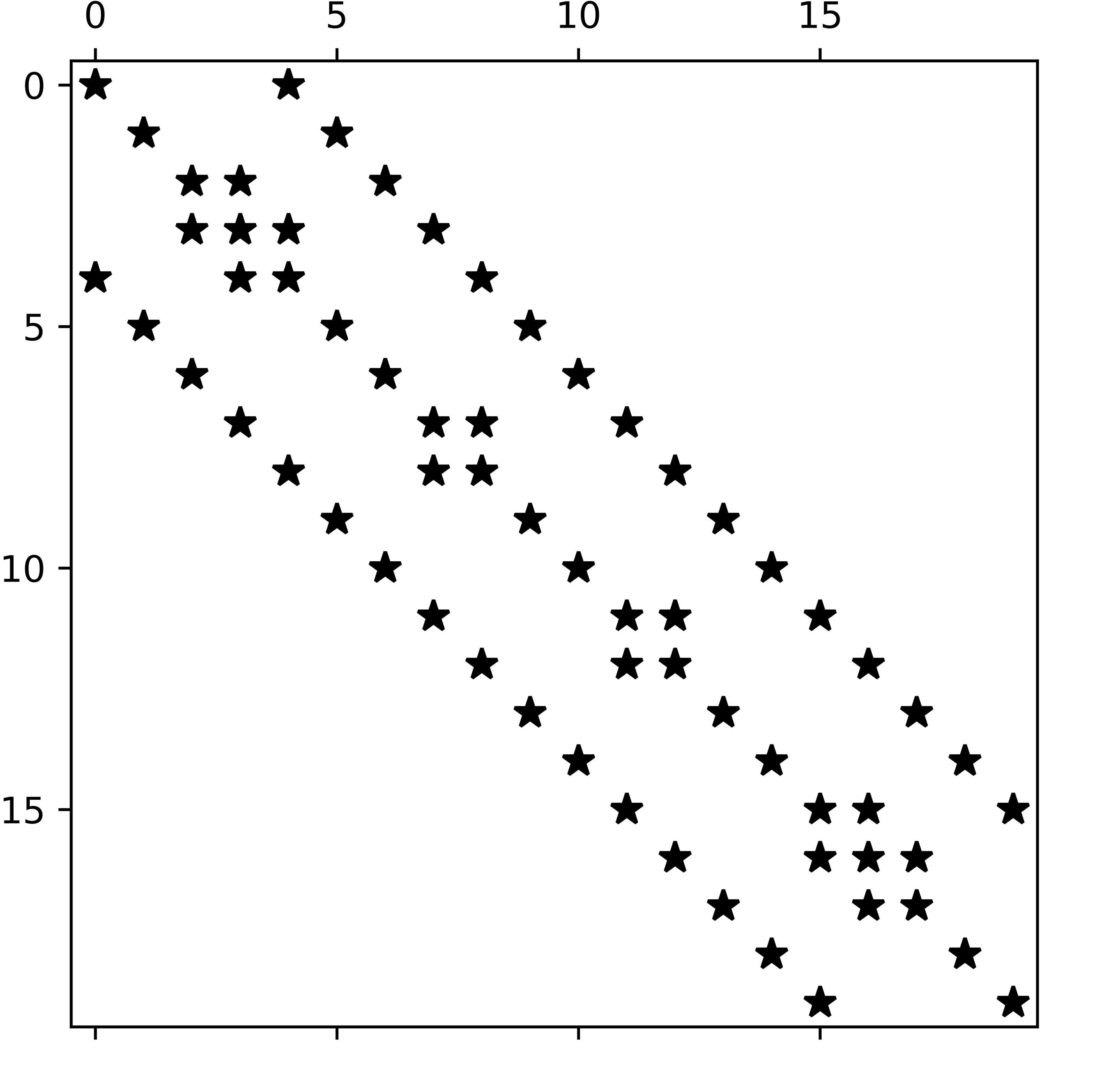}\label{fig:stencil_matrix}}
	\hfill
	\subfloat[Permuted graph]{\raisebox{6.5mm}{%
	%  \tikzsetnextfilename{#2}%
	\edef\scaleFac{0.75}
\edef\bright{60}
%\begin{document}
\begin{tikzpicture}[
	l0/.style={circle,draw,fill=col0!\bright!white,minimum size=12.5, inner sep=0.5pt},
	l1/.style={circle,draw,fill=col1!\bright!white,minimum size=12.5, inner sep=0.5pt},
	l2/.style={circle,draw,fill=col2!\bright!white,minimum size=12.5, inner sep=0.5pt},
	l3/.style={circle,draw,fill=col3!\bright!white,minimum size=12.5, inner sep=0.5pt},
	l4/.style={circle,draw,fill=col4!\bright!white,minimum size=12.5, inner sep=0.5pt},
	l5/.style={circle,draw,fill=col5!\bright!white,minimum size=12.5, inner sep=0.5pt},
	l6/.style={circle,draw,fill=col6!\bright!white,minimum size=12.5, inner sep=0.5pt},
	l7/.style={circle,draw,fill=col7!\bright!white,minimum size=12.5, inner sep=0.5pt},
	l8/.style={circle,draw,fill=col8!\bright!white,minimum size=12.5, inner sep=0.5pt},
	l9/.style={circle,draw,fill=col9!\bright!white,minimum size=12.5, inner sep=0.5pt},
	l10/.style={circle,draw,fill=col10!\bright!white,minimum size=12.5, inner sep=0.5pt},
	l13/.style={circle,draw,fill=col13!\bright!white,minimum size=12.5, inner sep=0.5pt},
	yscale=-1
	]
	
	\foreach \x in {0,...,4}
	\foreach \y in {0,...,3} 
	{\pgfmathtruncatemacro{\label}{\x\y}
		\node [l0]  (\x\y) at (\scaleFac*\x,\scaleFac*\y) {};
	} 
	
	\node [l0]  (00) at (\scaleFac*0,\scaleFac*0) {\fontsize{7.4}{9}\selectfont{0}};
	\node [l13]  (01) at (\scaleFac*0,\scaleFac*1) {\fontsize{7.4}{9}\selectfont{18}};
	\node [l5]  (02) at (\scaleFac*0,\scaleFac*2) {\fontsize{7.4}{9}\selectfont{10}};
	\node [l5]  (03) at (\scaleFac*0,\scaleFac*3) {\fontsize{7.4}{9}\selectfont{11}};
	\node [l9]  (10) at (\scaleFac*1,\scaleFac*0) {\fontsize{7.4}{9}\selectfont{19}};
	\node [l4]  (11) at (\scaleFac*1,\scaleFac*1) {\fontsize{7.4}{9}\selectfont{7}};
	\node [l4]  (12) at (\scaleFac*1,\scaleFac*2) {\fontsize{7.4}{9}\selectfont{8}};
	\node [l4]  (13) at (\scaleFac*1,\scaleFac*3) {\fontsize{7.4}{9}\selectfont{9}};
	\node [l3]  (20) at (\scaleFac*2,\scaleFac*0) {\fontsize{7.4}{9}\selectfont{4}};
	\node [l3]  (21) at (\scaleFac*2,\scaleFac*1) {\fontsize{7.4}{9}\selectfont{5}};
	\node [l3]  (22) at (\scaleFac*2,\scaleFac*2) {\fontsize{7.4}{9}\selectfont{6}};
	\node [l5]  (23) at (\scaleFac*2,\scaleFac*3) {\fontsize{7.4}{9}\selectfont{12}};
	\node [l2]  (30) at (\scaleFac*3,\scaleFac*0) {\fontsize{7.4}{9}\selectfont{2}};
	\node [l2]  (31) at (\scaleFac*3,\scaleFac*1) {\fontsize{7.4}{9}\selectfont{3}};
	\node [l6]  (32) at (\scaleFac*3,\scaleFac*2) {\fontsize{7.4}{9}\selectfont{13}};
	\node [l7]  (33) at (\scaleFac*3,\scaleFac*3) {\fontsize{7.4}{9}\selectfont{17}};
	\node [l1]  (40) at (\scaleFac*4,\scaleFac*0) {\fontsize{7.4}{9}\selectfont{1}};
	\node [l7]  (41) at (\scaleFac*4,\scaleFac*1) {\fontsize{7.4}{9}\selectfont{16}};
	\node [l6]  (42) at (\scaleFac*4,\scaleFac*2) {\fontsize{7.4}{9}\selectfont{14}};
	\node [l6]  (43) at (\scaleFac*4,\scaleFac*3) {\fontsize{7.4}{9}\selectfont{15}};
	\draw[-] (10) to [out=180+45, in=45, looseness=0](01);
	\draw[-] (20) to [out=180+45, in=45, looseness=0](11);
	\draw[-] (30) to [out=180+45, in=45, looseness=0](21);
	\draw[-] (40) to [out=180+45, in=45, looseness=0](31);
	\draw[-] (20) to [out=0, in=180, looseness=0](30);
	\draw[-] (30) to [out=0, in=180, looseness=0](40);
	
	\draw[-] (11) to [out=180+45, in=45, looseness=0](02);
	\draw[-] (21) to [out=180+45, in=45, looseness=0](12);
	\draw[-] (31) to [out=180+45, in=45, looseness=0](22);
	\draw[-] (41) to [out=180+45, in=45, looseness=0](32);
	\draw[-] (21) to [out=0, in=180, looseness=0](31);
	
	\draw[-] (12) to [out=180+45, in=45, looseness=0](03);
	\draw[-] (22) to [out=180+45, in=45, looseness=0](13);
	\draw[-] (32) to [out=180+45, in=45, looseness=0](23);
	\draw[-] (42) to [out=180+45, in=45, looseness=0](33);
	\draw[-] (12) to [out=0, in=180, looseness=0](22);	
	
	\draw[-] (03) to [out=0, in=180, looseness=0](13);
	\draw[-] (13) to [out=0, in=180, looseness=0](23);
	
	\draw[-] (40) to [out=-90, in=-90, looseness=.6](00);
	\draw[-] (41) to [out=-90, in=-90, looseness=.6](01);
	\draw[-] (42) to [out=-90, in=-90, looseness=.6](02);
	\draw[-] (43) to [out=-90, in=-90, looseness=.6](03);
	
\end{tikzpicture}%
%\includegraphics{tikz_cache/paper-figure\theplotCtr.pdf}
%\stepcounter{plotCtr}
}
		\label{fig:stencil_graph_permuted}}
		\hfill
	\subfloat[Permuted matrix]{\includegraphics[width=.5\columnwidth]{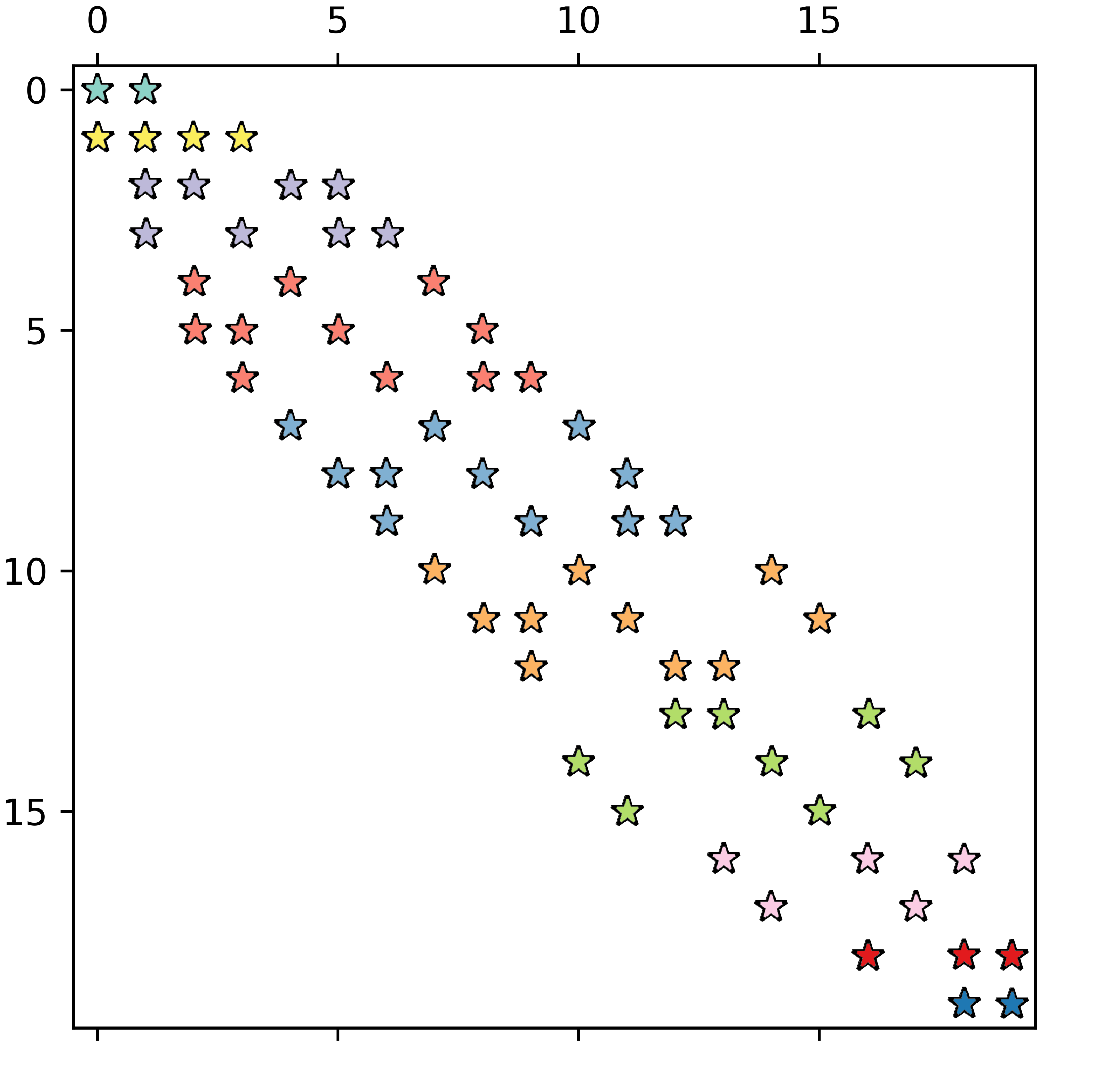}\protect\label{fig:stencil_matrix_permuted}}
	\protect\caption{
		Graph~(a) and sparsity pattern~(b) of the matrix associated with a modified 5-point stencil. 
		Graph~(c) shows the permuted graph and (d) the sparsity pattern of the matrix after applying Breadth First Search (BFS) reordering. The vertices (rows) of the graph (matrix) that belong to a level are 
		represented with the same color.\label{fig:entire_stencil_figure}
	}
\end{figure*}
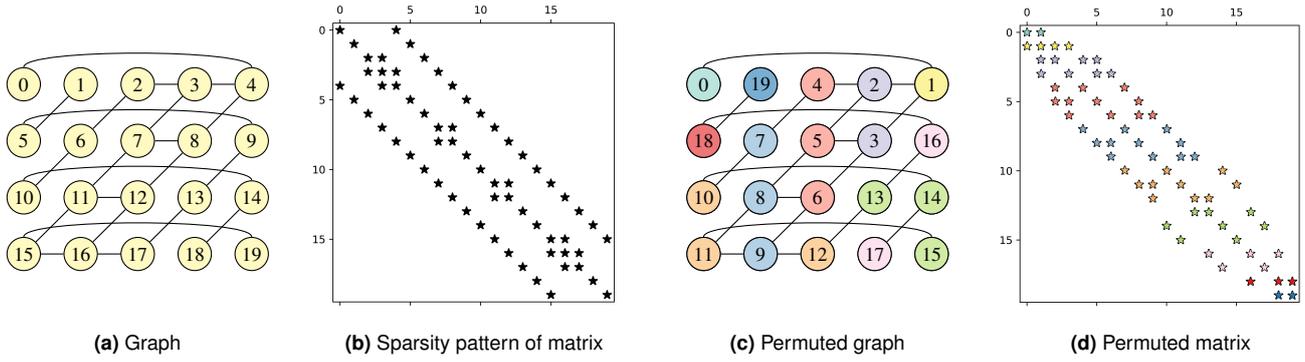

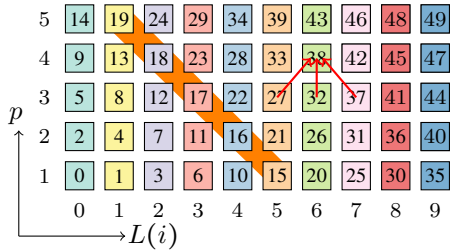
\begin{figure}[tbp]
	\centering
	%
	%  \tikzsetnextfilename{#2}%
	\edef\scaleFac{0.52}
\edef\bright{60}
%power=6
%\def\sumlevelarray{{0,1,3,6,10,15,21,27,33,39,45,51,57,63,69,74,78,81,83,84}}
%power=5
\def\sumlevelarray{{0,1,3,6,10,15,20,25,30,35,40,45,50,55,60,64,67,69,70}}
%\begin{document}
	\begin{tikzpicture}[l0/.style={rectangle,draw,fill=col0!\bright!white,minimum size=10.6, inner sep=0.5pt},
	l1/.style={rectangle,draw,fill=col1!\bright!white,minimum size=10.6, inner sep=0.5pt},
	l2/.style={rectangle,draw,fill=col2!\bright!white,minimum size=10.6, inner sep=0.5pt},
	l3/.style={rectangle,draw,fill=col3!\bright!white,minimum size=10.6, inner sep=0.5pt},
	l4/.style={rectangle,draw,fill=col4!\bright!white,minimum size=10.6, inner sep=0.5pt},
	l5/.style={rectangle,draw,fill=col5!\bright!white,minimum size=10.6, inner sep=0.5pt},
	l6/.style={rectangle,draw,fill=col6!\bright!white,minimum size=10.6, inner sep=0.5pt},
	l7/.style={rectangle,draw,fill=col7!\bright!white,minimum size=10.6, inner sep=0.5pt},
	l8/.style={rectangle,draw,fill=col13!\bright!white,minimum size=10.6, inner sep=0.5pt},
	l9/.style={rectangle,draw,fill=col9!\bright!white,minimum size=10.6, inner sep=0.5pt}]

%	\draw[draw=yellow,fill=yellow!\bright!white] (\scaleFac*9.6,-0.4*\scaleFac) rectangle (\scaleFac*10.4,\scaleFac*4.4);
	\draw[fill=orange, draw=none] (4.8*\scaleFac,-0.3*\scaleFac) -- (5.2*\scaleFac, 0.3*\scaleFac) -- (1.2*\scaleFac, 4.3*\scaleFac) -- (0.8*\scaleFac, 3.7*\scaleFac) --  (4.8*\scaleFac,-0.2*\scaleFac); 
				
	\foreach \x in {0,...,9}
	\foreach \y in {0,...,4} 
	{
	%	\pgfmathtruncatemacro{\labelu}{89 - ((14 - \y - \x)*(14 - \y-\x + 1)*0.5+ 7-\y)
		\pgfmathtruncatemacro{\level}{\y + \x  }
		\pgfmathtruncatemacro{\label}{\sumlevelarray[\level]+\y}
		\node [l\x]  (\x\y) at (\scaleFac*\x,\scaleFac*\y) {\fontsize{7.6}{9}\selectfont{\label}};
	} 
	
	\node [l9] at (\scaleFac*9,\scaleFac*1) {\fontsize{7.6}{9}\selectfont{$40$}};
	\node [l8] at (\scaleFac*8,\scaleFac*2) {\fontsize{7.6}{9}\selectfont{$41$}};
	\node [l7] at (\scaleFac*7,\scaleFac*3) {\fontsize{7.6}{9}\selectfont{$42$}};
	\node [l6] at (\scaleFac*6,\scaleFac*4) {\fontsize{7.6}{9}\selectfont{$43$}};
	\node [l9] at (\scaleFac*9,\scaleFac*2) {\fontsize{7.6}{9}\selectfont{$44$}};
	\node [l8] at (\scaleFac*8,\scaleFac*3) {\fontsize{7.6}{9}\selectfont{$45$}};
	\node [l7] at (\scaleFac*7,\scaleFac*4) {\fontsize{7.6}{9}\selectfont{$46$}};
	\node [l9] at (\scaleFac*9,\scaleFac*3) {\fontsize{7.6}{9}\selectfont{$47$}};
	\node [l8] at (\scaleFac*8,\scaleFac*4) {\fontsize{7.6}{9}\selectfont{$48$}};
	\node [l9] at (\scaleFac*9,\scaleFac*4) {\fontsize{7.6}{9}\selectfont{$49$}};

	\foreach \x in {0,...,9}
	{
		\node at (\scaleFac*\x,-0.46) {\fontsize{7.6}{9}\selectfont$\x$};
	}
%	\node at (\scaleFac*0,-0.46) {\fontsize{7.6}{9}\selectfont$0$};
%	\node at (\scaleFac*3,-0.46) {\fontsize{7.6}{9}\selectfont$\cdots$};
%	\node at (\scaleFac*6,-0.46) {\fontsize{7.6}{9}\selectfont$6$};
%	\node at (\scaleFac*10,-0.46) {\fontsize{7.6}{9}\selectfont$\cdots$};
%	\node at (\scaleFac*14,-0.46) {\fontsize{7.6}{9}\selectfont$14$};
	
	\node at (-0.46,0*\scaleFac) {\fontsize{7.6}{9}\selectfont$1$};
	\node at (-0.46,1*\scaleFac) {\fontsize{7.6}{9}\selectfont$2$};
	\node at (-0.46,2*\scaleFac) {\fontsize{7.6}{9}\selectfont$3$};
	\node at (-0.46,3*\scaleFac) {\fontsize{7.6}{9}\selectfont$4$};
	\node at (-0.46,4*\scaleFac) {\fontsize{7.6}{9}\selectfont$5$};
%	\node at (-0.46,5*\scaleFac) {\fontsize{7.6}{9}\selectfont$ 5$};

	\draw[->] (-0.8,-0.8) -- (0.6,-0.8);
	\node at (0.97,-0.8) {$L(i)$};
	\draw[->] (-0.8,-0.8) -- (-0.8,0.6);
	\node at (-0.84,0.8) {$p$};
	
	%dependency arrows
	\draw[->, red, thick] (\scaleFac*6,\scaleFac*2) -- (\scaleFac*6,\scaleFac*3);
	\draw[->, red, thick] (\scaleFac*5,\scaleFac*2) -- (\scaleFac*5.85,\scaleFac*3);
	\draw[->, red, thick] (\scaleFac*7,\scaleFac*2) -- (\scaleFac*6.15,\scaleFac*3);

%	\draw [white] (0.1,-0.5) rectangle (0.2,-1.5);
	\end{tikzpicture}
%\end{document}  
%
%\includegraphics{tikz_cache/paper-figure\theplotCtr.pdf}
%\stepcounter{plotCtr}

	\caption{$Lp$ diagram with 10 levels ($L(0),\ldots,L(9)$) and a maximum power of $p_m=5$. Level colors are the same as in Figure~\ref{fig:stencil_graph_permuted}.
		Each node in the $Lp$ diagram is numbered according to the execution order.
		For $p=4$ and level $L(6)$, the explicit dependencies to levels at $p=3$ are indicated with red arrows. 
		The nodes highlighted in orange fulfill $i+p=6$ (``diagonal'').
		%	The nodes highlighted in orange background belong to one diagonal plane of the $Lp$ diagram.
		\protect\label{fig:lp_basic}}
\end{figure}

For a given square sparse matrix $A$ and dense vector $x$, MPK computes all vectors $y_p \gets A^px$ for each power $p=1,\dots,p_m$, and stores the result into $p_m$ dense vectors. As mentioned before, this is traditionally implemented as a series of back-to-back SpMVs, using the output vector from the previous iteration as the input vector $x$; e.g., at the $k$-th SpMV invocation, $y_k \gets Ax$ where $x = y_{k-1}$.
SpMV is the central kernel of MPK, whose traditional implementation will be limited by the same bottleneck as SpMV. For matrices $A$ that do not fit into the cache  on modern CPUs (so-called ``memory-resident" matrices), the limiting performance bottleneck is the main memory load bandwidth.

The key observation when cache blocking the MPK is that we can compute
$A^{p}x$ on a subset of rows without waiting for the entire $A^{p-1}x$
computation to finish first for all rows of $A$. The only data
dependency for ``promoting'' a row $v$ from $A^{p-1}x$ to $A^px$,
i.e., executing the $p$-th SpMV operation on it, is that the rows that
correspond to the column indices of the non-zero elements in row $v$
have already been promoted to $A^{p-1}x$. When using LB-MPK, cache
blocking is achieved by detecting the dependencies between successive
SpMV invocations using the level-based SpMV formation within RACE. The
degree to which this fact can be exploited strongly depends on the
sparsity pattern of the matrix. To understand the cache-blocking
scheme in the shared memory setting, we describe this level-based
formulation here.

Given a matrix $A$, there exists a one-to-one correspondence with a graph $G(V,E)$. The set of vertices $V$ represents the rows of $A$, and the set of edges $E$ represents the non-zero elements. If row $v$ in $A$ has a non-zero element at column $j$, then there exists a corresponding edge from vertex $j$ to vertex $v$ in $G(V,E)$. In order to make the correspondence more immediate, we use $G(A)$ to denote the matrix which has $A$ as its adjacency matrix. For the purposes of this work, the values of the non-zero entries of the corresponding matrix $A$ are not considered in the graph. An example of such a correspondence is given by the sparse matrix representing a modified 5pt stencil in Figure~\ref{fig:stencil_matrix} and the associated graph shown in Figure~\ref{fig:stencil_graph}. If vertex $v$ is in the set of ``neighbors" of $u$, $$ N(u) = \{ v \in V : \{ u,v \} \in E \},$$ then $v$ is said to be ``distance 1" from $u$. We say that a vertex $q$ is ``distance $k$" from a vertex $u$ when $q$ is in the ``$k$th neighborhood" of $u$, where we recursively define $$N^k(u) = N^{k-1}(N(u)), \dots, N^2(u) = N(N(u)).$$

RACE will start a Breadth-First Search (BFS) at some ``root vertex," typically at row index 0. In the next step, all vertices that have an edge connected to this root vertex (i.e., its neighbors) are collected into a structure which we call a ``level." In general for a graph $G(V,E)$, we can define the $i$-th level as:
\begin{align*}
	& L(i) = 
	\begin{cases}
		\text{root vertex} \text{ if } i = 0, \\
		u \in N(L(i-1)) \text{ if } i = 1, \\
		u \in N(L(i-1)) \cap \overline{N(L(i-2))} \cap \overline{L(i-2)} \text{ else.}
	\end{cases}
\end{align*}
 At each successive step in the search, all vertices in the current level are scanned, and all neighbors of these vertices that have not yet been touched are collected into the next level. The process continues until the graph is fully traversed, at which point every vertex is collected into a mutually exclusive level\endnote{RACE internally handles non-symmetric matrices as symmetric, filling in non-symmetric entries to aid the collection of vertices into levels. These filled-in elements do not appear on the actual matrix, and it is therefore sufficient to discuss only symmetric matrices.}. Once the graph is fully traversed and each vertex is assigned to a level, RACE then permutes the matrix $A$ in a symmetric manner (rows and columns) based on the levels collected. The symmetric permutation, referred to as ``BFS reordering," improves the temporal locality on the RHS x-vector and avoids irregular accesses to matrix elements. An example of this reordering is given for our 5pt stencil matrix in Figure~\ref{fig:stencil_matrix_permuted} and the associated graph in Figure~\ref{fig:stencil_graph_permuted}. %Details about level-set reorderings can be found in \cite{yousefSaad}.

We can visualize the dependencies and traversal order of LB-MPK with an ``$Lp$ diagram" given in Figure~\ref{fig:lp_basic}. The x-axis is the index of the level $L$, and the y-axis is the power $p$ in $A^px$. An important property of levels is that neighbors of $L(i)$ are contained in $\{L(i-1), L(i), L(i+1)\}$. This means in order to compute $Ax$ on $L(i)$, $x$ has to be known only on $\{L(i-1), L(i), L(i+1)\}$. More generally, to compute $A^px=AA^{p-1}x$ on the vertices of $L(i)$, $A^{p-1}x$ computations on the vertices of $L(i-1)$, $L(i)$, and $L(i+1)$ must have already been completed. One particular example is featured in Figure~\ref{fig:lp_basic} for the computation of $A^4x$ at $L(6)$, where the dependencies lie on $p=3$ at $\{L(5), L(6), L(7)\}$. 

One way to ensure these dependencies are fulfilled at any point in time is to traverse the $Lp$ diagram such that each diagonal defined by $i+p:=\text{const}$ carries out computations in a ``bottom-right to top-left" fashion for increasing values of ``const" (i.e., $i+p=1, i+p=2, \dots$). This execution order is given by the numbered boxed in Figure~\ref{fig:lp_basic}, and emphasized by the highlighted diagonal for $i+p=6$. 

\begin{figure*}[tbp]
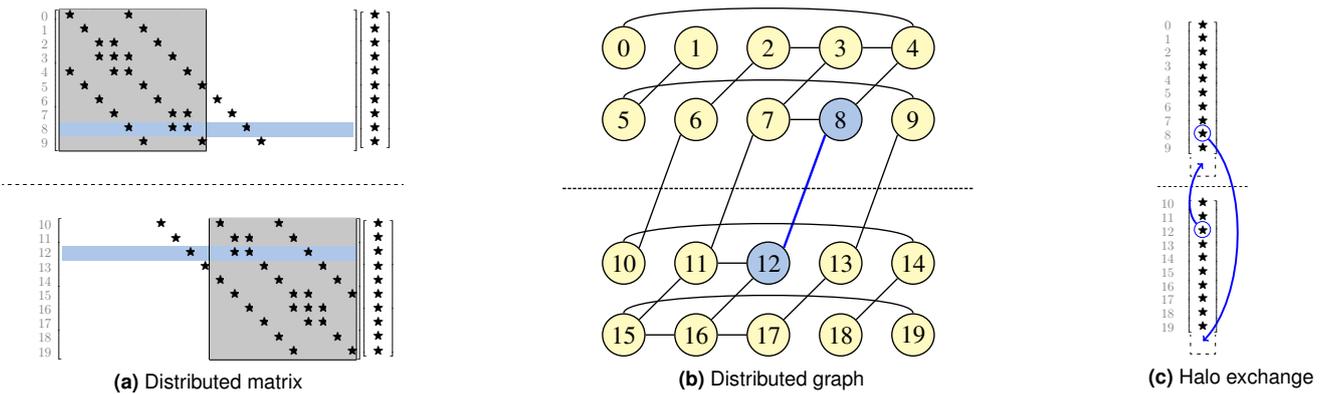

	\centering
	\captionsetup[subfigure]{justification=centering}
	\subfloat[Distributed matrix\label{fig:Distributed adjacency matrix}]{\raisebox{1.75mm}
		{\resizebox{.3\linewidth}{!}{%
	%  \tikzsetnextfilename{#2}%
	\input{figures/dist_explaination/seg_matrix.tex}%
%\includegraphics{tikz_cache/paper-figure\theplotCtr.pdf}
%\stepcounter{plotCtr}
}}
	}
	\hfill
	\subfloat[Distributed graph\label{fig:Distributed graph}]{\raisebox{-20.75mm}
		{\resizebox{.325\linewidth}{!}{%
	%  \tikzsetnextfilename{#2}%
	%\begin{tikzpicture}[node distance={15mm}, thick, main/.style = {draw, circle}]
%	\node[main] (0) {0};
%	\node[main] (1) [above right of=0] {1};
%	\node[main] (2) [above right of=1] {2};
%	\node[main, fill = lblue] (3) [above left of=2] {3};
%	\node[main] (4) [above left of=3] {4};
%	\node[main, fill = lblue] (5) [below left of=4] {5};
%	\node[main, fill = lorange] (6) [below left of=5] {6};
%	\node[main, fill = lblue] (7) [below right of=6] {7};
%	
%	%       \draw[<-] (0) to [out=0+240, in=0+300, looseness=5](0);
%	\draw[-] (0) to [out=90, in=270, looseness=1](3);
%	%       \draw[<-] (1) to [out=45+240, in=45+300, looseness=5](1);
%	\draw[-] (1) to [out=90+30, in=270, looseness=1](4);
%	\draw[-] (2) to [out=180+60, in=30, looseness=1](1);
%	%       \draw[<-] (3) to [out=45+330, in=45+30, looseness=5](3);
%	\draw[-, blue] (3) to [out=180+15+5, in=-20, looseness=1](5);
%	\draw[-] (4) to [out=60+180, in=45-25, looseness=1](5);
%	%       \draw[<-] (5) to [out=45+60, in=45+120, looseness=5](5);
%	\draw[-, orange] (5) to [out=200-15, in=90-15](6);
%	\draw[-] (6) to [out=15, in=180-15, looseness=1](2);
%	\draw[-] (6) to [out=270+45-25, in=90+45+25, looseness=1](7);
%	%       \draw[<-] (7) to [out=45+150, in=45+210, looseness=5](7);
%	\draw[-, blue] (7) to [out=45-15, in=45+180+15, looseness=1](3);
%	
%	\draw[line width=.25mm, dashed] let \p1 = (0), \p2 = (4) in (\x1 - 15, \y1 - 20) -- (\x2 - 15, \y2 + 10);% node [above](textNode){MPI Boundary};
%\end{tikzpicture}

\edef\scaleFac{0.75}
%\begin{document}
\begin{tikzpicture}[darkstyle/.style={circle,draw,fill=yellow!30!white,minimum size=12.5, inner sep=0.5pt}, stencilstyle/.style={circle,draw,red,fill=red!30!white,minimum size=13.2, inner sep=0.5pt}, 
	yscale=-1]

	\node [darkstyle]  (00) at (\scaleFac*0,\scaleFac*0) {\fontsize{7.4}{9}\selectfont{0}}; 
	\node [darkstyle]  (10) at (\scaleFac*1,\scaleFac*0) {\fontsize{7.4}{9}\selectfont{1}};
	\node [darkstyle]  (20) at (\scaleFac*2,\scaleFac*0) {\fontsize{7.4}{9}\selectfont{2}};
	\node [darkstyle]  (30) at (\scaleFac*3,\scaleFac*0) {\fontsize{7.4}{9}\selectfont{3}};
	\node [darkstyle]  (40) at (\scaleFac*4,\scaleFac*0) {\fontsize{7.4}{9}\selectfont{4}};
	\node [darkstyle]  (01) at (\scaleFac*0,\scaleFac*1) {\fontsize{7.4}{9}\selectfont{5}};
	\node [darkstyle]  (11) at (\scaleFac*1,\scaleFac*1) {\fontsize{7.4}{9}\selectfont{6}};
	\node [darkstyle]  (21) at (\scaleFac*2,\scaleFac*1) {\fontsize{7.4}{9}\selectfont{7}};
	\node [darkstyle, fill=lblue]  (31) at (\scaleFac*3,\scaleFac*1) {\fontsize{7.4}{9}\selectfont{8}};
	\node [darkstyle]  (41) at (\scaleFac*4,\scaleFac*1) {\fontsize{7.4}{9}\selectfont{9}};
	
	\node [darkstyle]  (02) at (\scaleFac*0,\scaleFac*3) {\fontsize{7.4}{9}\selectfont{10}}; 
	\node [darkstyle]  (12) at (\scaleFac*1,\scaleFac*3) {\fontsize{7.4}{9}\selectfont{11}};
	\node [darkstyle, fill=lblue]  (22) at (\scaleFac*2,\scaleFac*3) {\fontsize{7.4}{9}\selectfont{12}};
	\node [darkstyle]  (32) at (\scaleFac*3,\scaleFac*3) {\fontsize{7.4}{9}\selectfont{13}};
	\node [darkstyle]  (42) at (\scaleFac*4,\scaleFac*3) {\fontsize{7.4}{9}\selectfont{14}};
	\node [darkstyle]  (03) at (\scaleFac*0,\scaleFac*4) {\fontsize{7.4}{9}\selectfont{15}};
	\node [darkstyle]  (13) at (\scaleFac*1,\scaleFac*4) {\fontsize{7.4}{9}\selectfont{16}};
	\node [darkstyle]  (23) at (\scaleFac*2,\scaleFac*4) {\fontsize{7.4}{9}\selectfont{17}};
	\node [darkstyle]  (33) at (\scaleFac*3,\scaleFac*4) {\fontsize{7.4}{9}\selectfont{18}};
	\node [darkstyle]  (43) at (\scaleFac*4,\scaleFac*4) {\fontsize{7.4}{9}\selectfont{19}};
	\draw[-] (10) to [out=180+45, in=45, looseness=0](01);
	\draw[-] (20) to [out=180+45, in=45, looseness=0](11);
	\draw[-] (30) to [out=180+45, in=45, looseness=0](21);
	\draw[-] (40) to [out=180+45, in=45, looseness=0](31);
	\draw[-] (20) to [out=0, in=180, looseness=0](30);
	\draw[-] (30) to [out=0, in=180, looseness=0](40);
	
	\draw[-] (11) to [out=180+45, in=45, looseness=0](02);
	\draw[-] (21) to [out=180+45, in=45, looseness=0](12);
	\draw[-, draw=blue, line width=.25mm] (31) to [out=180+45, in=45, looseness=0](22);
	\draw[-] (41) to [out=180+45, in=45, looseness=0](32);
	\draw[-] (21) to [out=0, in=180, looseness=0](31);
	
	\draw[-] (12) to [out=180+45, in=45, looseness=0](03);
	\draw[-] (22) to [out=180+45, in=45, looseness=0](13);
	\draw[-] (32) to [out=180+45, in=45, looseness=0](23);
	\draw[-] (42) to [out=180+45, in=45, looseness=0](33);
	\draw[-] (12) to [out=0, in=180, looseness=0](22);	
	
	\draw[-] (03) to [out=0, in=180, looseness=0](13);
	\draw[-] (13) to [out=0, in=180, looseness=0](23);
	
	\draw[-] (40) to [out=-90, in=-90, looseness=.6](00);
	\draw[-] (41) to [out=-90, in=-90, looseness=.6](01);
	\draw[-] (42) to [out=-90, in=-90, looseness=.6](02);
	\draw[-] (43) to [out=-90, in=-90, looseness=.6](03);
	
	\draw[line width=.15mm, dash pattern=on 1pt off 0.5pt] let \p1 = (01), \p2 = (41) in (\x1-18, \y1+20.5) -- (\x2+18, \y2+20.5);
	
\end{tikzpicture}%
%\includegraphics{tikz_cache/paper-figure\theplotCtr.pdf}
%\stepcounter{plotCtr}
}}
	}
	\hfill
	\subfloat[Halo exchange\label{fig:Halo Comm}]{\raisebox{1.5mm}
		{\resizebox{.143\linewidth}{!}{%
	%  \tikzsetnextfilename{#2}%
	\input{figures/dist_explaination/halo_comm.tex}%
%\includegraphics{tikz_cache/paper-figure\theplotCtr.pdf}
%\stepcounter{plotCtr}
}}
	}
	\protect\caption{
		The global matrix $A$ from Figure~\ref{fig:stencil_matrix} and some RHS vector $x$ are partitioned in a row-wise manner over two MPI processes in (a). The gray boxed-out regions show, on each MPI process, which elements are ``local" (inside the gray region).
		The edge corresponding to the remote data dependency, i.e., the edge crossing the MPI boundary, is highlighted in blue in (b).
		The rows at global indices $8$ and $12$ are highlighted as examples of rows which contain remote data dependencies for the SpMV. In order to fulfill these data dependency, another MPI process must supply the appropriate ``halo elements." Shown in (c) is the process of data exchange on the x-vector for our two example rows, where incoming halo elements are received into an appropriately resized buffer.}
	\label{fig:Distributing matrix data over two processes}
\end{figure*}

With the aid of the $Lp$ diagram, the idea behind level-based cache blocking can now be briefly introduced. As LB-MPK diagonally traverses the levels as described above, levels (and therefore matrix entries) are reused after $p_m + 1$ execution steps (after the wind-up phase on the left end, and before the wind-down phase at the right end of the $Lp$ diagram). If all the non-zero matrix entries associated with these $p_m + 1$ levels accessed between two computations of the same $L(i)$ can be held in cache, then all matrix data for the following computation with $L(i)$ will be accessed from cache (with the exception of $p = 1$, which has a compulsory cache miss and must come from main memory). As an explicit example, see the level $L(5)$ which is used in the 15th step in the execution of LB-MPK. If all the matrix data corresponding to the six levels $L(1)$--$L(6)$ can be held in cache, then the vertices of $L(5)$ are reused in the 21st step in the execution of LB-MPK when computing $p=2$. 

\section{Challenges in the Distributed Setting}
\label{sec:Distributed_Challenges}

Distributing MPK for level-based cache blocking across multiple MPI processes is not as easy as just executing LB-MPK locally on each MPI process.
To understand this non-triviality, we first investigate the dependencies that arise from the ``traditional'' distributed MPK (TRAD).
Just as in the shared memory setting, a distributed MPK is traditionally constructed from back-to-back SpMV invocations.

In the distributed setting the matrix is partitioned among the available MPI processes.
The conventional approach, which we use, often employs row-based partitioning wherein both matrix and vector entries corresponding to a subset of rows are physically assigned to individual MPI processes.
Figure~\ref{fig:Distributed adjacency matrix} illustrates distributing the matrix $A$ from Figure~\ref{fig:stencil_matrix} across two MPI processes.
The dotted lines represent the MPI ``boundary," i.e., where the data is physically disjointed.
The corresponding graph of the matrix ($G(A)$) in the distributed setting is shown in Figure~\ref{fig:Distributed graph}.
The crux of the problem lies in the distributed nature of the x-vector.
During SpMV computations on a given MPI process, there may be non-zero matrix elements that do not have their corresponding RHS x-vector elements for the dot product locally on the process, necessitating their retrieval from remote MPI processes.
For instance, in Figure~\ref{fig:Distributed adjacency matrix}, row $8$ belonging to the first MPI process contains a non-zero element at column index $12$.
While the non-zero elements with column indices $4$, $7$, and $8$ in row $8$ can be multiplied with local x-vector data for the corresponding dot product, the non-zero element at column index $12$ lacks the requisite data on this MPI process's x-vector and must be fetched from the second MPI process.
Similarly, on the second MPI process, row index $12$ necessitates x-vector data corresponding to row index $8$, which resides on the first MPI process.

\begin{algorithm}[b]
	\scriptsize 
	\caption{Traditional Distributed MPK}\label{alg:Traditional Distributed Matrix Power Kernel}
	\begin{algorithmic}
		\Require{ \\
			$\texttt{double}\ x[N_{r,i}+N_{h,i}]$;\\
			$\texttt{sparseMatrix}\ A_i$\;\\
			$\texttt{int}\ p_m$\;
		}
		\Ensure{ \\
			$\texttt{double}\ y[N_{r,i}+N_{h,i}, p_m]$\; \\ \ \\
		}
		$y[:, 0] \gets x$\texttt{;}\;
		\For{$p \gets 1,\dots,p_m$} \\
		\qquad $y[:, p-1] \gets$ \texttt{haloComm}($y[:, p-1]$)\texttt{;}\; \\
		\qquad $y[:, p] \gets$ \texttt{SpMV}($y[:, p-1], A_i[:,:]$)\texttt{;}\;
		\EndFor
	\end{algorithmic}
\end{algorithm}
Transferring remote elements on-demand is feasible, but would result in significant performance overhead due to the high latency of MPI communications.
Consequently, a common strategy involves bulk transfer of all required remote elements before executing SpMV operations. 
These elements are then stored consecutively, typically at the end of the x-vector, forming what is commonly known as the ``halo region/buffer."
Figure~\ref{fig:Halo Comm} illustrates this halo region and the process of populating it with remote elements.
Algorithm~\ref{alg:Traditional Distributed Matrix Power Kernel} presents the pseudocode for traditional distributed MPK computing $A^{p_m}x$. Here, we assume a matrix $A$ has already been partitioned row-wise and distributed to each of the $n$ processes so that $A_i$ resides on process $i$.
The algorithm utilizes two subroutines: the \texttt{haloComm} routine that populates the halo region, and the sparse matrix-vector product \texttt{SpMV}. The local vector size is the local number of rows $N_{r,i}$, plus the number of remote elements the process $i$ needs to receive into its halo buffer $N_{h,i}$.
The ``MPI overhead" $O_{\text{MPI}}$ is understood to be the ratio of these halo rows on each x-vector across all the $n$ MPI processes to the total number of rows $N_r$\,($=\sum_{i=0}^n N_{i,r}$), i.e., 
\begin{equation}
  O_{\text{MPI}} := \frac{\sum_{i=0}^n N_{h,i}}{N_r}.\label{eq:mpi_oheads}
\end{equation}

\begin{figure*}[tbp]
	\centering
	\captionsetup[subfigure]{justification=centering}
	\subfloat[TRAD\label{fig:dot diagram trad_mpk}]{%
		\resizebox{.3\linewidth}{!}{%
	%  \tikzsetnextfilename{#2}%
	\edef\scaleFac{0.6}
\begin{tikzpicture}[darkstyle/.style={circle,draw,fill=yellow!30!white,minimum size=12.5, inner sep=0.5pt},     
	group0/.style={circle,fill=black, opacity = .75,minimum size=12.5, inner sep=0.5pt}, 
	group1/.style={circle,draw,blue,fill=blue!30!white,minimum size=12.5, inner sep=0.5pt},
	group2/.style={circle,draw,orange,fill=orange!30!white,minimum size=12.5, inner sep=0.5pt},
	group3/.style={circle,draw,green,fill=green!30!white,minimum size=12.5, inner sep=0.5pt},
	group4/.style={circle,draw,brown,fill=brown!30!white,minimum size=12.5, inner sep=0.5pt},
	group5/.style={circle,draw,cyan,fill=cyan!30!white,minimum size=12.5, inner sep=0.5pt},
	buffer/.style={circle,draw,gray,fill=gray!30!white,minimum size=12.5, inner sep=0.5pt},
	stencilstyle/.style={circle,draw,red,fill=red!30!white,minimum size=12.5, inner sep=0.5pt},
	yscale=-1]
	
	\foreach \x in {0,...,8}
	\foreach \y in {0,...,3} 
	{\pgfmathtruncatemacro{\label}{\y*8 + \x  }
		\node [darkstyle]  (\x\y) at (\scaleFac*\x,\scaleFac*\y) {};}
	%	\node [darkstyle]  (\x\y) at (\scaleFac*\x,\scaleFac*\y) {\x\y};}

%	\draw[->] (-0.5,-0.5) -- (1.5,-0.5);
%	\node at (1.5+0.25,-0.5) {$x_i$};
%	\draw[->] (-0.5,-0.5) -- (-0.5,1.5);
%	\node at (-0.5,1.5+0.25) {$p$};

	\draw[line width=.25mm, dashed] let \p1 = (40), \p2 = (43) in (\x1+8.5, \y1-8) -- (\x2+8.5, \y2+8);
	
	% y-axis labels
	\node [draw=none, fill=none, left of = 00, node distance = 6mm] (P3-label) {$3$};
	\node [draw=none, fill=none, left of = 01, node distance = 6mm] (P2-label) {$2$};
	\node [draw=none, fill=none, left of = 02, node distance = 6mm] (P1-label) {$1$};
	\node [draw=none, fill=none, left of = 03, node distance = 6mm] (P0-label) {$0$};
	
	% x-axis labels
	\node [draw=none, fill=none, below of = 03, node distance = 6mm] (X0-label) {$0$};
	\node [draw=none, fill=none, below of = 13, node distance = 6mm] (X1-label) {$1$};
	\node [draw=none, fill=none, below of = 23, node distance = 6mm] (X2-label) {$2$};
	\node [draw=none, fill=none, below of = 33, node distance = 6mm] (X3-label) {$3$};
	\node [draw=none, fill=none, above of = 40, node distance = 6mm] (B-label) {$B$};
	\node [draw=none, fill=none, below of = 53, node distance = 6mm] (X4-label) {$4$};
	\node [draw=none, fill=none, below of = 63, node distance = 6mm] (X5-label) {$5$};
	\node [draw=none, fill=none, below of = 73, node distance = 6mm] (X6-label) {$6$};
	\node [draw=none, fill=none, below of = 83, node distance = 6mm] (X7-label) {$7$};

	% Label axes
	\node [draw=none, fill=none, below of = X0-label, node distance = 4mm] (x-label-start) {};
	\node [draw=none, fill=none, below of = X1-label, node distance = 4mm] (x-label-end) {};
	\node [draw=none, fill=none, left of = P0-label, node distance = 4mm] (y-label-start) {};
	\node [draw=none, fill=none, left of = P1-label, node distance = 4mm] (y-label-end) {};
	\node [draw=none, fill=none, left of = P1-label, node distance = 4mm] (y-label-end) {};

	\node (labels-start) at ($(y-label-start.center)+(0,1)$) {};
	\draw[->] let \p1=(labels-start.center) in (\x1,\y1) -- (\x1+40,\y1);
	\draw[->] let \p1=(labels-start.center) in (\x1,\y1) -- (\x1,\y1-40);
	\node at ($(y-label-start.center)+(0,-0.73)$) {$p$};
	\node at ($(y-label-start.center)+(1.65,1)$) {$x$};

	% Halo buffer

	\draw [rounded corners, gray,fill=gray!10!white]  let \p1 = (43), \p2 = (40) in (\x1-7, \y1+7) rectangle (\x2+7, \y2-7) {};
	\node [buffer] at (41) {};
	\node [buffer] at (42) {};
	\node [buffer] at (43) {};

	% groups
	\node [group0] at (03) {};
	\node [group0] at (13) {};
	\node [group0] at (23) {};
	\node [group0] at (33) {};
	
	\node [group1] at (02) {};
	\node [group1] at (12) {};
	\node [group1] at (22) {};
	\node [group1] at (32) {};
	
	\node [group2] at (01) {};
	\node [group2] at (11) {};
	\node [group2] at (21) {};
	\node [group2] at (31) {};
	
	\node [group3] at (00) {};
	\node [group3] at (10) {};
	\node [group3] at (20) {};
	\node [group3] at (30) {};
	
	% Give execution order
	\node [text=black] at (\scaleFac*4,\scaleFac*3) {$0$};
	
%	\foreach \x in {0,...,3}
%	\foreach \y in {2} 
%	{\node [text=black] at (\scaleFac*\x,\scaleFac*\y) {$1$};}
	\node [text=black] at (0*\scaleFac,2*\scaleFac) {$1$};
	\node [text=black] at (1*\scaleFac,2*\scaleFac) {$2$};
	\node [text=black] at (2*\scaleFac,2*\scaleFac) {$3$};
	\node [text=black] at (3*\scaleFac,2*\scaleFac) {$4$};
	
	\node [text=black] at (\scaleFac*4,\scaleFac*2) {$5$};
	
%	\foreach \x in {0,...,3}
%	\foreach \y in {1} 
%	{\node [text=black] at (\scaleFac*\x,\scaleFac*\y) {$3$};}	
%	
	\node [text=black] at (0*\scaleFac,1*\scaleFac) {$6$};
	\node [text=black] at (1*\scaleFac,1*\scaleFac) {$7$};
	\node [text=black] at (2*\scaleFac,1*\scaleFac) {$8$};
	\node [text=black] at (3*\scaleFac,1*\scaleFac) {$9$};
	
	\node [text=black] at (\scaleFac*4,\scaleFac*1) {$10$};
	
%	\foreach \x in {0,...,3}
%	\foreach \y in {0} 
%	{\node [text=black] at (\scaleFac*\x,\scaleFac*\y) {$5$};}	
	\node [text=black] at (0*\scaleFac,0*\scaleFac) {$11$};
	\node [text=black] at (1*\scaleFac,0*\scaleFac) {$12$};
	\node [text=black] at (2*\scaleFac,0*\scaleFac) {$13$};
	\node [text=black] at (3*\scaleFac,0*\scaleFac) {$14$};	
	% Halo lines
	\draw[line width=.35mm, <-, draw=blue] let \p1=(43), \p2=(53) in (\x1+2,\y1) -- (\x2-2,\y2);
	\draw[line width=.35mm, <-, draw=blue] let \p1=(42), \p2=(52) in (\x1+2,\y1) -- (\x2-2,\y2);
	\draw[line width=.35mm, <-, draw=blue] let \p1=(41), \p2=(51) in (\x1+2,\y1) -- (\x2-2,\y2);
	
	\draw[line width=.35mm, <-, red] let \p1=(20), \p2=(11) in (\x1-2,\y1+2) -- (\x2+2,\y2-2);
	\draw[line width=.35mm, <-, red] let \p1=(20), \p2=(21) in (\x1,\y1+2) -- (\x2,\y2-2);
	\draw[line width=.35mm, <-, red] let \p1=(20), \p2=(31) in (\x1+2,\y1+2) -- (\x2-2,\y2-2);

\end{tikzpicture}%
%\includegraphics{tikz_cache/paper-figure\theplotCtr.pdf}
%\stepcounter{plotCtr}
}
	}
	\hfill
	\subfloat[CA-MPK\label{fig:dot diagram hoemmen_mpk}]{%
		\resizebox{.35\linewidth}{!}{%
	%  \tikzsetnextfilename{#2}%
	\edef\scaleFac{0.6}
\begin{tikzpicture}[darkstyle/.style={circle,draw,fill=yellow!30!white,minimum size=12.5, inner sep=0.75pt},     
	group0/.style={circle,fill=black, opacity = .75,minimum size=12.5, inner sep=0.5pt}, 
	group1/.style={circle,draw,blue,fill=blue!30!white,minimum size=12.5, inner sep=0.5pt},
	group2/.style={circle,draw,orange,fill=orange!30!white,minimum size=12.5, inner sep=0.5pt},
	group3/.style={circle,draw,green,fill=green!30!white,minimum size=12.5, inner sep=0.5pt},
	group4/.style={circle,draw,brown,fill=brown!30!white,minimum size=12.5, inner sep=0.5pt},
	group5/.style={circle,draw,cyan,fill=cyan!30!white,minimum size=12.5, inner sep=0.5pt},
	buffer/.style={circle,draw,gray,fill=gray!30!white,minimum size=12.5, inner sep=0.5pt},
	stencilstyle/.style={circle,draw,violet,fill=violet!30!white,minimum size=12.5, inner sep=0.5pt},
	yscale=-1]
	
	\foreach \x in {0,...,10}
	\foreach \y in {0,...,3} 
	{\pgfmathtruncatemacro{\label}{\y*8 + \x  }
		\node [darkstyle]  (\x\y) at (\scaleFac*\x,\scaleFac*\y) {};}
	%	\node [darkstyle]  (\x\y) at (\scaleFac*\x,\scaleFac*\y) {\x\y};}

\draw[line width=.25mm, dashed] let \p1 = (60), \p2 = (63) in (\x1+8.5, \y1-7.5) -- (\x2+8.5, \y2+7.5);

% y-axis labels
\node [draw=none, fill=none, left of = 00, node distance = 6mm] (P3-label) {$3$};
\node [draw=none, fill=none, left of = 01, node distance = 6mm] (P2-label) {$2$};
\node [draw=none, fill=none, left of = 02, node distance = 6mm] (P1-label) {$1$};
\node [draw=none, fill=none, left of = 03, node distance = 6mm] (P0-label) {$0$};

% x-axis labels
\node [draw=none, fill=none, below of = 03, node distance = 6mm] (X0-label) {$0$};
\node [draw=none, fill=none, below of = 13, node distance = 6mm] (X1-label) {$1$};
\node [draw=none, fill=none, below of = 23, node distance = 6mm] (X2-label) {$2$};
\node [draw=none, fill=none, below of = 33, node distance = 6mm] (X3-label) {$3$};
\node [draw=none, fill=none, below of = 43, node distance = 6mm] (B1-label) {};
\node [draw=none, fill=none, above of = 40, node distance = 6mm] (B2-label) {$B_{}$};
\node [draw=none, fill=none, above of = 50, node distance = 6mm] (E1-label) {$E_1$};
\node [draw=none, fill=none, above of = 60, node distance = 6mm] (E2-label) {$E_2$};
\node [draw=none, fill=none, below of = 63, node distance = 6mm] (B3-label) {};
\node [draw=none, fill=none, below of = 73, node distance = 6mm] (X4-label) {$4$};
\node [draw=none, fill=none, below of = 83, node distance = 6mm] (X5-label) {$5$};
\node [draw=none, fill=none, below of = 93, node distance = 6mm] (X6-label) {$6$};
\node [draw=none, fill=none, below of = 103, node distance = 6mm] (X7-label) {$7$};

% Halo buffer
\draw [rounded corners, gray,fill=gray!10!white]  let \p1 = (43), \p2 = (60) in (\x1-7, \y1+7.5) rectangle (\x2+7, \y2-7.5) {};

\node [stencilstyle] at (41) {};
%\node [draw=white, fill=white, overlay,minimum size=10, inner sep=0.5pt] at (51) {};
%\node [draw=white, fill=white, overlay,minimum size=10, inner sep=0.5pt] at (61) {};
\node [stencilstyle] at (42) {};
\node [stencilstyle] at (52) {};
%\node [draw=white, fill=white, overlay,minimum size=10, inner sep=0.5pt] at (62) {};
\node [buffer] at (63) {};
\node [buffer] at (53) {};
\node [buffer] at (43) {};

% Halo lines
\draw[line width=.35mm, <-, draw=blue] let \p1=(53), \p2=(83) in (\x1+2,\y1+2) to[out=90-60,in=180-20] (\x2-2,\y2+2);
\draw[line width=.35mm, <-, draw=blue] let \p1=(63), \p2=(93) in (\x1+2,\y1+2) to[out=90-60,in=180-20] (\x2-2,\y2+2);
\draw[line width=.35mm, <-, draw=blue] let \p1=(43), \p2=(73) in (\x1+2,\y1+2) to[out=90-60,in=180-20] (\x2-2,\y2+2);

% Spacing Hack
\node [draw=none, fill=none, left of = P0-label, node distance = 4mm] (y-label-start) {};
\phantom{
	\node (labels-start) at ($(y-label-start.center)+(0,1)$) {};
	\draw[->] let \p1=(labels-start.center) in (\x1,\y1) -- (\x1+40,\y1);
	\draw[->] let \p1=(labels-start.center) in (\x1,\y1) -- (\x1,\y1-40);
	\node at ($(y-label-start.center)+(0,-0.73)$) {$p$};
	\node at ($(y-label-start.center)+(1.65,1)$) {$x_i$};
}

%Outline local elems

\draw [rounded corners, lblue, fill=lblue!60!white]  let \p1 = (02), \p2 = (00) in (\x1-7, \y1+7) rectangle (\x2+7, \y2-7) {};
\draw [rounded corners, lblue, fill=lblue!60!white]  let \p1 = (02), \p2 = (11) in (\x1-7, \y1+7) rectangle (\x2+7, \y2-7) {};
\draw [rounded corners, lblue, fill=lblue!60!white]  let \p1 = (02), \p2 = (22) in (\x1-7, \y1+7) rectangle (\x2+7, \y2-7) {};
\draw [rounded corners, lblue, fill=lblue!60!white, draw=none]  let \p1 = (02), \p2 = (11) in (\x1-7, \y1+7) rectangle (\x2+7, \y2-7) {};
\draw [rounded corners, lblue, fill=lblue!60!white, draw=none]  let \p1 = (02), \p2 = (00) in (\x1-7, \y1+7) rectangle (\x2+7, \y2-7) {};

% Groups
\node [group0] at (03) {};
\node [group0] at (13) {};
\node [group0] at (23) {};
\node [group0] at (33) {};

\node [group1] at (00) {};
\node [group1] at (01) {};
\node [group1] at (02) {};
\node [group1] at (11) {};
\node [group1] at (02) {};
\node [group1] at (12) {};
\node [group1] at (22) {};
\node [group1] at (11) {};

\node [group2] at (32) {};

\node [group2] at (31) {};
\node [group2] at (21) {};

\node [group2] at (30) {};
\node [group2] at (20) {};
\node [group2] at (10) {};

%\node [group1] at (03) {};
%\node [group1] at (02) {};
%\node [group1] at (01) {};
%\node [group1] at (00) {};
%\node [group1] at (11) {};
%\node [group1] at (22) {};
%\node [group1] at (23) {};
%\node [group1] at (12) {};
%\node [group1] at (13) {};
%\node [group1] at (03) {};
%
%\node [group2] at (20) {};
%\node [group2] at (11) {};
%\node [group2] at (02) {};
%
%
%\node [group3] at (01) {};
%\node [group3] at (10) {};
%
%\node [group4] at (00) {};

	% Give execution order
\node [text=black] at (\scaleFac*4,\scaleFac*3) {$0$};
\node [text=black] at (\scaleFac*5,\scaleFac*3) {$1$};
\node [text=black] at (\scaleFac*6,\scaleFac*3) {$2$};

\node [text=black] at (\scaleFac*0,\scaleFac*2) {$3$};
\node [text=black] at (\scaleFac*1,\scaleFac*2) {$4$};
\node [text=black] at (\scaleFac*2,\scaleFac*2) {$6$};
\node [text=black] at (\scaleFac*0,\scaleFac*1) {$5$};
\node [text=black] at (\scaleFac*1,\scaleFac*1) {$7$};
\node [text=black] at (\scaleFac*0,\scaleFac*0) {$8$};

\node [text=black] at (\scaleFac*3,\scaleFac*2) {$9$};
\node [text=black] at (\scaleFac*4,\scaleFac*2) {$12$};
\node [text=black] at (\scaleFac*5,\scaleFac*2) {$15$};

\node [text=black] at (\scaleFac*2,\scaleFac*1) {$10$};
\node [text=black] at (\scaleFac*3,\scaleFac*1) {$13$};
\node [text=black] at (\scaleFac*4,\scaleFac*1) {$16$};

\node [text=black] at (\scaleFac*1,\scaleFac*0) {$11$};
\node [text=black] at (\scaleFac*2,\scaleFac*0) {$14$};
\node [text=black] at (\scaleFac*3,\scaleFac*0) {$17$};

% Recompute lines
\draw[violet, line width=.35mm, <-] let \p1=(42), \p2=(53) in (\x1+2,\y1+2) -- (\x2-2,\y2-2);
\draw[violet, line width=.35mm, <-] let \p1=(52), \p2=(63) in (\x1+2,\y1+2) -- (\x2-2,\y2-2);
\draw[violet, line width=.35mm, <-] let \p1=(41), \p2=(52) in (\x1+2,\y1+2) -- (\x2-2,\y2-2);

\draw[line width=.35mm, <-, red] let \p1=(20), \p2=(11) in (\x1-2,\y1+2) -- (\x2+2,\y2-2);
\draw[line width=.35mm, <-, red] let \p1=(20), \p2=(21) in (\x1,\y1+2) -- (\x2,\y2-2);
\draw[line width=.35mm, <-, red] let \p1=(20), \p2=(31) in (\x1+2,\y1+2) -- (\x2-2,\y2-2);

\end{tikzpicture}%
%\includegraphics{tikz_cache/paper-figure\theplotCtr.pdf}
%\stepcounter{plotCtr}
}
	}
	\hfill
	\subfloat[DLB-MPK\label{fig:dot diagram dlb_mpk}]{%
		\resizebox{.3\linewidth}{!}{%
	%  \tikzsetnextfilename{#2}%
	\edef\scaleFac{0.6}
\begin{tikzpicture}[darkstyle/.style={circle,draw,fill=yellow!30!white,minimum size=12.5, inner sep=0.75pt},     
	group0/.style={circle,fill=black, opacity = .75,minimum size=12.5, inner sep=0.5pt}, 
	group1/.style={circle,draw,blue,fill=blue!30!white,minimum size=12.5, inner sep=0.5pt},
	group2/.style={circle,draw,orange,fill=orange!30!white,minimum size=12.5, inner sep=0.5pt},
	group3/.style={circle,draw,green,fill=green!30!white,minimum size=12.5, inner sep=0.5pt},
	group4/.style={circle,draw,brown,fill=brown!30!white,minimum size=12.5, inner sep=0.5pt},
	group5/.style={circle,draw,cyan,fill=cyan!30!white,minimum size=12.5, inner sep=0.5pt},
	buffer/.style={circle,draw,gray,fill=gray!30!white,minimum size=12.5, inner sep=0.5pt},
	stencilstyle/.style={circle,draw,red,fill=red!30!white,minimum size=12.5, inner sep=0.5pt},
	yscale=-1]
	
	\foreach \x in {0,...,8}
	\foreach \y in {0,...,3} 
	{\pgfmathtruncatemacro{\label}{\y*8 + \x  }
		\node [darkstyle]  (\x\y) at (\scaleFac*\x,\scaleFac*\y) {};}
	%	\node [darkstyle]  (\x\y) at (\scaleFac*\x,\scaleFac*\y) {\x\y};}

\draw[line width=.25mm, dashed] let \p1 = (40), \p2 = (43) in (\x1+8.5, \y1-8) -- (\x2+8.5, \y2+8);

% y-axis labels
\node [draw=none, fill=none, left of = 00, node distance = 6mm] (P3-label) {$3$};
\node [draw=none, fill=none, left of = 01, node distance = 6mm] (P2-label) {$2$};
\node [draw=none, fill=none, left of = 02, node distance = 6mm] (P1-label) {$1$};
\node [draw=none, fill=none, left of = 03, node distance = 6mm] (P0-label) {$0$};

% x-axis labels
\node [draw=none, fill=none, below of = 03, node distance = 6mm] (X0-label) {$0$};
\node [draw=none, fill=none, below of = 13, node distance = 6mm] (X1-label) {$1$};
\node [draw=none, fill=none, below of = 23, node distance = 6mm] (X2-label) {$2$};
\node [draw=none, fill=none, below of = 33, node distance = 6mm] (X3-label) {$3$};
\node [draw=none, fill=none, above of = 40, node distance = 6mm] (B-label) {$B_{}$};
\node [draw=none, fill=none, above of = 30, node distance = 6mm] (I1-label) {$I_1$};
\node [draw=none, fill=none, above of = 20, node distance = 6mm] (I2-label) {$I_2$};
\node [draw=none, fill=none, below of = 53, node distance = 6mm] (X4-label) {$4$};
\node [draw=none, fill=none, below of = 63, node distance = 6mm] (X5-label) {$5$};
\node [draw=none, fill=none, below of = 73, node distance = 6mm] (X6-label) {$6$};
\node [draw=none, fill=none, below of = 83, node distance = 6mm] (X7-label) {$7$};

% Halo buffer

\draw [rounded corners, gray,fill=gray!10!white]  let \p1 = (43), \p2 = (40) in (\x1-7, \y1+7) rectangle (\x2+7, \y2-7) {};
\node [buffer] at (41) {};
\node [buffer] at (42) {};
\node [buffer] at (43) {};

% Cache blocked groups
\node [group0] at (03) {};
\node [group0] at (13) {};
\node [group0] at (23) {};
\node [group0] at (33) {};

\node [group1] at (32) {};
\node [group1] at (21) {};
\node [group1] at (10) {};
\node [group1] at (00) {};
\node [group1] at (01) {};
\node [group1] at (02) {};
\node [group1] at (12) {};
\node [group1] at (11) {};
\node [group1] at (22) {};

\node [group2] at (31) {};
\node [group2] at (20) {};

\node [group3] at (30) {};

% Spacing Hack
\node [draw=none, fill=none, left of = P0-label, node distance = 4mm] (y-label-start) {};
\phantom{
	\node (labels-start) at ($(y-label-start.center)+(0,1)$) {};
	\draw[->] let \p1=(labels-start.center) in (\x1,\y1) -- (\x1+40,\y1);
	\draw[->] let \p1=(labels-start.center) in (\x1,\y1) -- (\x1,\y1-40);
	\node at ($(y-label-start.center)+(0,-0.73)$) {$p$};
	\node at ($(y-label-start.center)+(1.65,1)$) {$x_i$};
}

	\node [text=black] at (\scaleFac*4,\scaleFac*1) {$13$};
	\node [text=black] at (\scaleFac*4,\scaleFac*2) {$10$};
	\node [text=black] at (\scaleFac*4,\scaleFac*3) {$0$};
	
	\node [text=black] at (\scaleFac*0,\scaleFac*2) {$1$};
	\node [text=black] at (\scaleFac*1,\scaleFac*2) {$2$};
	\node [text=black] at (\scaleFac*2,\scaleFac*2) {$4$};
	\node [text=black] at (\scaleFac*3,\scaleFac*2) {$7$};
	\node [text=black] at (\scaleFac*0,\scaleFac*1) {$3$};
	\node [text=black] at (\scaleFac*1,\scaleFac*1) {$5$};
	\node [text=black] at (\scaleFac*2,\scaleFac*1) {$8$};
	\node [text=black] at (\scaleFac*0,\scaleFac*0) {$6$};
	\node [text=black] at (\scaleFac*1,\scaleFac*0) {$9$};
	
	\node [text=black] at (\scaleFac*3,\scaleFac*1) {$11$};
	\node [text=black] at (\scaleFac*2,\scaleFac*0) {$12$};

	\node [text=black] at (\scaleFac*3,\scaleFac*0) {$14$};

% Halo lines
\draw[line width=.35mm, <-, draw=blue] let \p1=(43), \p2=(53) in (\x1+2,\y1) -- (\x2-2,\y2);
\draw[line width=.35mm, <-, draw=blue] let \p1=(42), \p2=(52) in (\x1+2,\y1) -- (\x2-2,\y2);
\draw[line width=.35mm, <-, draw=blue] let \p1=(41), \p2=(51) in (\x1+2,\y1) -- (\x2-2,\y2);

\draw[line width=.35mm, <-, red] let \p1=(20), \p2=(11) in (\x1-2,\y1+2) -- (\x2+2,\y2-2);
\draw[line width=.35mm, <-, red] let \p1=(20), \p2=(21) in (\x1,\y1+2) -- (\x2,\y2-2);
\draw[line width=.35mm, <-, red] let \p1=(20), \p2=(31) in (\x1+2,\y1+2) -- (\x2-2,\y2-2);

\end{tikzpicture}%
%\includegraphics{tikz_cache/paper-figure\theplotCtr.pdf}
%\stepcounter{plotCtr}
}
	}
	\protect\caption{
		Comparison of three MPK implementations for the computation of $A^3x$ on a 1D tri-diagonal stencil matrix, distributed across two MPI processes, where the execution order is written in each node. The traditional MPK implementation of back-to-back SpMVs is shown in (a), the ``Communication Avoiding" MPK with redundant SpMVs in (b), and our implementation of DLB-MPK is shown in (c). Each dot represents a vertex of $G(A)$ for the respective power. In each of the three diagrams, the dashed vertical line denotes the MPI boundary, the process-local x-vector data is shown in black on the bottom layer and is assumed to already be present, and the halo buffer is shown in gray. The x-axis represents the index of the RHS x-vector. The red arrows indicate that the data dependencies are the same in each MPK version, regardless of the execution order.\label{fig:dot_diagram_1D_tridiag_comparison}}
\end{figure*}
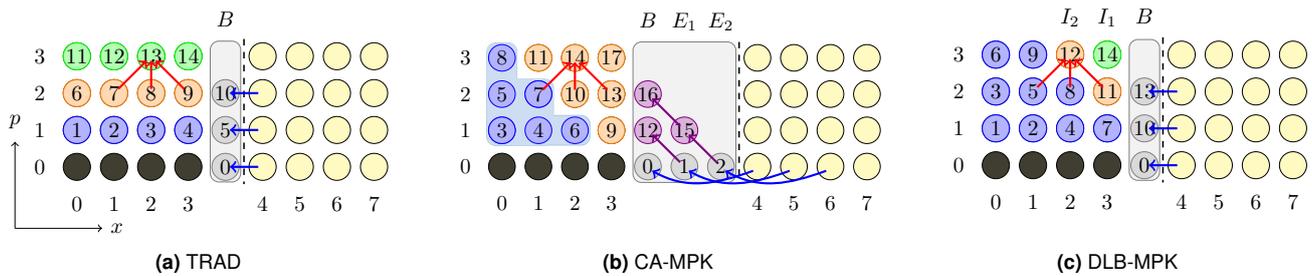

Figure~\ref{fig:dot diagram trad_mpk} illustrates the distributed TRAD MPK approach and shows the required halo communication.
The number on each vertex represents the execution order for computing SpMV on the particular vertex $x_i$. The TRAD approach necessitates a complete SpMV operation to be carried out before initiating the subsequent halo communication routine.
This poses a challenge to cache blocking, particularly when dealing with large in-memory matrices, as the cache may not be able to accommodate all matrix elements loaded during the entire SpMV computation.
In Section~\ref{sec:MPK RACE}, we have seen that caching can be realized by the LB-MPK approach on shared memory.
This necessitates that all the $p_m$ SpMV computations required to raise the local matrix $A_i$ to power $p_m$ be carried out consecutively in one kernel.
This requirement renders the basic halo communication scheme explained above inadequate for distributed-memory parallelization since only the halos necessary for a single SpMV are communicated in this step.
However, for each halo element, we now require the values of all $p_m-1$ powers, i.e., $A^{p}x$ for all $p$ in the range $[0, p_m-1]$.
Complicating matters further, these values (for $p \geq 1$)
are not yet available at this stage because SpMV computations have not been performed. 

For example, consider Figure~\ref{fig:Distributing matrix data over two processes}: employing a \texttt{haloComm} routine ensures that the first MPI process gains access to $x$ values corresponding to all rows which require remote elements (in this case, rows $6$, $7$, $8$ and $9$). However, when conducting LB-MPK with $p_m=2$, the first MPI process requires 
$Ax$ values (as opposed to $x$) at, e.g., the $12$th row when computing $A^{2}x$ for row $8$, but the $Ax$ value at the $12$th row has yet to be computed by the other process.

One potential solution to this problem, as explained by \cite{10.1145/1654059.1654096}, is known as communication-avoiding MPK (CA-MPK).
In this approach, all the necessary values of halo elements, $A^{p}x$
for all $p$ in the range $[1, p_m-1]$, are computed locally on each MPI process. To achieve this, each MPI process conducts additional SpMVs on the halo elements.
However, as discussed in Section~\ref{sec:MPK RACE}, computing $A^{p}x$ by an SpMV operation necessitates updating its neighbors to the 
$A^{p-1}x$ value, which in turn requires updating its neighbors to 
$A^{p-2}x$, and so on until it reaches the input vector $A^{0}x=x$. Consequently, to raise boundary halos $B$ to the 
$p_m-1$  power, all its distance-($p_m-1$) neighbors must also be updated. 
Given that these neighbors often reside on different MPI processes, remote elements must be brought into the current MPI process, thereby requiring additional halo elements. Figure~\ref{fig:dot diagram hoemmen_mpk} illustrates the additional halos required by the CA-MPK approach on a 1D tri-diagonal stencil example. Additional SpMVs take place within the halo buffer, i.e., vertices that are ``external" to the process-local data. In our example, these redundant SpMVs occur at execution stages $12$, $15$, and $16$. To compute $A^{p}x$, CA-MPK requires $p-1$ groups of these external vertices. In general, the halos are organized based on their distance from the boundary $B$, where $E_k$ represents the set of external vertices that are at a distance of $k$ from $B$. The boundary halo elements $B=E_0$ are elevated to power $p_m-1$, while the remaining halo elements $E_k$ are elevated to power $p_m-1-k$ to fulfill the dependencies.

To facilitate cache blocking, a diagonal-style execution order, similar to that in LB-MPK (see Section~\ref{sec:MPK RACE}), can be employed. 
The name ``communication-avoiding" stems from the ability of the CA-MPK approach to overlap communication and computations.
In Figure~\ref{fig:dot diagram hoemmen_mpk} the purely local part (outlined in blue boundary) can be overlapped with the communication of the remote elements. Although the CA-MPK approach enables cache blocking, the overheads resulting from additional halo communication and SpMV computations escalate with the power $p_m$ and the number of MPI processes $n$. 
It is important to note that these extra SpMV computations on halos are redundant, as the MPI process possessing the element locally also conducts SpMVs on these elements.
Particularly with irregular sparse matrices, these overheads can be substantial and may lead to limited speedups as shown in \cite{6877272}.

One way to eliminate redundant computations involves a fine-grained synchronization mechanism, wherein the other process transmits the 
$Ax$ value of halo elements once computations are completed, 
and the other process waits to receive this data. 
However, this entails significant synchronization overhead and the transmission of small MPI messages, ultimately resulting in substantial performance degradation due to the high latency of MPI communications. 
In the following section, we will introduce a savvy new approach to mitigate these performance pitfalls.

\section{DLB-MPK Methodology}
\label{sec:DLB-MPK Methodology}
The DLB-MPK approach enables cache blocking while mitigating the drawbacks associated with CA-MPK, namely the need for additional communication and computations. 
DLB-MPK achieves this by utilizing the same halo communication routine as in the traditional approach (TRAD), but with a reordering of computations and communications to facilitate cache blocking.

In our algorithm, following the initial halo communication, LB-MPK is executed on the local vertices. However, not all local vertices can be elevated to power $p_m$ immediately due to dependencies with the halo elements in $B$, which contain only the input value $x$. Internal vertices that are distance-1 neighbors to $B$
can only be promoted to $Ax$ ($p=1$), 
while their neighbors can only be promoted up to $A^2x$, and so forth.
In general, internal vertices at a distance of $k$ from the boundary $B$, 
denoted as $I_k$, can only be elevated up to $A^kx$.
This implies that, at this stage of DLB-MPK, computations are incomplete on internal vertices $I_k$ where $1 \leq k < p_m$.
The final step of the DLB-MPK method is an iterative process ensuring the completion of SpMV computations on the incomplete internal vertices. The iterative post-computation phase begins with synchronization followed by a call to the halo communication routine to update halo boundaries $B$
with the next power value ($Ax$ in the first iteration). 
This enables all incomplete internal vertices $I_k$ to perform SpMVs, advancing their power computations by one step.
This remainder phase is repeated for a total of $p_m-1$ times to ensure all internal vertices reach power $p_m$.
Figure~\ref{fig:dot diagram dlb_mpk} illustrates the DLB-MPK approach using a 1D tri-diagonal stencil example.

\begin{figure}
	\centering
	\includegraphics[width=\linewidth]{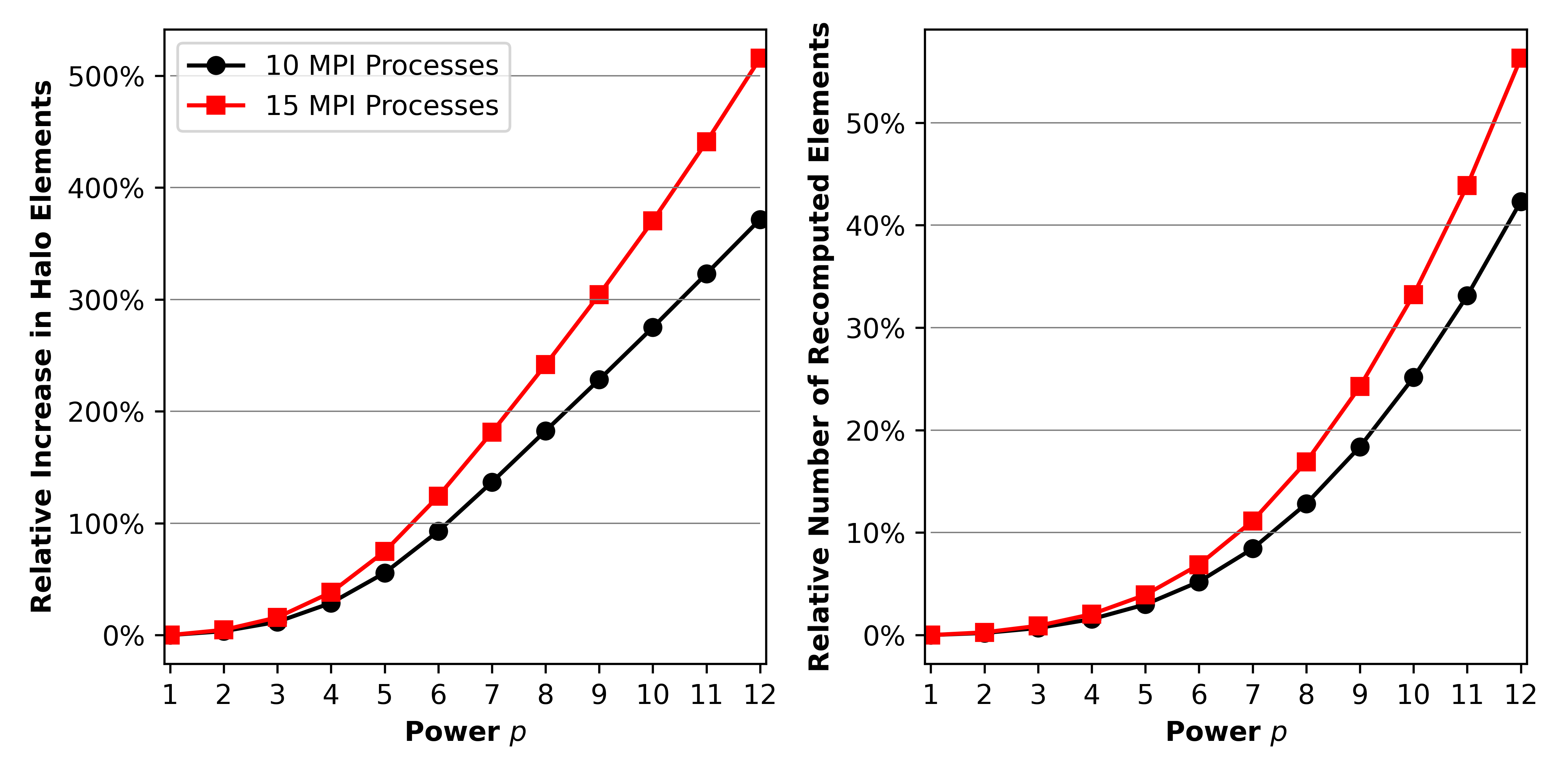}
	\caption{Overheads in CA-MPK associated with the \texttt{Serena} matrix, partitioned over 10 and 15 MPI processes for powers $p \in \{1,2,\dots,12\}$. \textit{Left}: additional halo elements relative to the total number of rows $N_r$. \textit{Right}: recomputed elements relative to the total number of non-zero elements $N_{nz}$.}
	\label{fig:CA-bad}
\end{figure}

As shown in Figure~\ref{fig:dot_diagram_1D_tridiag_comparison}, DLB-MPK requires the same halos as TRAD while benefiting from cache blocking advantages similar to CA-MPK due to its diagonal-style execution; refer to Section~\ref{sec:MPK RACE} for details.
Figure~\ref{fig:CA-bad} quantifies the advantages of reduced halo elements and zero redundant computations for DLB-MPK for an irregular sparse matrix (\texttt{Serena}), which is partitioned row-wise over 10 and 15 MPI processes, respectively. In order to minimize communication and optimize load balance, METIS by \cite{Karypis_1998} was chosen as the global partitioner. The left subfigure in Figure~\ref{fig:CA-bad} shows the relative number of halo elements incurred by CA-MPK in addition to what is caused by DLB-MPK accumulated over all MPI processes, while the right subfigure shows the relative number of required redundant computations for CA-MPK subject to the same global partitioning. Despite the banded sparsity pattern of the matrix, the halo elements required for CA-MPK grow significantly with the power $p$ and the number of MPI processes.%, whereas the halo elements required by DLB-MPK stay constant.

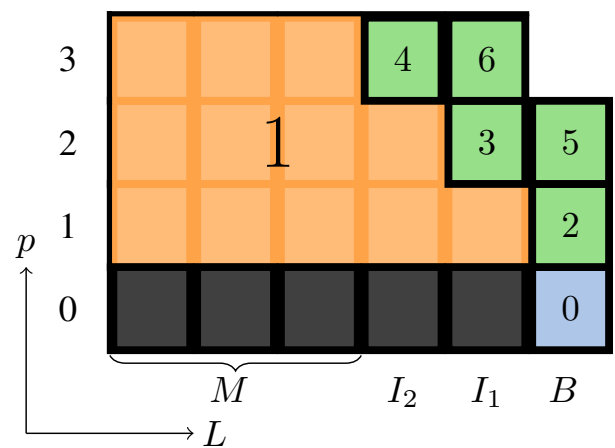
\begin{figure}
	\resizebox{\columnwidth}{!}{%
	%  \tikzsetnextfilename{#2}%
	\edef\bright{60}
\def\sumlevelarray{{0,1,3,6,10,15,20,25,30,35,40,45,50,55,60,64,67,69,70}}
\edef\scaleFac{0.52}
	\begin{tikzpicture}[l0/.style={rectangle,draw,fill=col0!\bright!white,minimum size=10.6, inner sep=0.5pt},
		l1/.style={rectangle,draw,fill=col1!\bright!white,minimum size=10.6, inner sep=0.5pt},
		l2/.style={rectangle,draw,fill=col2!\bright!white,minimum size=10.6, inner sep=0.5pt},
		l3/.style={rectangle,draw,fill=col3!\bright!white,minimum size=10.6, inner sep=0.5pt},
		l4/.style={rectangle,draw,fill=col4!\bright!white,minimum size=10.6, inner sep=0.5pt},
		l5/.style={rectangle,draw,fill=col5!\bright!white,minimum size=10.6, inner sep=0.5pt},
		l6/.style={rectangle,draw,fill=col6!\bright!white,minimum size=10.6, inner sep=0.5pt},
		l7/.style={rectangle,draw,fill=col7!\bright!white,minimum size=10.6, inner sep=0.5pt},
		l8/.style={rectangle,draw,fill=col13!\bright!white,minimum size=10.6, inner sep=0.5pt},
		l9/.style={rectangle,draw,fill=col9!\bright!white,minimum size=10.6, inner sep=0.5pt}]
	\tikzstyle{box} = [text width=5mm, line width=.75mm, minimum height=8mm,draw, rectangle, align=center]
	
	\begin{scope}[x=8mm, y=8mm]
		%		\draw[->] (-1,-1) -- (0,-1)  node [right, draw=none, fill=none] {$R$};
		%		\draw[->] (-1,-1) -- (-1,0)  node [above, draw=none, fill=none] {$p$};
		\draw[->] (-0.5,-0.5) -- (1.5,-0.5);
		\node at (1.5+0.25,-0.5) {$L$};
		\draw[->] (-0.5,-0.5) -- (-0.5,1.5);
		\node at (-0.5,1.5+0.25) {$p$};

		\node (BL-P4) at (0+1,4+2-1) { }; % hack for brace

		\node (P4-label) at (0+1,3+3-1) {};
		\node [draw=none, fill=none] (P3-label) at (0,3+2-1) {3};
		\node [draw=none, fill=none] (P2-label) at (0,2+2-1) {2};
		\node [draw=none, fill=none] (P1-label) at (0,1+2-1) {1};
		\node [draw=none, fill=none] (P0-label) at (0,0+2-1) {0};
		
		\node [draw=none, fill=none] (L5-label) at (5+3-2,0+1-1) {$B_{\phantom{0}}$};
		\node [draw=none, fill=none] (L4-label) at (4+3-2,0+1-1) {$I_1$};
		\node [draw=none, fill=none] (L3-label) at (3+3-2,0+1-1) {$I_2$};
		\node [draw=none, fill=none] (L1-label) at (2,0+1-1) {$M_{\phantom{0}}$};
		\node [draw=none, fill=none] (L2-label) at (2+3-2,0+1-1) {};

		\node [draw=none, fill=none] (L00-label) at (0+2-2,0+1-1) {};

		\node [box, fill=lblue, text=black] (BR-P0) at (5+3-2,0+2-1) {$0$};

		\node [box, fill=lorange, draw=lorange!60!orange] (L3-P1) at (3+3-2,1+2-1) {};
		\node [box, fill=lorange, draw=lorange!60!orange] (L4-P1) at (4+3-2,1+2-1) {};
		\node [box, fill=lorange, draw=lorange!60!orange] (L2-P1) at (2+3-2,1+2-1) {};
		\node [box, fill=lorange, draw=lorange!60!orange] (L2-P2) at (2+3-2,2+2-1) {};
		\node [box, fill=lorange, draw=lorange!60!orange] (L2-P3) at (2+3-2,3+2-1) {};
		\node [box, fill=lorange, draw=lorange!60!orange] (L3-P2) at (3+3-2,2+2-1) {};

		\node [box, fill=lorange, draw=lorange!60!orange] (1) at (2+5-3-2,1+2-1) {};
		
		\node [box, fill=lorange, draw=lorange!60!orange] (2)  at (2+5-3-2,2+2-1) {};
		\node [box, fill=lorange, draw=lorange!60!orange]  (3) at (2+5-3-2,3+2-1) {};

		\node [box, fill=lorange, draw=lorange!60!orange] (4) at (2+4-3-2,1+2-1) {};
		\node [box, fill=lorange, draw=lorange!60!orange] (5) at (2+4-3-2,2+2-1) {};
		\node [box, fill=lorange, draw=lorange!60!orange] (6) at (2+4-3-2,3+2-1) {};
		
		\node [box, fill=black, draw=black, fill opacity = .75] (bl) at (2+4-3-2,2-1) {};
		\node [box, fill=black, draw=black, fill opacity = .75] at (2+5-3-2,2-1) {};
		\node [box, fill=black, draw=black, fill opacity = .75] (L2-P0) at (2+3-2,0+2-1) {};
		\node [box, fill=black, draw=black, fill opacity = .75] (L3-P0) at (3+3-2,0+2-1) {};
		\node [box, fill=black, draw=black, fill opacity = .75] (L4-P0) at (4+3-2,0+2-1) {};

		\node [box, fill=lgreen, text=black] (BR-P1) at (5+3-2,1+2-1) {$2$};
		
		\node [box, fill=lgreen, text=black] (L4-P2) at (4+3-2,2+2-1) {$3$};
		
		\node [box, fill=lgreen, text=black] (L3-P3) at (3+3-2,3+2-1) {$4$};
		
		\node [box, fill=lgreen, text=black] (BR-P2) at (5+3-2,2+2-1) {$5$};
		\node [box, fill=lgreen, text=black] (L4-P3) at (4+3-2,3+2-1) {$6$};

		%RACE part
		
%		\draw[fill=orange, rotate around={-17:(2,2)}, draw=none, opacity = .5] (4.8*\scaleFac+.75,-0.3*\scaleFac+2) -- (5.2*\scaleFac+.75, 0.3*\scaleFac+2) -- (1.2*\scaleFac+.75, 4.3*\scaleFac+2) -- (0.8*\scaleFac+.75, 3.7*\scaleFac+2) --  (4.8*\scaleFac+.75,-0.2*\scaleFac+2); 
	%\draw [fill=orange, rotate around={27:(2.65,2.5)}, draw=orange, opacity = .25] (1.75, 2.25) rectangle (2, 4.65) {};
			
		\draw[line width=.5mm] (3.north west) |- (3.north east);
		\draw[line width=.5mm] (6.north west) |- (6.north east);
		\draw[line width=.5mm] (L2-P3.north west) |- (L2-P3.north east);
		\draw[line width=.5mm] (L4-P3.north west) |- (L4-P3.north east);
		\draw[line width=.5mm] (L3-P3.north west) |- (L3-P3.north east);
		\draw[line width=.5mm] (4.south west) |- (4.north west);
		\draw[line width=.5mm] (5.south west) |- (5.north west);
		\draw[line width=.5mm] let \p1=(6.south west), \p2=(6.north west) in (\x1,\y1) |- (\x2,\y2+.75);
		\draw[line width=.5mm] (bl.south west) |- (bl.north west);
		
%		\foreach \x in {0,...,4}
%		\foreach \y in {0,...,2} 
%		{
%			%	\pgfmathtruncatemacro{\labelu}{89 - ((14 - \y - \x)*(14 - \y-\x + 1)*0.5+ 7-\y)
%				\pgfmathtruncatemacro{\level}{\y + \x  }
%				\pgfmathtruncatemacro{\label}{\sumlevelarray[\level]+\y}
%				\node [l\x, opacity = .25]  (\x\y) at (\scaleFac*\x + 1,\scaleFac*2*\y+2) {};
%			
%		}
%		\foreach \x in {5,...,6}
%		\foreach \y in {0,...,1} 
%		{
%			%	\pgfmathtruncatemacro{\labelu}{89 - ((14 - \y - \x)*(14 - \y-\x + 1)*0.5+ 7-\y)
%				\pgfmathtruncatemacro{\level}{\y + \x  }
%				\pgfmathtruncatemacro{\label}{\sumlevelarray[\level]+\y}
%				\node [l\x, opacity = .25]  (\x\y) at (\scaleFac*\x + 1,\scaleFac*2*\y+2) {};
%				
%			}
%		\foreach \x in {5,...,6}
%		\foreach \y in {0,...,1} 
%		{
%			%	\pgfmathtruncatemacro{\labelu}{89 - ((14 - \y - \x)*(14 - \y-\x + 1)*0.5+ 7-\y)
%				\pgfmathtruncatemacro{\level}{\y + \x  }
%				\pgfmathtruncatemacro{\label}{\sumlevelarray[\level]+\y}
%				\node [l\x, opacity = .25]  (\x\y) at (\scaleFac*\x + 1,\scaleFac*2*\y+2) {};
%				
%			}
		
		\node [text=black] at (2.5,3) {\huge$1$};

		\draw [decorate,decoration={brace,amplitude=1.5mm, mirror, raise=-3.5mm},xshift=0mm,yshift=0mm] (1-.5,0) -- (3+.5,0);

	\end{scope}
\end{tikzpicture}%
%\includegraphics{tikz_cache/paper-figure\theplotCtr.pdf}
%\stepcounter{plotCtr}
}
	\caption{\label{fig:A3x DLB-MPK} An adapted $Lp$ diagram for DLB-MPK executing $A_i^3x$ on some MPI process $i$. The numbers indicate the order of execution of DLB-MPK. The colors of the boxes indicate the phase of DLB-MPK in which they are executed. The blue box corresponds to the first phase, the orange to the LB-MPK phase, and the green to the iterative third phase.}
\end{figure}

The implementation of DLB-MPK can be straightforwardly derived from the execution order illustrated in Figure~\ref{fig:dot diagram dlb_mpk} for a 1D tri-diagonal example. 
However, when dealing with a general sparse matrix, the internal boundary vertices
$I_k$ for $k<p_m$ may not be ordered consecutively.
Therefore, an efficient implementation will require gathering these boundary vertices and reordering the matrix during preprocessing to ensure that these vertices (rows in the matrix) appear consecutively. All vertices which are a distance of $p_m$ or greater from the boundary, that is all vertices in $I_k$ for $k\geq p_m$, are collected into a single main ``bulk structure" $M$, which is large in practice.

The algorithm is separated into three distinct phases: i) execute the initial halo communication, ii) use the cache blocking capabilities of RACE to fully promote all levels in $M$ to $p_m$, and each $I_k$ level to $k$, and iii) iteratively finish remaining computations and communications. We use a modified $Lp$ diagram in Figure~\ref{fig:A3x DLB-MPK} as an example of DLB-MPK executing $A_i^3x$ on some MPI process $i$. The color of the box indicates in which phase it is executed. The blue box corresponds to the first phase, the orange to the LB-MPK phase, and the green to the iterative third phase. As previously mentioned, $B = I_0$ is the halo buffer, while $I_1$ and $I_2$ are all vertices that are distances 1 and 2, respectively, away from the MPI boundary. Instead of labeling individual levels, Figure~\ref{fig:A3x DLB-MPK} represents the main bulk structure by $M$. It is here that RACE can safely perform cache blocking.

\begin{algorithm}
	\scriptsize 
	\caption{Distributed Level-Blocked MPK
	}\label{alg:Distributed Level-Blocked Matrix Power Kernel}
	\begin{algorithmic}
		\Require{ \\
			$\texttt{double}\ x[N_{r,i}+N_{h,i}]$\;\\
			% \nonl $\ \texttt{double}\ \texttt{val}[N_{nz}]$\tcp*{CRS arrays}\\
			% \nonl $\ \texttt{int}\ \texttt{col}[N_{nz}], \ \texttt{rowPtr}[N_r]$\; \\
			% \nonl $\ \texttt{int}\ p_m$\tcp*{highest power to compute $A^{p_m}x$}
			% $\ \texttt{int}\ \texttt{distFromRemotePtr}[p_m+1]$\tcp*{boundary level array} \\
			$\texttt{commFuncType}$ \texttt{haloComm}\;\\
			%$\texttt{commFuncArgType}$ \texttt{commArgs}\;\\
			$\texttt{spmvFuncType}$ \texttt{SpMV}\;\\
			$\texttt{levelPointer}\ I$\;\\
			%$\texttt{spmvFuncArgType}$ \texttt{spmvArgs}\;
			$\texttt{sparseMatrix}\ A_i$\;\\
			$\texttt{int}\ p_m$\;
		}
		\Ensure{ \\
			$\texttt{double}\ y[N_{r,i}+N_{h,i}, p_m]$\; 
		}
		\\ \ \\
		%$\texttt{userInput} \gets \texttt{commFunc, commArgs, spmvFunc, spmvArgs;}$\;\\
		%	
		$y[:, 0] \gets x$\texttt{;}\;\\
		%$[x,y] \gets $ \texttt{mpiPre($x$, $y$, haloComm);}\;\\
		\hspace*{-\fboxsep}\colorbox{lblue!50!white}{$y[:, 0] \gets$ \texttt{haloComm}($y[:, 0]$)\texttt{;}\;} \\
		\hspace*{-\fboxsep}\colorbox{lorange!50!white}{$[x,y] \gets $ \texttt{localLBMPK($x, y, \text{\texttt{SpMV}}$);}\;}
		\For{$p \gets 1,\dots,p_m-1$}\\
		\qquad\hspace*{-\fboxsep}\colorbox{lgreen!50!white}{$y[:, p] \gets$ \texttt{haloComm}($y[:, p]$)\texttt{;}}\;
		\qquad\For{$k \gets 1,\dots,p_m-p$} \\
		\qquad \qquad\colorbox{lgreen!50!white}{$y[I[k], p+1] \gets$ \texttt{SpMV}($y[I[k], p], A_i[I[k],:]$)\texttt{;}}\;
		\EndFor
		\EndFor
	\end{algorithmic}
\end{algorithm}

A benefit of DLB-MPK is that we can use the same MPI routines as for TRAD. Hence, we are able to easily integrate our algorithm into external libraries with existing SpMV and halo communication routines. This is shown in Algorithm~\ref{alg:Distributed Level-Blocked Matrix Power Kernel}, which gives a high-level overview of DLB-MPK. The call-back functions \texttt{haloComm} and \texttt{SpMV} are both provided by the user. The structure $I$ contains the first $p_m-1$ levels of $A_i$, where again $I[0]=B$ contains the boundary vertices. The initial halo exchange takes place in the first phase, highlighted in blue. The cache-blocking second phase is executed during \texttt{localLBMPK}, highlighted in orange. Finally, the iterative third phase, represented by the nested for-loops, it highlighted in green.

The percentage of vertices that fall outside of the bulk structure is considered as the ``local overhead" $O_{\text{DLB-MPK},i}$ of DLB-MPK. While not an ``overhead" per se, it is a useful quantity for our investigation as it expresses the efficiency of cache blocking. With $M_i$ denoting the bulk structure level on MPI process $i$, we can define this overhead as
\begin{equation} O_{\text{DLB-MPK},i} := 1 - \frac{|M_i|}{N_{i,r}}. \label{eq:local_oheads}\end{equation}
To have a single number which represents the ``global overhead" from cache blocking, we collect the local overheads in Equation~\eqref{eq:local_oheads} from each of the $n$ processes and normalize them over the total number of rows, yielding
\begin{equation} O_{\text{DLB-MPK}} := \frac{\sum_{i=0}^n \left(N_{i,r} \cdot O_{\text{DLB-MPK},i} \right) }{N_r}.\label{eq:global_oheads}\end{equation}

\section{Results}
\label{sec:Results}
In this section, we investigate the performance and scaling characteristics of DLB-MPK and how they compare to TRAD in a variety of scenarios on a selection of modern multicore CPUs.
To gain a deeper understanding of the performance of our level-based cache-blocked MPK, we first establish a theoretical roof{}line-based upper performance prediction for the SpMV kernel, the main kernel used in MPK.

It is well known that SpMV (and by extension traditional MPK) is usually a memory-bound kernel on modern hardware for sparse matrices from science and engineering, as described by \cite{doi:10.1137/130930352}. According to the roof{}line model, in the memory bound regime with the CRS matrix storage format\endnote{We are not bound to any particular matrix format, but choose CRS for its ubiquity in the literature.} using 8 bytes for the matrix values and 4 bytes for the column indices and row pointer, performance is limited by \begin{equation} P = \frac{b_s}{6\,\byte+14\,\byte/N_{nzr}},\label{eq:roofline}\end{equation}
where $b_s$ denotes the saturated main memory bandwidth, and $N_{nzr} = N_{nz}/N_r$ denotes the average number of non-zero elements per row.

\subsection{Experimental Setup}
\label{sec:Experimental Setup}
The relevant hardware and software environment used for the measurements is explained in the following.

\subsubsection{Hardware}
\label{sec:Hardware}
\begin{table}[]
	
	\begin{center}
		\caption{Single-Socket Hardware Configurations}
		\label{tab:Hardware Config.}
		\resizebox{\columnwidth}{!}{%
			\begin{tabular}{l|c|c|c}
				Architecture & ICL & SPR & MIL \\
				\hline
				Chip Model & Xeon Platinum 8360Y & Xeon Platinum 8470 & AMD EPYC 7763 \\
				Microarchitecture & Sunny Cove & Golden Cove & Zen 3\\
				Cores & 36 & 52 & 64\\
				ccNUMA domains & 2 & 4 & 4\\
				Max. SIMD width & 512 bits & 512 bits & 256 bits \\
				L1D cache capacity & $36\times48$ KiB & $52\times48$ KiB & $64\times32$ KiB\\
				L2 cache capacity & $36\times1.25$ MiB & $52\times2$ MiB & $64\times512$ KiB \\
				L3 cache capacity & $54$ MiB & $105$ MiB & $8\times32$ MiB \\
				%L2+L3 cache/ccNUMA domain & 49 MiB & 52 MiB & 72 MiB \\
				L3 Load Bandwidth & 452 GB/s & 826 GB/s & 2642 GB/s \\
				Mem. Configuration & 8 ch. DDR4-3200 & 8 ch. DDR5-4400 & 8 ch. DDR4-3200 \\
				Mem. Load Bandwidth & 180 GB/s & 241 GB/s & 179 GB/s \\
			\end{tabular}
		}
	\end{center}
\end{table}

In this work, all experiments were conducted on dual-socket nodes of either Intel Ice Lake (ICL), Intel Sapphire Rapids (SPR), or AMD Epyc Zen3 (MIL). Table~\ref{tab:Hardware Config.} details the important aspects of each architecture. 
ICL and SPR are both capable of performing AVX-512 instructions, while MIL supports only AVX-2. Sub-NUMA Clustering with the maximum possible number of ccNUMA domains was enabled on the Intel systems (two on ICL and four on SPR), and NPS=4 was set on MIL. 
In order to reflect practical use case scenarios, turbo mode was enabled for all experiments.

All of these machines have three levels of cache: private, inclusive L1 and L2, and a victim-style L3 cache. This means that we can consider the sum of L2 and L3 cache as the total size for which we can use RACE to cache block. RACE excels when blocking for outer level caches, as shown by \cite{RACEMPK}.

\begin{figure*}[tp]
	\captionsetup[subfigure]{justification=centering}
	\subfloat[ICL\label{fig:ICL Bandwidths}]{%
		\includegraphics[width=.3\textwidth]{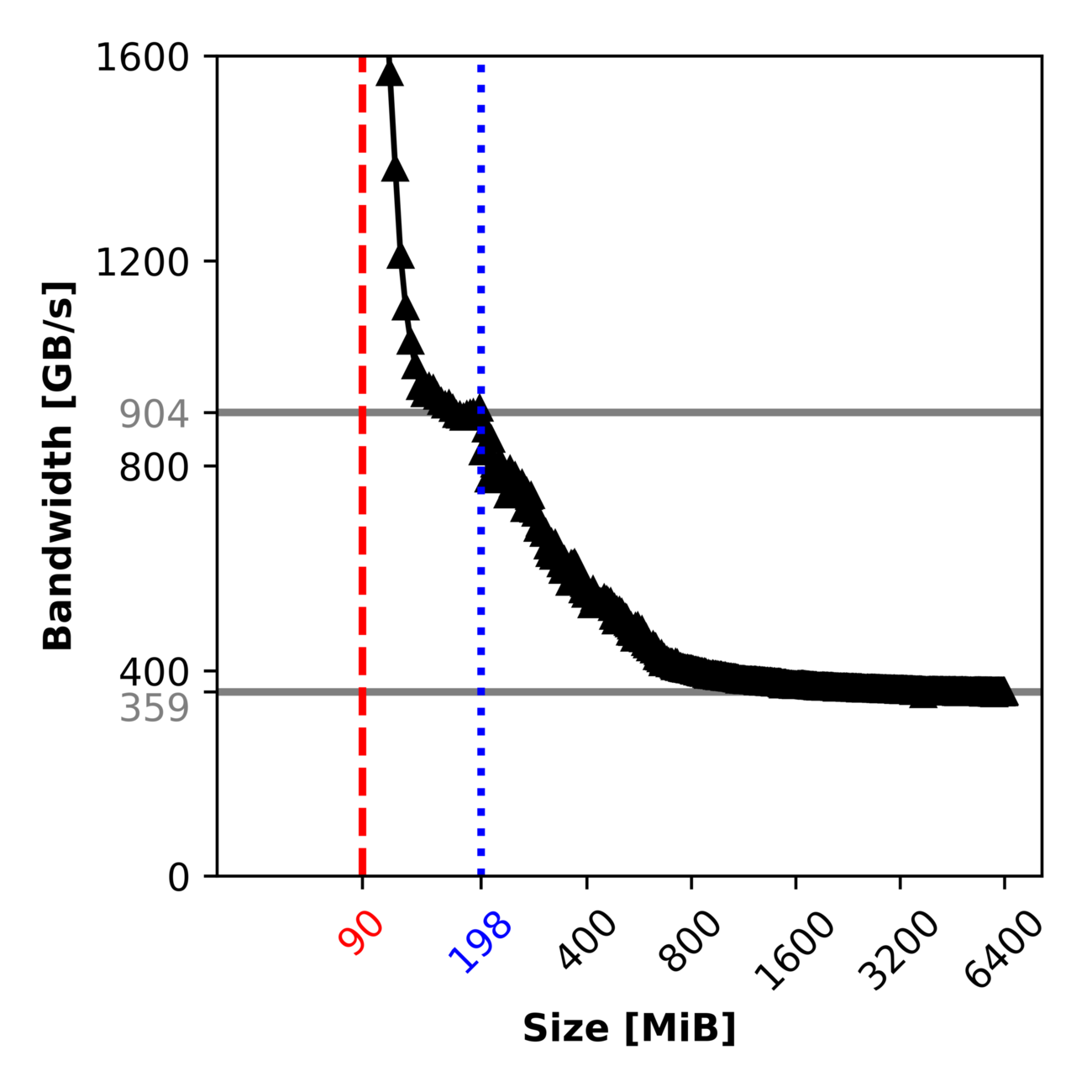}
	}
	\hfill
	\subfloat[SPR\label{fig:SPR Bandwidths}]{%
		\includegraphics[width=.3\textwidth]{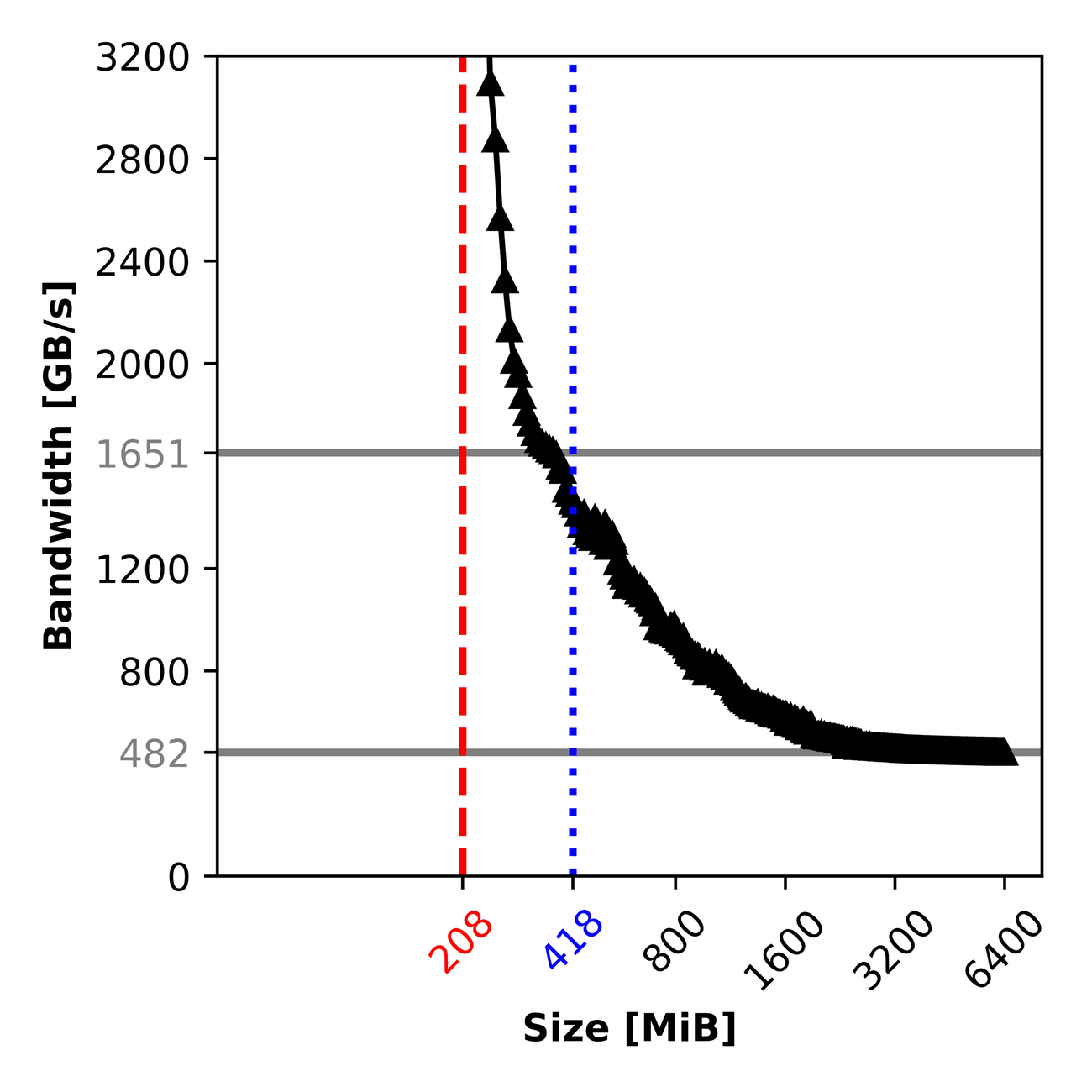}
	}
	\hfill
	\subfloat[MIL\label{fig:MIL Bandwidths}]{%
		\includegraphics[width=.3\textwidth]{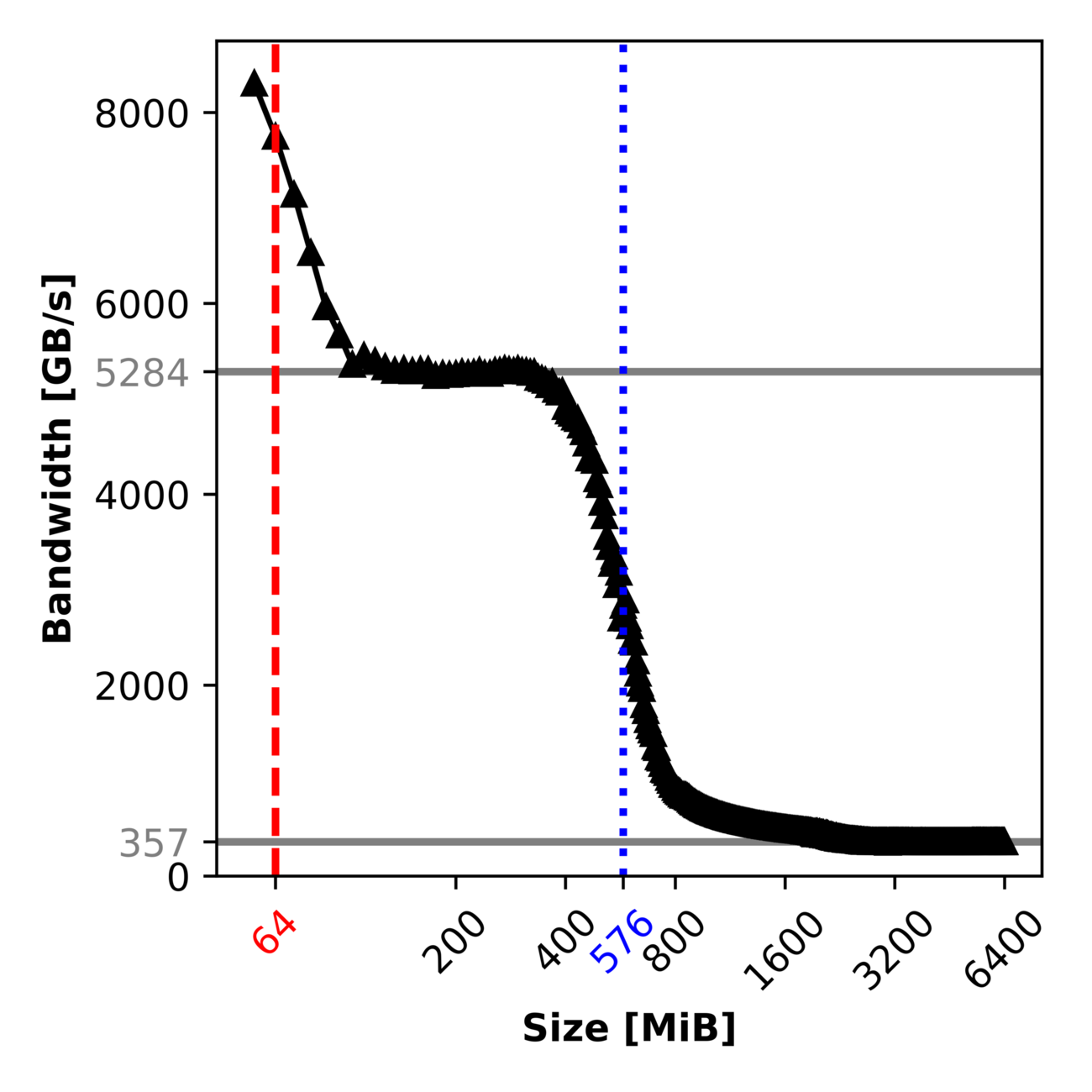}
	}
	\caption{Full-node measured load bandwidths in GB/s (y-axis) vs.\ data set size. The higher solid horizontal line represents the estimated L3 cache bandwidths and the lower one represents the estimated bandwidth from main memory. The dashed red line marks the overall L2 cache size for the entire node, while the dotted blue line represents the aggregate L2+L3 cache size for the entire node.
		%Data is loaded across ccNUMA domains equally to ensure optimal shared last-level cache utilization. 
		The widest SIMD registers are used on each machine for the load instructions. }
	\label{fig:All Bandwidths}
\end{figure*}

Since the achievable bandwidth of the hardware plays a vital role in determining the performance of SpMV-like kernels (see Equation~\eqref{eq:roofline}), we investigate the bandwidths on each of the machines in Figure~\ref{fig:All Bandwidths}. The load-only kernel from \likwidbench\ by \cite{5599200} is used here to determine the bandwidth as it reflects the predominant behavior of the SpMV kernel. The most striking contrast between the three plots in Figure~\ref{fig:All Bandwidths} is the difference in scale of the y-axes. ICL has about half the L3 bandwidth of SPR, and only about a sixth of the L3 bandwidth of MIL\@. In terms of main memory bandwidth, however, ICL narrowly surpasses MIL, while SPR beats both by at least 30\%.

The cache ``plateaus" at which we estimate the L3 bandwidths vary in behavior. Both ICL and SPR display a gradual degradation of bandwidth after the data set exceeds a cache size, which is due to Intel's ``dynamic replacement policy." It has been shown by \cite{10.1007/978-3-030-50743-5_21} that this policy makes intelligent use of the cache for data sets that exceed the cache size.
AMD's cache replacement policy is different and leads to faster bandwidth degradation as can be seen in Figure~\ref{fig:MIL Bandwidths}. The wide plateau in Figure~\ref{fig:MIL Bandwidths} can be explained by MIL having the largest ratio between L2+L3 and L2 sizes out of the three architectures considered, which is due to its massive L3 cache. Figure~\ref{fig:All Bandwidths} indicates that we should expect strong residual caching effects for matrices up to about $800$ MiB on ICL, and up to about $2400$ MiB for SPR and MIL.

\subsubsection{Software}
\label{sec:Software}

\begin{table}[]
	
	\begin{center}
		\caption{Software Configurations and Compiler Flags}
		\label{tab:Software Config.}
		\resizebox{\columnwidth}{!}{%
			\begin{tabular}{l|c|c|c}
				Architecture & ICL & SPR & MIL \\
				\hline
				OS & AlmaLinux 8.8 & AlmaLinux 8.8 & RHEL 8.8 \\
				MPI library version & Intel MPI 2021.10 & Intel MPI 2021.10 & Intel MPI 2023.03 \\
				Compiler & icx 2023.2.0 & icx 2023.2.0 & icx 2023.0.3 \\
				Flags & & & \\
				\null\hfill Opt. level & -Ofast & -Ofast & -Ofast \\
				\null\hfill Arch & -xhost & -xhost & -march=core-avx2\\
				& & & -mtune=core-avx2 \\
				\null\hfill Downfall fix & -Xclang -target-feature  & -Xclang -target-feature & -Xclang -target-feature\\
				& -Xclang +prefer-no-gather & -Xclang +prefer-no-gather & -Xclang +prefer-no-gather \\
				\null\hfill Force AVX512& -xCORE-AVX512 & -xCORE-AVX512 & \\
				& -qopt-zmm-usage=high & -qopt-zmm-usage=high & \\
				\null\hfill Misc. & -std=c++14 -fopenmp & -std=c++14 -fopenmp & -std=c++14 -fopenmp 
				
			\end{tabular}
		}
	\end{center}
\end{table}

Table \ref{tab:Benchmark Matrices} lists the matrices used for benchmarking with their number of rows $N_r$, number of non-zero elements $N_{nz}$, average number of non-zero elements per row $N_{nzr} = N_{nz}/N_r$, and the size of the matrix data in CRS format. The total size of a matrix is  $(4N_r+12N_{nz})$\,B. Here, matrix sizes are rounded to the nearest whole number in MiB.

Our selection of benchmark matrices are commonly used in the literature for performance investigations. They show the performance of DLB-MPK compared to TRAD across a wide variety of sparsity patterns while keeping data sets generally large enough to not be completely cache resident (thus, eliminating the need for cache blocking). Most matrices are freely available from the Suite Sparse matrix collection, with the exception of the \texttt{Lynx} matrices, which come from a finite-volume code for Cardiac Arrhythmia simulations over unstructured meshes as described by \cite{7155461,9005849}.

Measures had to be taken against the patch for the ``Downfall" security bug as explained by \cite{moghimi2023downfall}, incurring a penalty for gather instructions on the architectures under consideration. The latest LLVM-based Intel compiler was required, with special compilation flags in order to avoid the expensive gather instructions. In Table \ref{tab:Software Config.}, these flags are given under ``Downfall fix." To ensure vectorization of the SpMV kernel, \texttt{\#pragma omp simd simdlen(VECLEN) reduction(+:sum)} is used on the innermost SpMV loop, where \texttt{VECLEN} is the maximum SIMD width on the respective hardware (see Table \ref{tab:Hardware Config.}), and \texttt{sum} is our accumulator for the SpMV. 
On the Intel architectures, the flags \texttt{-xCORE-AVX512} and \texttt{-qopt-zmm-usage=high} shown in Table~\ref{tab:Software Config.} were also required so that the compiler would generate instructions using the 512-bit wide \texttt{zmm} registers. 

The same affinity is used for benchmarking on each architecture.
Each MPI processes is pinned to one ccNUMA domain, process $i+1$ is mapped physically as close as possible to process $i$, and OpenMP threads are also pinned compactly to the physical cores. Simultaneous Multithreading (SMT) was disabled across all the systems.
While not a primary focus of this work, RACE allows users to specify a maximum recursion stage $s_m$ which enables the breaking down of ``bulky" levels for increasing cache blocking efficiency. 
This maximum recursion stage is set to $s_m = 50$ for all matrices except \texttt{Lynx1151}, where it is set to $s_m = 80$.

What aim to understand the performance gained from cache blocking, not from improved data accesses on the RHS x-vector through the local symmetric BFS permutations (see Section~\ref{sec:MPK RACE}). In an effort to not conflate the two, TRAD is executed with and without local symmetric BFS permutations and the representative performance metric is taken as the maximum of the two. Similarly, we take the maximum performance of DLB-MPK with and without recursion as the representative performance metric.

All numerical results are validated against Intel's Math Kernel Library\endnote{\url{https://software.intel.com/en-us/mkl}}.
Benchmarks are repeated several times, and the median performance is taken as the representative performance metric.
Error bars are excluded from our plots as run-to-run deviations are less than 5\%. 
\begin{table}
	\begin{center}
		\caption{Benchmark Matrices}
		\label{tab:Benchmark Matrices}
		\resizebox{\columnwidth}{!}{%
			\begin{tabular}{rcccc}
				Matrix & $N_r$ & $N_{nz}$ & $N_{nzr}$ & CRS Size [MiB] \\
				\hline
				\texttt{inline\_1} 			& $503,712$ & $36,816,342$ & $73.0$ & $423$ \\
				\texttt{Emilia\_923} 		& $923,136$ & $41,005,206$ & $44.4$ & $473$ \\
				\texttt{ldoor} 				& $952,203$ & $46,522,475$ & $48.8$ & $536$ \\
				\texttt{af\_shell10} 		& $1,508,065$ & $52,672,325$ & $34.9$ & $609$ \\
				\texttt{Hook\_1498} 		& $1,498,023$ & $60,917,445$ & $40.6$ & $703$ \\
				\texttt{Geo\_1438} 			& $1,437,960$ & $63,156,690$ & $43.9$ & $728$ \\
				\texttt{Serena} 			& $1,391,349$ & $64,531,701$ & $46.3$ & $744$ \\
				\texttt{bone010}			& $986,703$ & $71,666,325$ & $72.6$ & $824$ \\
				\texttt{audikw\_1} 			& $943,695$ & $77,651,847$ & $82.2$ & $892$ \\
				\texttt{channel-500x100}	& $4,802,000$ & $85,362,744$ & $17.7$ & $995$ \\
				\texttt{Long\_Coup\_dt0}	& $1,470,152$ & $87,088,992$ & $59.2$ & $1,002$ \\
				\texttt{dielFilterV3real} 	& $1,102,824$ & $89,306,020$ & $80.9$ & $1,026$ \\
				\texttt{nlpkkt120} 			& $3,542,400$ & $96,845,792$ & $27.3$ & $1,122$ \\
				\texttt{ML\_Geer} 			& $1,504,002$ & $110,879,972$ & $73.7$ & $1,275$ \\
				\texttt{Lynx68} 			& $6,811,350$ & $111,560,826$ & $16.3$ & $1,303$ \\
				\texttt{Flan\_1565} 		& $1,564,794$ & $117,406,044$ & $75.0$ & $1,350$ \\
				\texttt{Cube\_Coup\_dt0} 	& $2,164,760$  & $127,206,144$ & $58.7$ & $1,464$ \\
				\texttt{Bump\_2911} 		& $2,911,419$ & $127,729,899$ & $43.9$ & $1,473$ \\
				\texttt{van\_stokes\_4M} 	& $4,382,246$ & $131,577,616$ & $30.0$ & $1,523$ \\
				\texttt{Queen\_4147} 		& $4,147,110$  & $329,499,284$ & $79.5$ & $3,787$ \\
				\texttt{nlpkkt200} 			& $16,240,000$ & $448,225,632$ & $27.6$ &  $5,191$ \\
				\texttt{nlpkkt240} 			& $27,993,600$ & $774,472,352$ & $27.6$ &  $8,970$ \\
				\texttt{Lynx649} 			& $64,950,632$ & $978,866,282$ & $15.0$ & $11,450$ \\
				\texttt{Lynx1151} 			& $115,187,228$ & $1,934,489,424$ & $16.8$ & $22,578$
			\end{tabular}
		}
	\end{center}
\end{table}

\subsection{Parameter Study}
\label{sec:Parameter Study}
RACE provides tuning parameters to optimize performance for the specific hardware under consideration.
In this section, we perform a parameter study on ICL with the matrix \texttt{ML\_Geer} to better understand the influence of these parameters on the performance of DLB-MPK.
This will also serve as an example of how one could perform such an investigation.

\begin{figure}
	\centering
	\includegraphics[width=.8\columnwidth]{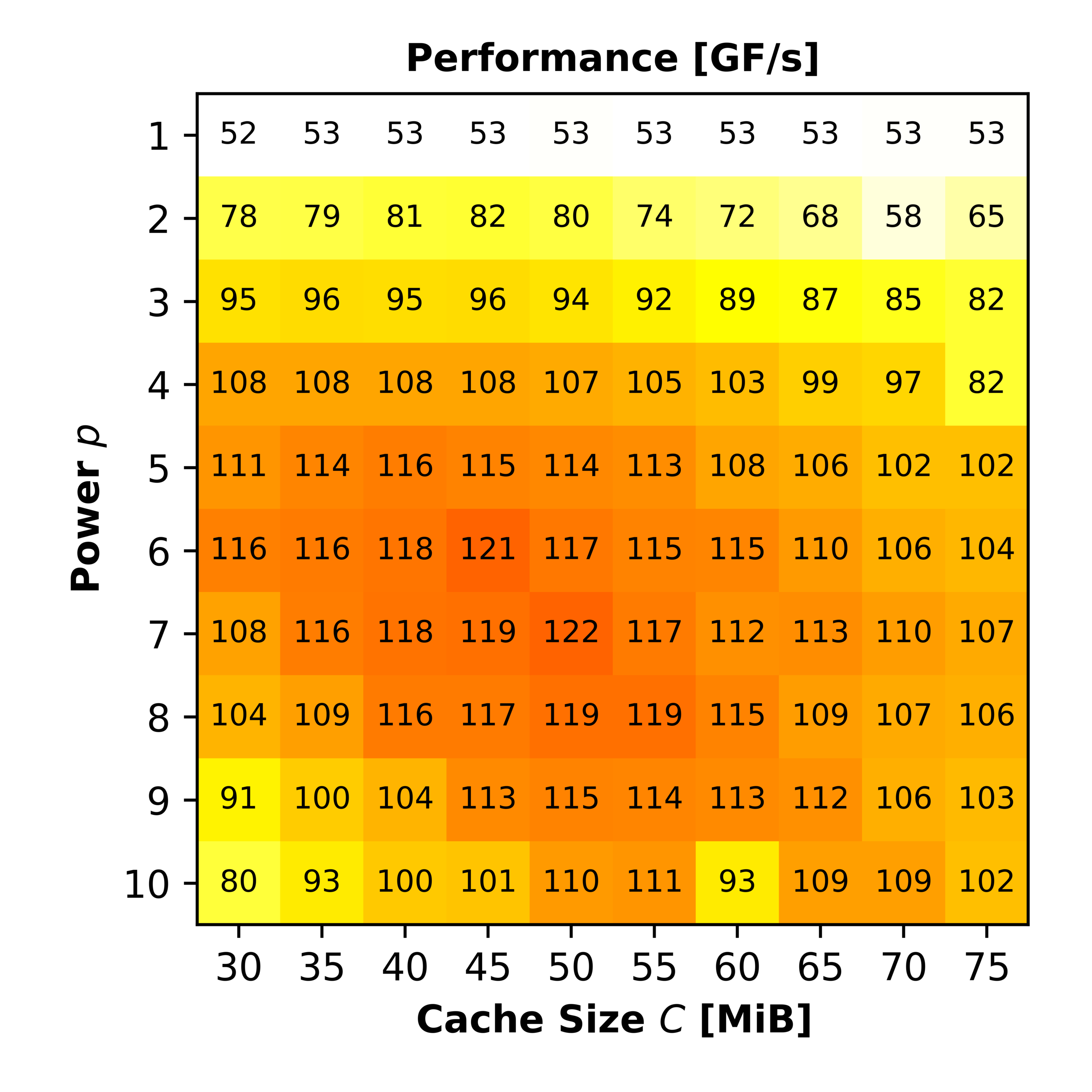}
	\caption{Parameter study with \texttt{ML\_Geer} on one ICL node, scanning $p \in \{1,2,\dots,10\}$ and $C \in \{30,35,\dots,75\}$. %Performance metrics have their decimal parts truncated for space concerns.
	}
	\label{fig:ML_Geer Heatmap}
\end{figure}

We focus here only on the parameters $p$ and $C$, since we have fixed the recursion depth $s_m$ as described before. 
In Figure~\ref{fig:ML_Geer Heatmap}, we scan various powers $p$ and cache sizes $C$ when performing DLB-MPK on \texttt{ML\_Geer}. We use METIS as the global partitioner, pinning one MPI process to each each of the four ccNUMA domains compactly. We see there is a local maximum at $p = 7$ and $C = 50$, after which performance degrades for higher values of $p$ and $C$. 
Higher $p$ values lead to smaller levels and higher synchronization costs between threads in RACE, whereas a $C$ larger than the total cache size can cause more cache misses; see~\cite{RACEMPK} for details.

From Table~\ref{tab:Hardware Config.}, we know that one ccNUMA domain on ICL has $49$ MiB L2+L3 aggregate cache, so we would expect an optimal value for the parameter $C$ to be around this range. The optimal $C$ does not always correspond directly with the amount of available cache per process, due to a safety factor internal to RACE. A user of DLB-MPK would tune these two parameters in order to achieve the best possible performance for their use case. Notice that the DLB-MPK performance for $p = 1$ stays roughly constant as cache size grows. This corroborates our claim from Section~\ref{sec:MPK RACE} that computing $y \gets A^px$ for $p=1$ can not make use of cache blocking.

\subsection{Performance Results Summary}
\label{sec:Summary Performance}
In this section, we give a concise high-level single-node performance summary of DLB-MPK and TRAD on our benchmark matrices.

\begin{figure}[tp]
	\captionsetup[subfigure]{justification=centering}
	\subfloat[ICL\label{fig:ICL Summary}]{%
		\includegraphics[width=\linewidth]{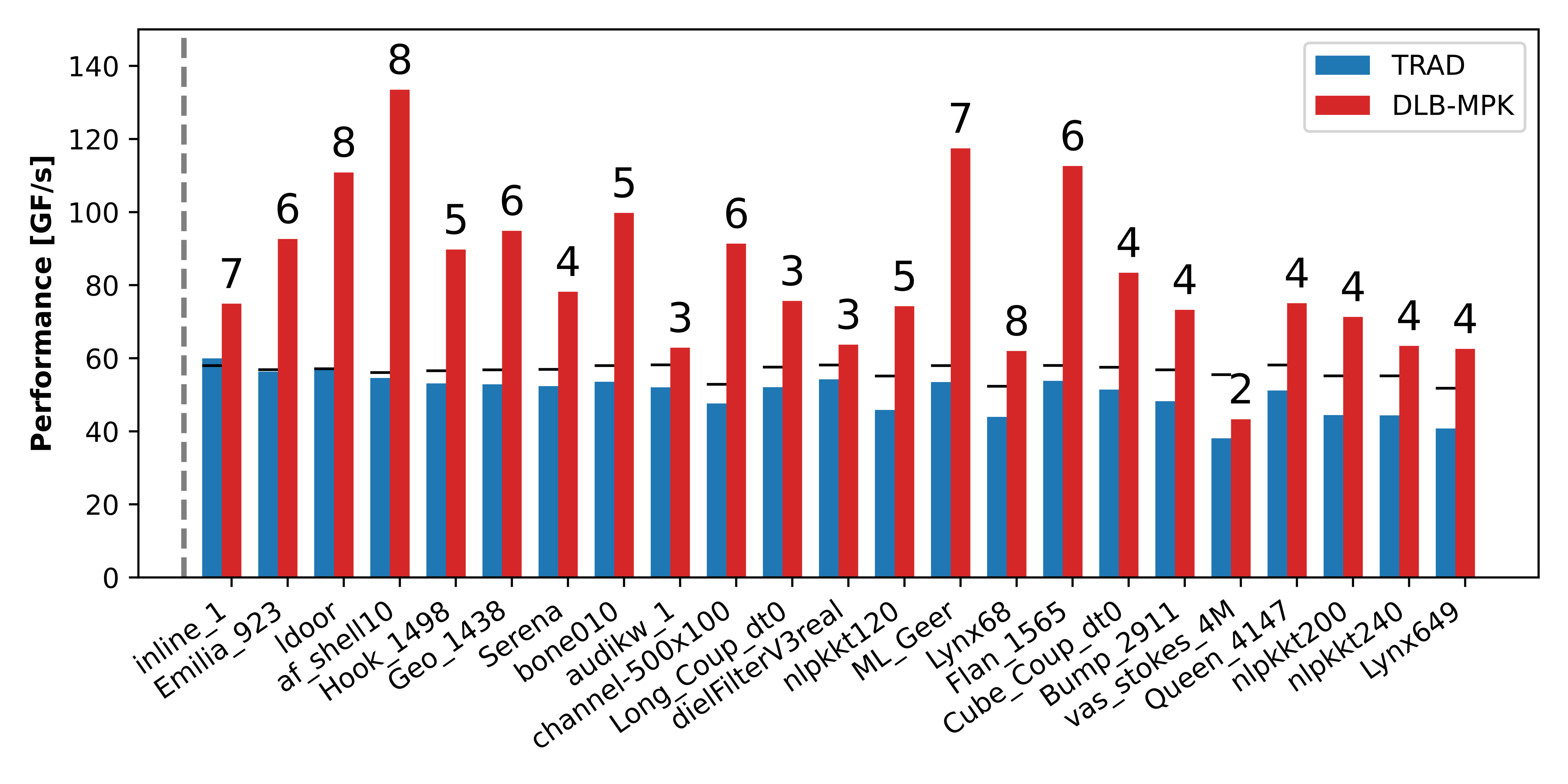}
	}
	
	\subfloat[SPR\label{fig:SPR Summary}]{%
		\includegraphics[width=\linewidth]{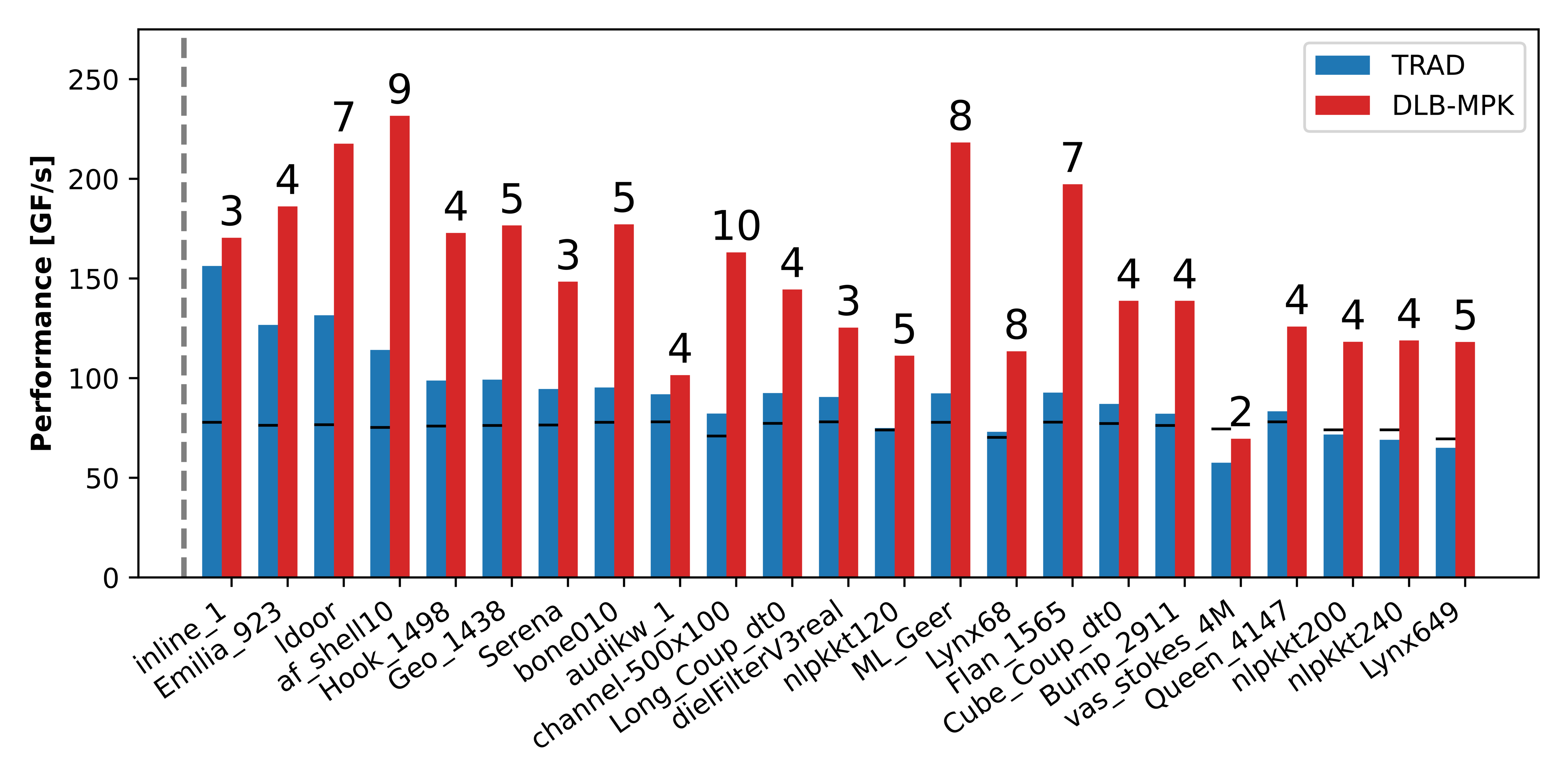}
	}
	
	\subfloat[MIL\label{fig:MIL Summary}]{%
		\includegraphics[width=\linewidth]{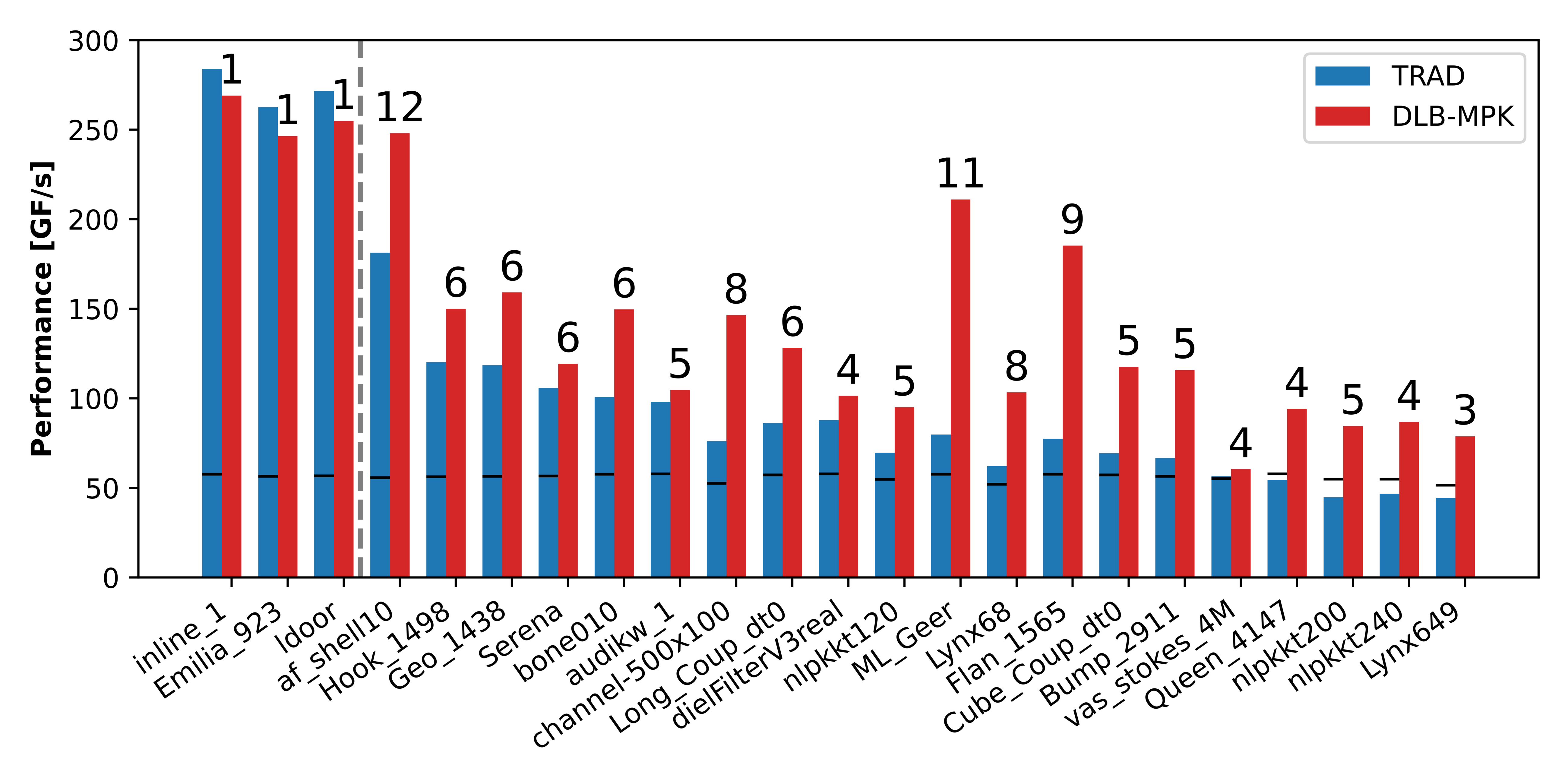}
	}
	
	\caption{Node-level performance summary for benchmark matrices in Table \ref{tab:Benchmark Matrices}, ordered by CRS size. For each matrix, the numbers above the bars denote the optimal power $p$ for which DLB-MPK was tuned. The horizontal black lines are the roof{}line predictions for TRAD according to Eq.~(\ref{eq:roofline}). The vertical dashed line represents the aggregate L2+L3 cache size.}
	\label{fig:Performance Summaries}
\end{figure}

Figure~\ref{fig:Performance Summaries} shows the node-level performance of DLB-MPK (red, right bars) as compared to TRAD (blue, left bars) for optimally tuned parameters $C$ and $p$. The matrices are ordered according to their size, and the vertical dashed line indicates the L2+L3 aggregate cache size of the  architecture.
On MIL there are some matrices that fit in the cache, i.e., left of the dashed vertical line. In this regime, DLB-MPK has no benefit compared to TRAD since the matrices already fit in cache and cache blocking is pointless.
The behavior is very similar with cache-resident matrices on ICL and SPR, although for this work, we chose large in-memory matrices to elucidate the situations in which DLB-MPK is advantageous to use.

The short black line in or above each TRAD bar is the memory-bound roof{}line performance limit of SpMV for the given matrix and hardware computed using Equation~\eqref{eq:roofline}.
As TRAD performs back-to-back SpMVs, ideally one would expect the performance of TRAD for large in-memory matrices to be below the roof{}line limit.
However, in many cases, close to the cache boundary (just to the right of the dashed vertical line), TRAD's performance exceeds the roof{}line limit by a small margin.
This is due to the residual caching effects as also observed in Figure~\ref{fig:All Bandwidths} for the load-only benchmark. As predicted for SPR and MIL, TRAD exhibits these residual caching effects until the matrix size is up to $2400$ MiB, i.e., until \texttt{van\_stokes\_4M}, after which point the performance of TRAD is almost always lower than the upper roof{}line bound.

In general, towards the right of the dashed vertical line (the in-memory matrices), DLB-MPK has a significant advantage over TRAD.
The performance of DLB-MPK is much higher than the roof{}line prediction and TRAD, due to cache blocking resulting in lower main memory traffic.
We observe an average (maximum) speedup
of $1.6\times$ ($2.5\times$), $1.7\times$ ($2.4\times$), and $1.6\times$ ($2.7\times$) for large in-memory datasets on ICL, SPR, and MIL, respectively.
The numbers annotated above DLB-MPK bars show the optimal power value tuned in the range of $p \in \{1,2,\dots,12\}$.
As shown by \cite{RACEMPK}, the preprocessing costs associated with RACE are typically equivalent to 5 to 50 SpMVs (increasing with the recursion stage $s_m$). The preprocessing costs associated with the introduction of MPI are minimal, since the only additional steps are the identification and collection of the boundary vertices. As this is equivalent to each MPI process scanning its local rows once, this overhead is equivalent to roughly 1 additional SpMV.

 %The optimal power $p$ tends to be an even number for memory-resident matrices due to the minor cache blocking procedures internal to RACE mentioned at the end of Section \ref{sec:DLB-MPK Methodology}.

\subsection{Strong Scaling}
\label{sec:Strong Scaling}
It is frequently more important to understand the scaling characteristics of performance rather than taking a snapshot for a single parameter configuration and input. 
We now investigate how the performance of DLB-MPK grows with increasing ccNUMA domains.
The experiment is conducted on eight nodes of SPR.
As in previous section, the power $p$ is tuned in the range $p \in \{1,2, \dots,12\}$.

\begin{figure}[tp]
	\captionsetup[subfigure]{justification=centering}
	\subfloat[Lynx1151 $A^4x$ and $A^6x$ strong scaling\label{fig:Lynx1151 Strong Scaling}]{%
		\includegraphics[width=.86\linewidth]{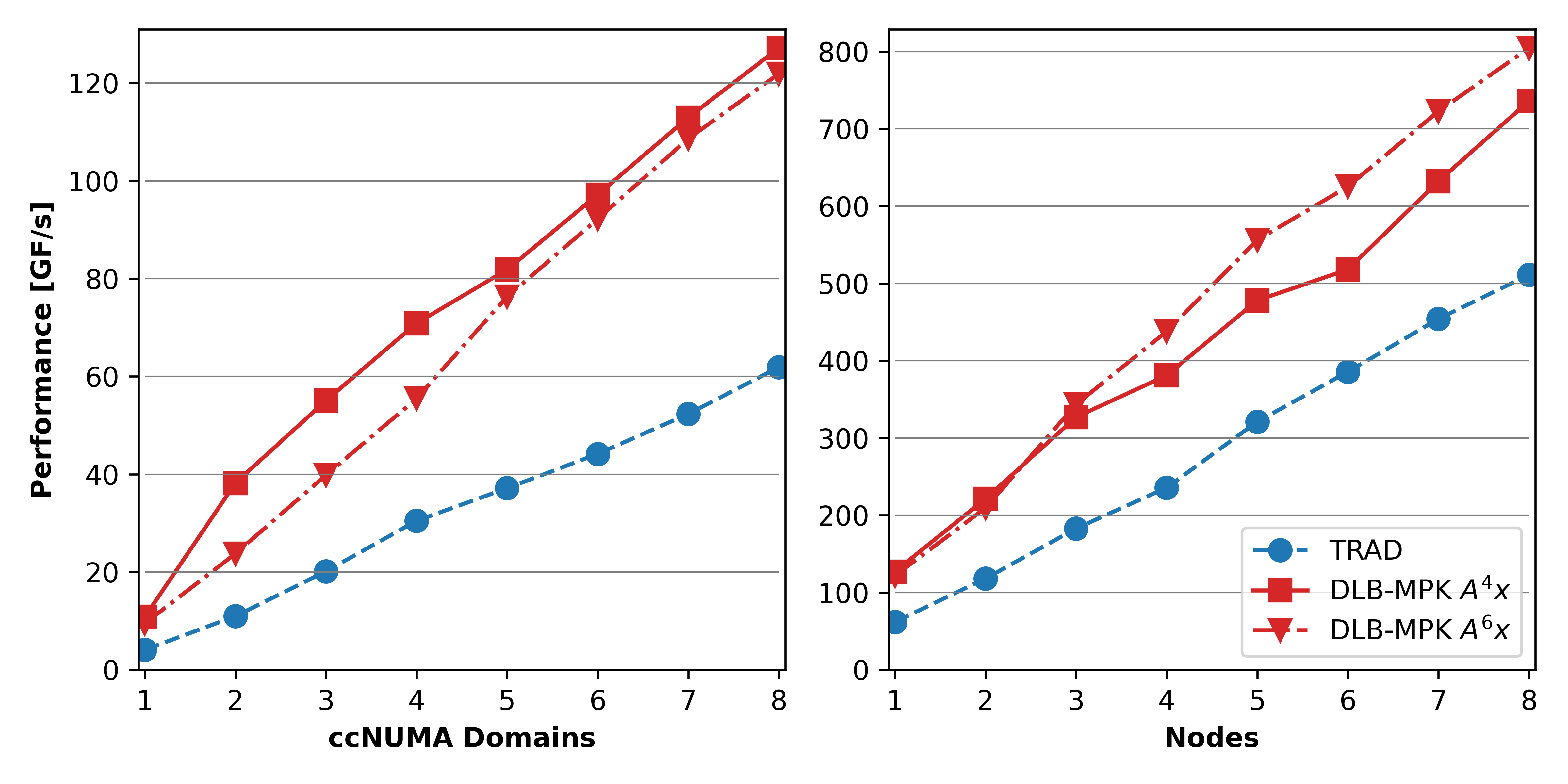}
	}
	\begin{center}
		\subfloat[Lynx1151 $A^4x$ overheads\label{fig:Lynx1151 A4x Overheads}]{%
			\includegraphics[width=.95\linewidth]{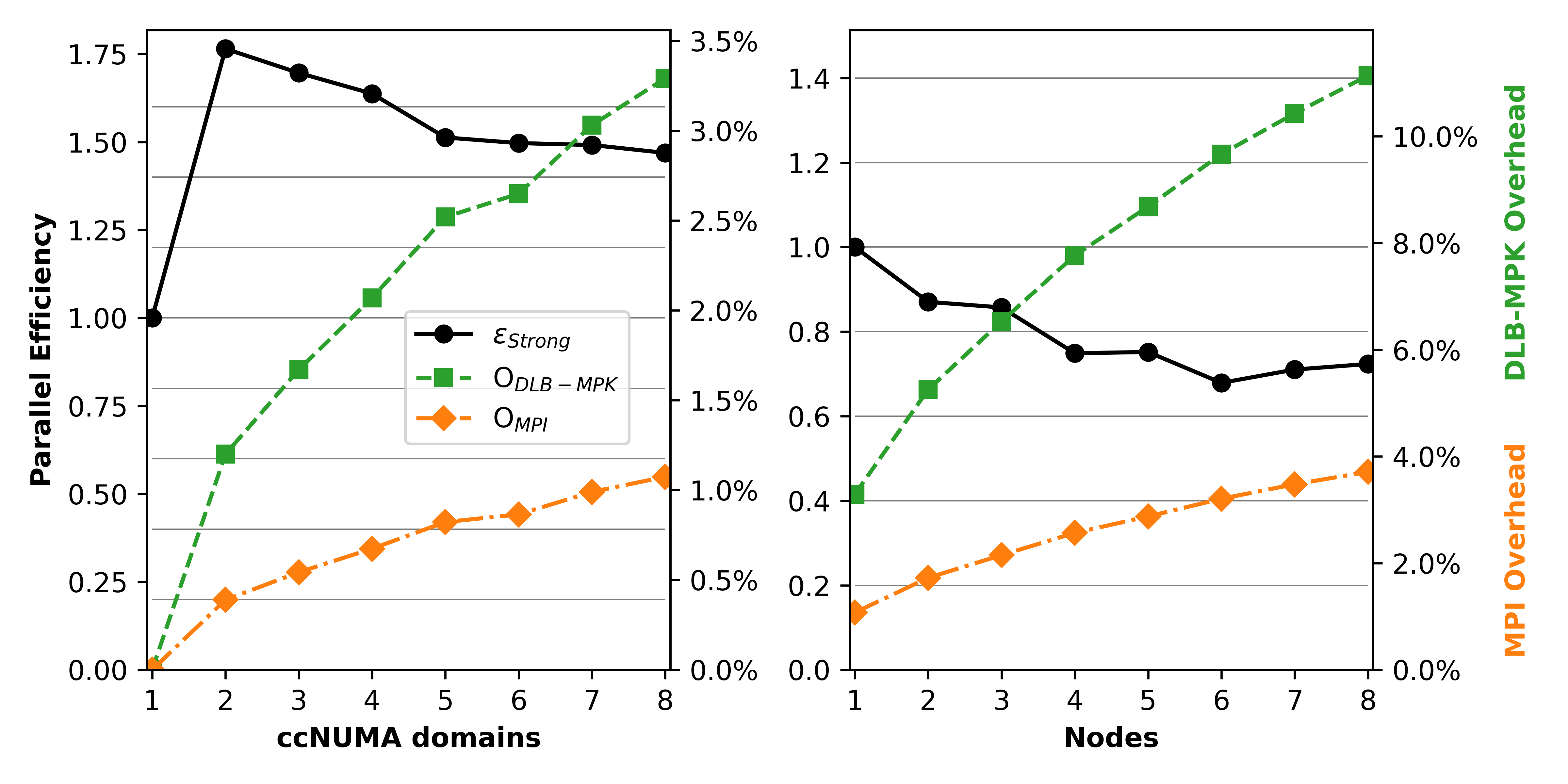}
		}
		
		\subfloat[Lynx1151 $A^6x$ overheads\label{fig:Lynx1151 A6x Overheads}]{%
			\includegraphics[width=.95\linewidth]{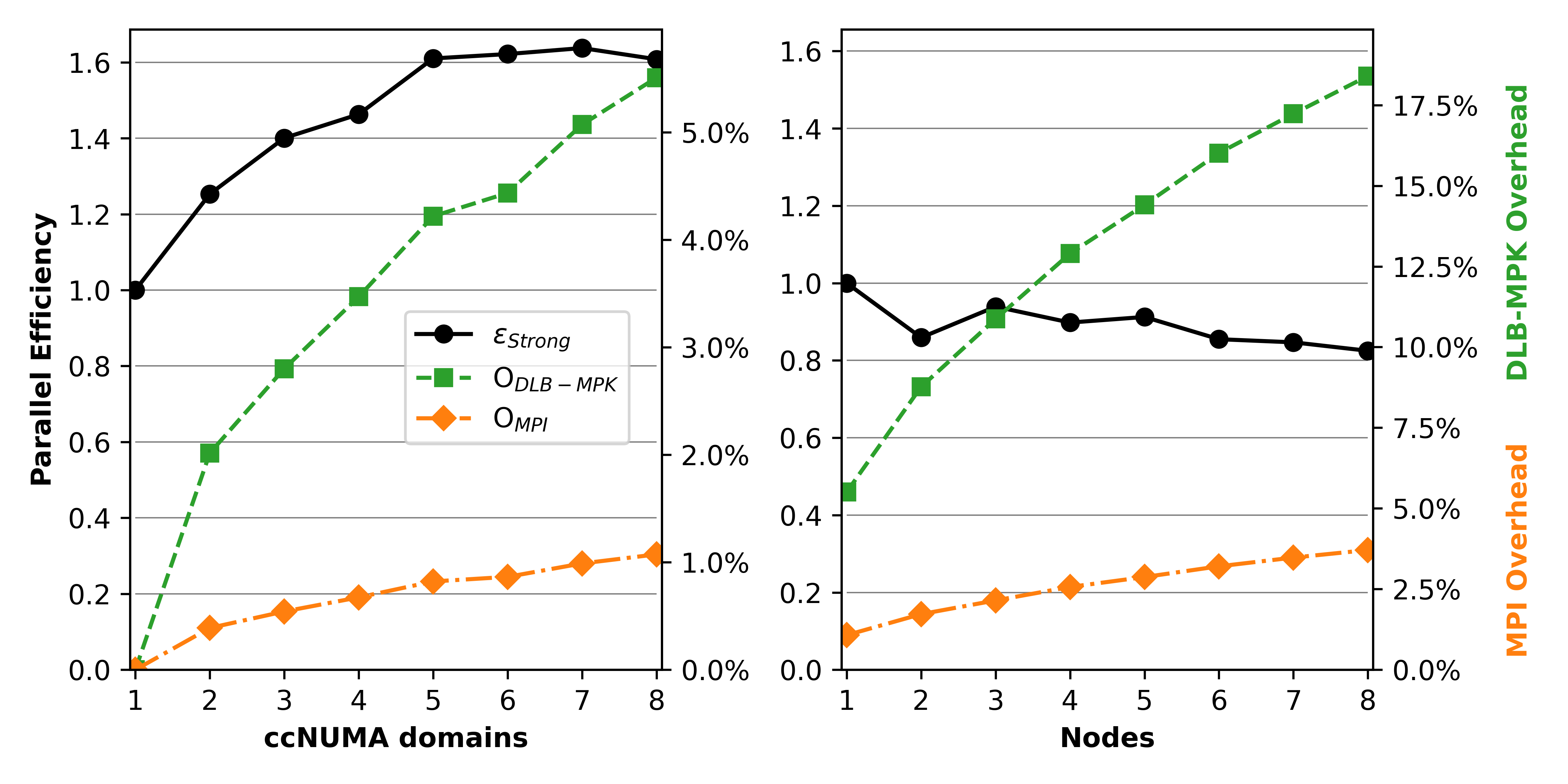}
		}
	\end{center}
	
	\subfloat[nlpkkt240 $A^4x$ strong scaling\label{fig:nlpkkt240 A4x Strong Scaling}]{%
		\includegraphics[width=.86\linewidth]{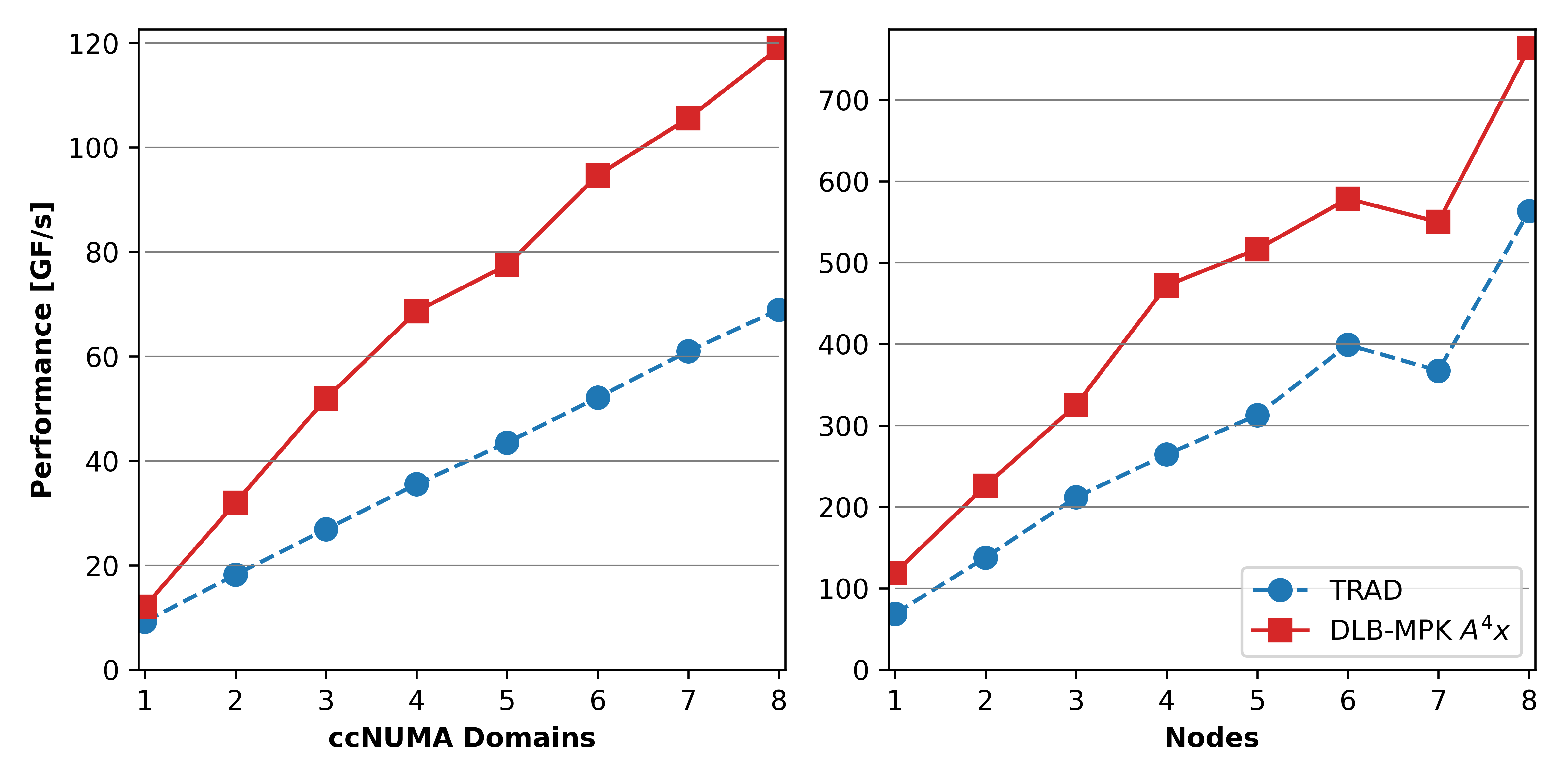}
	}
	
	\begin{center}
		\subfloat[nlpkkt $A^4x$ overheads\label{fig:nlpkkt A4x Overheads}]{%
			\includegraphics[width=.95\linewidth]{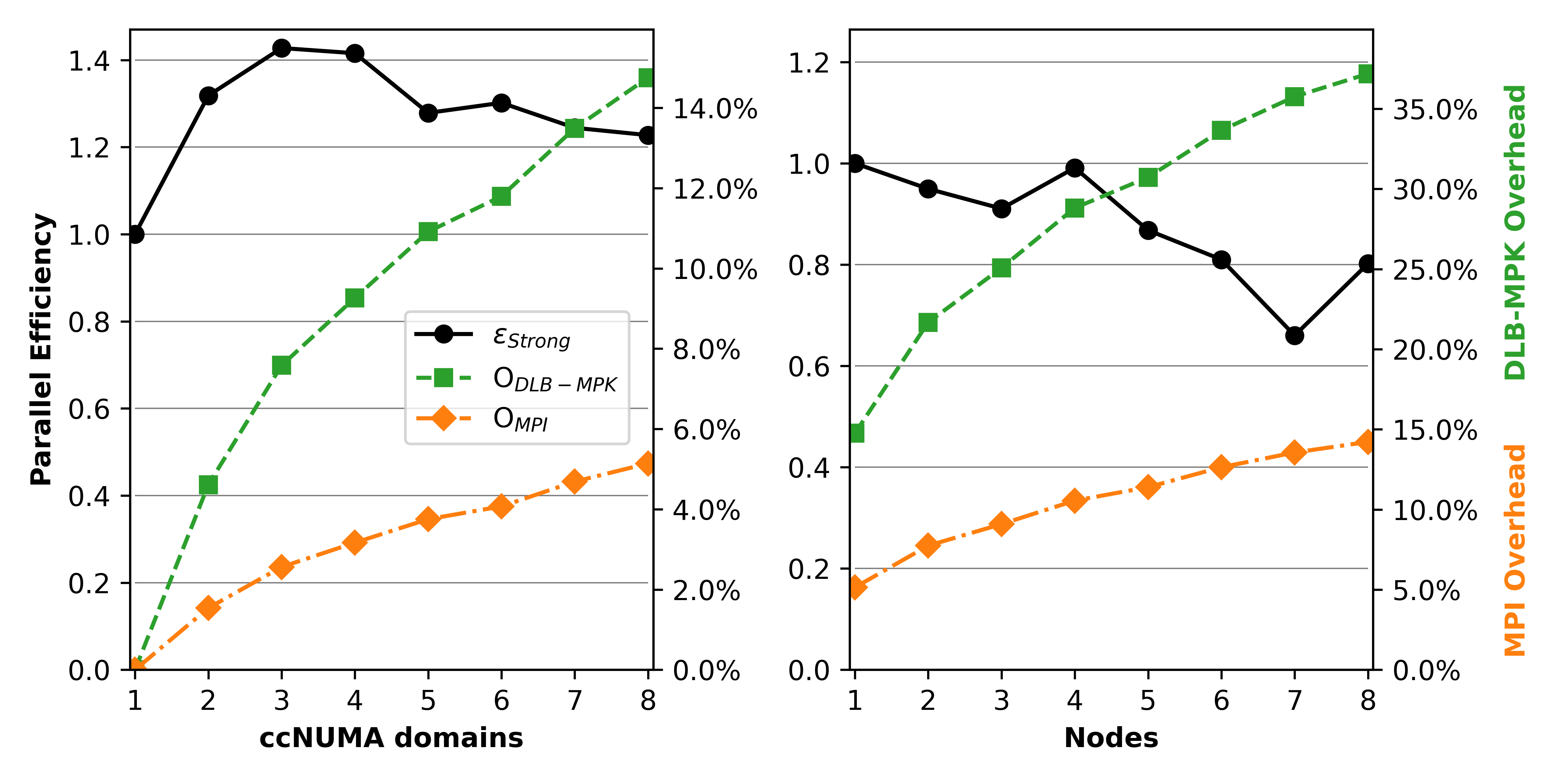}
		}
	\end{center}
	
	\caption{Single- (left) and multi-node (right) strong scaling performance and overhead results for \texttt{Lynx1151} and \texttt{nlpkkt240} on SPR nodes.}
	\label{tab:Strong Scaling Results}
\end{figure}

Figure~\ref{fig:Lynx1151 Strong Scaling} shows the performance of TRAD versus DLB-MPK for both $p = 4$ and $p = 6$ on \texttt{Lynx1151}. The reason that the performance for both $A^4x$ and $A^6x$ is shown is that both powers are optimal for \texttt{Lynx1151}, depending on the scale one considers. For a single SPR node, $p=4$ performs better than $p=6$. However, with more cache becoming available when scaling to  multiple nodes, $p=6$ performs better.

Figure~\ref{fig:Lynx1151 A4x Overheads} and \ref{fig:Lynx1151 A6x Overheads} show on the right y-axis how the two overheads introduced in Section \ref{sec:Distributed_Challenges} and \ref{sec:DLB-MPK Methodology} -- MPI overhead ($O_{\text{MPI}}$ from Equation~\eqref{eq:mpi_oheads}) and DLB-MPK overhead ($O_{\text{DLB-MPK}}$ from Equation~\eqref{eq:global_oheads}) -- scale for \texttt{Lynx1151} with a growing number of processes. On the left y-axis we show parallel efficiency for strong scaling $\varepsilon_{\text{strong}} := T_1/(nT_n)$ for DLB-MPK, where $T_1$ is the time required by DLB-MPK for a single process and is $T_n$ the time required by $n$ processes. 
Since we have a fixed workload, we can choose $T_n = 1/P_n$, where $P_n$ is the performance of DLB-MPK on $n$ ccNUMA domains.% The parallel efficiency is normalized to DLB-MPK executed on a single MPI processes (so essentially, just LB-MPK). 

Since we are blocking for a higher power in Figure~\ref{fig:Lynx1151 A6x Overheads}, it makes sense that $O_{\text{DLB-MPK}}$ will be higher than in Figure~\ref{fig:Lynx1151 A4x Overheads}, since there will be fewer vertices contained in the bulk structure $M$ as described in Section~\ref{sec:DLB-MPK Methodology}. MPI overhead will be the same for both $p=4$ and $p=6$, since $O_{\text{MPI}}$ depends only on the matrix structure and number of MPI processes.

We see $\varepsilon_{\text{strong}} \geq 1$ in the intra-node regime in the left subfigure in Figure~\ref{fig:Lynx1151 A4x Overheads} where we normalize $\varepsilon_{\text{strong}}$ against the time taken by DLB-MPK on one ccNUMA domain. The sharp increase in $\varepsilon_{\text{strong}}$ from $1$ to $2$ processes is due to the additional cache available with the second ccNUMA domain. As the number of processes increases, we gain access to more cache, yet the MPI costs grow as we communicate with other processes which are physically farther away. Alternatively, in the right subfigure in Figure~\ref{fig:Lynx1151 A4x Overheads} $\varepsilon_{\text{strong}} \leq 1$ for the inter-node regime, where we normalize $\varepsilon_{\text{strong}}$ against the time taken by DLB-MPK on one entire node. We see the impact of MPI on a larger scale here, as inter-node communication latency is much higher and bandwidth is lower than within a single node. 
Parallel efficiency reaches a higher maximum with $p=4$ for the intra-node case as shown in Figure~\ref{fig:Lynx1151 A4x Overheads}, but is sustained for larger MPI process for the inter-node case with $p=6$ as shown in Figure~\ref{fig:Lynx1151 A6x Overheads}.

Figures~\ref{fig:nlpkkt240 A4x Strong Scaling} and \ref{fig:nlpkkt A4x Overheads} show how the performance and overheads of DLB-MPK scale for \texttt{nlpkkt240}. Although the maximum performance attained is roughly the same as for \texttt{Lynx1151} on all 8 nodes, \texttt{nlpkkt240} exhibits different scaling behavior. There are two reasons for the strange scaling behavior of \texttt{nlpkkt240}.

First, the matrix structure is much ``worse," i.e., the sparsity pattern is not banded, and there are many non-zero elements that are far from the diagonal. This will not only increase DLB-MPK overhead as there are fewer levels (i.e., fewer vertices inside the bulk structure $M$), but it will also increase the MPI overhead as there are more halo elements on each process. The second reason is that we recognize residual caching effects after around 4--5 nodes by the sharp jumps in the performance of both TRAD and DLB-MPK. 
From Table \ref{tab:Benchmark Matrices}, we can compute that if \texttt{Lynx1151} is partitioned roughly equally across 8 nodes, about $2.8$ GiB of matrix data lies on each node. Since this is above $2400$ MiB, we will not see any residual caching effects. But if we partition \texttt{nlpkkt240} in the same manner, only about $1.1$ GiB of matrix data will reside on each node. %This is the reason why we see residual caching effects in Figure \ref{fig:nlpkkt240 A4x Strong Scaling} and \ref{fig:nlpkkt A4x Overheads} from 4 nodes onward.

This is not uncommon and poses a difficulty when performing scaling studies with DLB-MPK. Most matrices from Suite Sparse are simply not large enough to fully take advantage of DLB-MPK. %Applications which allow the problem size to increase proportional to resources are most suited for DLB-MPK, as this enables the matrix data to stay completely memory-resident.% An example of such an application is given in the following section.

\section{Application: Chebyshev Time Propagation}
\label{sec:Application}
A common application that can benefit from the DLB-MPK is the Chebyshev method for the time evolution of quantum states as shown by \cite{ChebyshevTimeEvolution,FEHSKE20092182}. 
In this section, we demonstrate the advantage of cache blocking in the context of this application and investigate the weak scaling characteristics of DLB-MPK.

Given a Hamiltonian $\hat{H}$ and an initial state $|\psi(0)\rangle$, the goal is to solve $|\psi (\tau)\rangle = e^{-i \tau \hat{H}} |\psi(0) \rangle$ for some target time $\tau$. This can be achieved by splitting the exponential into multiple small time steps $\delta \tau$ and approximating each as a polynomial in $\hat{H}$. 
Using an expansion in Chebyshev polynomials and keeping the first $\mathcal{M}+1$ terms leads to the following approximation for a single time step:
\begin{align}
	|\psi(\tau + \delta \tau) \rangle  &= e^{-i \delta \tau \hat{H}} |\psi(\tau) \rangle \nonumber \\ &\approx  J_0(\delta \tau) |v_0\rangle + 2\sum_{k=1}^\mathcal{M} (-i)^k J_k(\delta \tau)  |v_k\rangle ,  
\end{align}
where $J_k(\delta \tau)$ is the Bessel function of the first kind of order $k$. The states $|v_k\rangle$ are calculated recursively using the relations
\begin{gather}
	|v_{k+1}\rangle = 2 \hat{H} |v_k\rangle - |v_{k-1}\rangle ,\label{eq:anderson_spmv} \\
	|v_0 \rangle = |\psi(\tau) \rangle , \ \ |v_1 \rangle = \hat{H} |\psi(\tau) \rangle ,
\end{gather}
which primarily amounts to a sequence of $\mathcal{M}$ SpMVs when $\hat{H}$ is given as a sparse matrix. 
Since these SpMVs are the computational hot spot of the algorithm, the Chebyshev time-propagation method can potentially be sped up significantly by using the DLB-MPK. 

\begin{figure}[!h]
	\centering
	\includegraphics[width=0.9\linewidth]{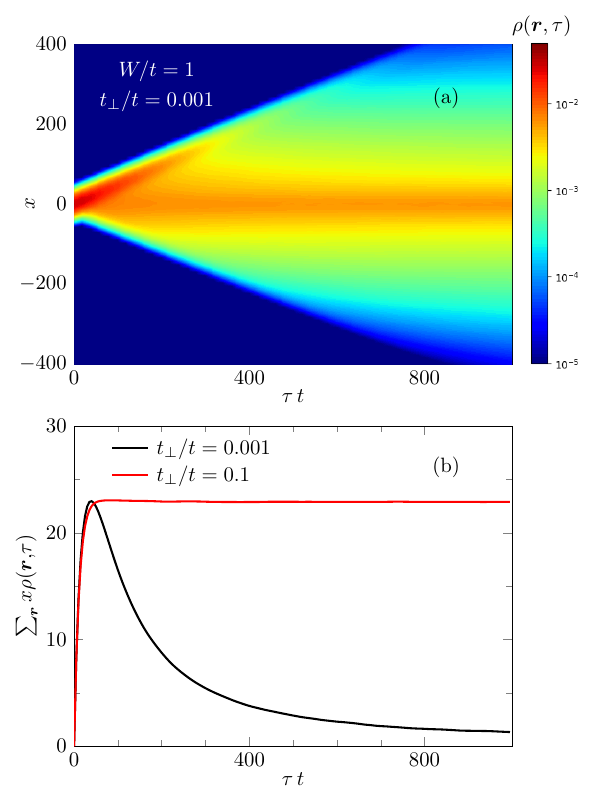}
	\caption{Time evolution of a wave packet (Eq.~\eqref{eq:wavepacket}) with width $\sigma = 20$ and momentum 
		$\bm{k}_0 = \frac{\pi}{2}\bm{e}_x$. 
		Panel (a) shows the time-dependent density distribution in the localized regime with parameters $t_{\perp}/t=0.001$ and $W/t = 1$. The center-of-mass motion is displayed in panel (b), which also includes data for a delocalized system with $t_{\perp}/t=0.1$.
		%While the center of mass approaches $x\approx 0$ for long times in the localized system (quantum boomerang effect), it remains at a finite displacement in the delocalized one.
		We used a finite rectangular system with dimensions $L_y=L_z=100$ and $L_x=3000$ for the simulations, and averaged the results over $50$ runs with different random potentials $w_{\bm{r}}$.}
	\label{fig:Anderson}
\end{figure}

We demonstrate the Chebyshev time propagation for the \texttt{Anderson} matrix. Physically, it represents a single-particle Hamiltonian for electrons in a disordered medium 
\begin{align}
	\hat{H} &= \frac{W}{2} \sum_{\bm{r}} w_{\bm{r}} |\bm{r} \rangle \langle \bm{r}| - t \sum_{\langle \bm{r}, \bm{r}' \rangle} |\bm{r} \rangle \langle \bm{r}' |,
	\label{eq:Anderson}
\end{align}
where the states $|\bm{r}\rangle$ with $\bm{r}=(x,y,z)\in \mathbb{Z}^3$ correspond to sites in a cubic lattice, and the second summation is over nearest-neighbor pairs. The parameter $W$ determines the strength of the disorder potential. Here, we assume an uncorrelated random potential, with $w_{\bm{r}}$ drawn uniformly from the interval $[-1,1]$. Equation~\eqref{eq:Anderson} is a paradigmatic model for the metal-insulator transition due to Anderson localization as described by~\cite{AndersonLocalization}: while the system is a conductor for small $W$, it becomes an insulator above some critical value $W_c$. For $W>W_c$, the eigenstates of $\hat{H}$ are localized, i.e., they are restricted to a finite region outside of which their weight decreases exponentially. As a consequence, an initially local state, e.g., a Gaussian wave packet 
\begin{align}
	|\psi (0 ) \rangle &\propto \sum_{\bm{r}} e^{-\frac{r^2}{2 \sigma^2} + i \bm{k}_0 \bm{r}} |\bm{r}\rangle
	\label{eq:wavepacket}
\end{align}
of width $\sigma$, does not diffuse and instead remains localized indefinitely. Moreover, it was recently shown that the density distribution $\rho(\bm{r},\tau) = |\langle \bm{r} | \psi (\tau) \rangle |^2$ at long times $\tau$ is insensitive to the initial momentum $\bm{k}_0$ of the wave packet, so that the center of mass of the wave packet must return to its origin.
\cite{QuantumBoomerang,QuantumBoomerangInteraction,QuantumBoomerangNoTRSym} have numerically investigated this ``quantum boomerang effect'' for various models using the Chebyshev time-propagation method. Here, we consider a variant of the Anderson model~\eqref{eq:Anderson} in which the hopping parameter $t$ along the $y$ and $z$ axis is replaced by $t_{\perp}<t$, i.e., a system of weakly coupled chains. By tuning $t_{\perp}$, a localization transition can be induced at fixed disorder $W$, as shown by \cite{PhysRevB.56.12221}. 
Figure~\ref{fig:Anderson} displays results for the time evolution of a wave packet moving in the $x$ direction with $\bm{k}_0 = \frac{\pi}{2}\bm{e}_x$. As expected, the center of mass approaches $x\approx 0$ for long times in the localized system with small $t_{\perp}/t = 0.001$, while it remains at a finite displacement in the delocalized one with $t_{\perp}/t = 0.1$. In addition, the density distribution $\rho(\bm{r},\tau)$ for $t_{\perp}/t = 0.001$ becomes stationary at long times $\tau$, i.e., the wave packet stops spreading, which is another signature of Anderson localization. 

	%As a weak scaling study, and to demonstrate the effectiveness of DLB-MPK on a real-world application, we perform Chebyshev time propagation for the Schr\"odinger equation 
	 To demonstrate the effectiveness of DLB-MPK on a real-world application, we now perform a weak scaling study on the above-described Chebyshev time propagation method. 
	 %for the Schr\"odinger equation
	%$$i \hbar \frac{\partial}{\partial t}\Psi(\mathbf{r},t) = \hat H \Psi(\mathbf{r},t).$$
	The \texttt{Anderson} matrix is generated using the ScaMaC matrix generator.\endnote{\url{https://alvbit.bitbucket.io/scamac_docs/index.html}}
	%choosing the \texttt{Anderson} matrix as our application scenario. \CAcomm{repeat.}

	Previous state-of-the-art implementations of the Chebyshev time propagation method perform back-to-back SpMVs to compute $|v_{k+1} \rangle$ for successive time steps.
	However, these SpMVs can be accelerated by cache blocking using the DLB-MPK scheme. 
	In order to be well outside of the residual caching effects on SPR, the study is constructed so that we always have about $342$ MiB of matrix data per ccNUMA domain. Compared with our observations in Figure~\ref{fig:All Bandwidths}, a matrix data size of $2,743$ MiB per node will be far outside of the cache. Specifically, we double the number of lattice sites in a selected direction ($x$, $y$, or $z$) in order to double the number of rows in the matrix. The \texttt{Anderson} matrix configurations used can be seen in Table \ref{tab:Weak Scale Matrices}.
	
	\begin{table}
		\begin{center}
			\caption{\texttt{Anderson} Matrix Configurations}
			\label{tab:Weak Scale Matrices}
			\resizebox{\columnwidth}{!}{%
				\begin{tabular}{cccccc}
					$\#$ ccNUMA Domains & (Lx, Ly, Lz) & $N_r$ & $N_{nz}$ & $N_{nzr}$ & CRS Size [MiB] \\
					\hline
					$1$ & $(160, 160, 160)$ & $4,096,000$ & $28,518,400$ & $7.0$ & $342$\\
					$2$ & $(320, 160, 160)$ & $8,192,000$ & $57,088,000$ & $7.0$ & $685$\\
					$4$ & $(320, 320, 160)$ & $16,384,000$ & $114,278,400$ & $7.0$ & $1,370$\\
					$8$ & $(320, 320, 320)$ & $32,768,000$ & $228,761,600$ & $7.0$ & $2,743$\\
					$16$ & $(640, 320, 320)$ & $65,536,000$ & $457,728,000$ & $7.0$ & $5,488$\\
					$32$ & $(640, 640, 320)$ & $131,072,000$ & $915,865,600$ & $7.0$ & $10,981$\\
					$64$ & $(640, 640, 640)$ & $262,144,000$ & $1,832,550,400$ & $7.0$ & $21,972$\\
				\end{tabular}
			}
		\end{center}
	\end{table}
	
\begin{figure}[tp]
	\captionsetup[subfigure]{justification=centering}
	\subfloat[Weak scaling performance\label{fig:app_weak_scale_perf}]{%
		\includegraphics[width=.91\linewidth]{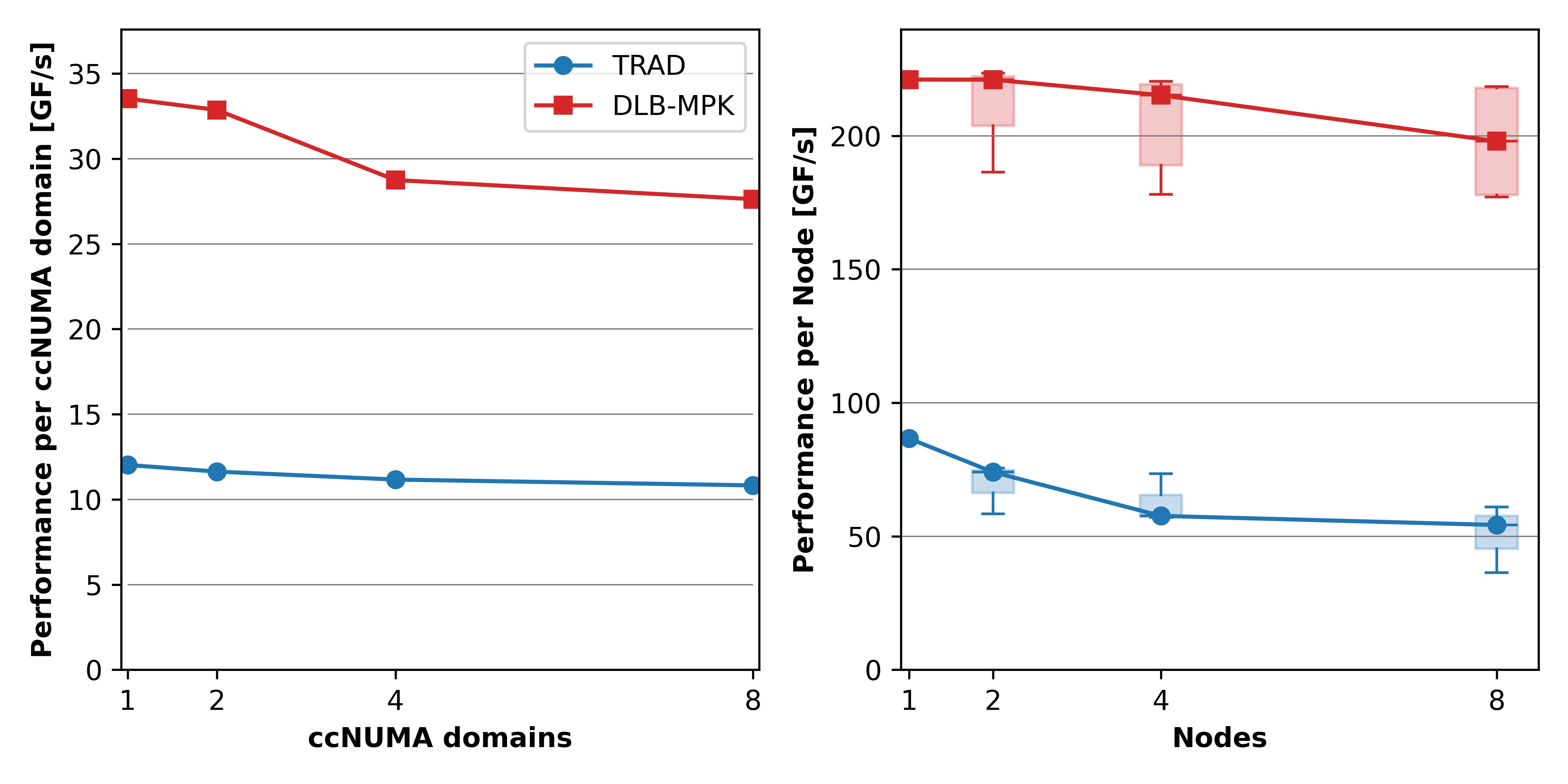}
	}%
	\hspace{2mm}%
	\subfloat[Weak scaling overheads\label{fig:app_weak_scale_oheads}]{%
		\includegraphics[width=\linewidth]{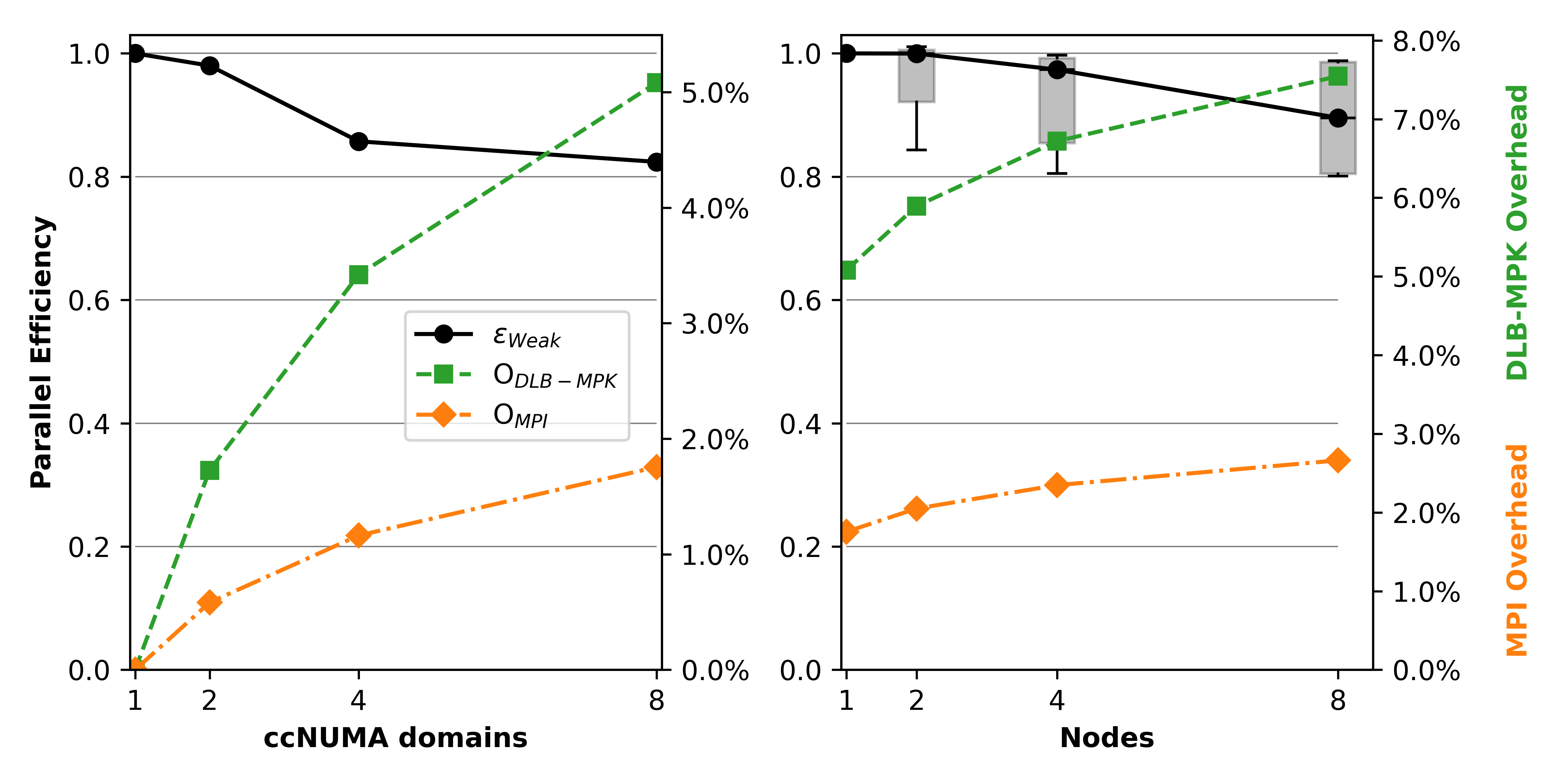}
	}
	%\end{center}
	
	\caption{Weak scaling investigation using the \texttt{Anderson} matrices in Table~\ref{tab:Weak Scale Matrices} on a single (left) and multiple (right) SPR node(s). Boxplots are given when performance fluctuates by more than 5\%.}
	\label{fig:Application Weak Scale}
\end{figure}
	As previously mentioned, Equation~\eqref{eq:anderson_spmv} represents a series of $\mathcal{M}$ SpMVs with the same matrix data $\hat{H}$. Since $\mathcal{M}$ is on the order of 100-1000s, we must choose a factor $p_m < \mathcal{M}$ by which we block the matrix data in the cache.
	
	In the same manner described in Section \ref{sec:Parameter Study}, we first tune $p_m$ and the cache size $C$ to obtain optimal node-wide performance. After such an investigation, DLB-MPK yields the highest performance on SPR at $p_m = 8$, $C=35$ MiB. Note that $C$ is much smaller than the available L2+L3 aggregate cache given for SPR in Table \ref{tab:Hardware Config.}. This is expected since there are other data structures in the application that will also occupy space in the cache. We define parallel efficiency in the weak scaling case as $\varepsilon_{\text{weak}} := T_1/T_n.$ Since our workload now increases with the number of processes, we choose $T_1 = 1/P_1$ and $T_n = n/P_n$, where $P_n$ is still the performance of DLB-MPK on $n$ ccNUMA domains. 
	
	Figure~\ref{fig:Application Weak Scale} shows the weak scaling performance per MPI process and the overheads of DLB-MPK applied to the Chebyshev time propagation method using various sizes of the \texttt{Anderson} matrix. We double the innermost spatial dimension last to respect layer conditions for cache blocking. In the single-node regime, we took the median of five executions of both TRAD and DLB-MPK as the representative performance, yet the fluctuations were less than 5\%. For the multi-node regime we included the box-and-whisker plots in addition to the median for both TRAD and DLB-MPK, since we noticed higher performance fluctuations when scaling to multiple nodes. The tip and tail of the whiskers are the minimum and maximum performance observed, and the boxes denote the interquartile range of the five executions. Our selected affinity is the same as described in Section~\ref{sec:Results}.
		
	DLB-MPK maintains a speed-up of about $2.8\times$ as compared against TRAD for $1$ and $2$ ccNUMA domains. When moving from $2$ to $4$, and then to $8$ ccNUMA domains, speed-up drops to about $2.5\times$. In the multi-node regime (i.e., past 8 ccNUMA domains) we maintain a speed-up of $2\times$ to $3.3\times$ for the worst performing DLB-MPK executions, and $2.5\times$ to $4\times$ for the best.
	
	%This can be explained by the fact that the MPI latency costs are higher for inter-socket communication.}  
		%first can be explained by the fact that SPR has a shared L3 cache between the 4 ccNUMA domains on a socket, and the second because MPI must now communicate over the QPI link connecting the two sockets on a node.} 

\section{Summary}
We have motivated and developed a novel cache-blocked MPI-parallel matrix power kernel based on the level-blocking capabilities of RACE. The resulting algorithm extends the ideas developed by \cite{RACEMPK} by first organizing local vertices on each MPI process by their distance $k$ from the halo buffer into levels $I_k$, and then interleaving a local cache blocking MPK with communication steps to fulfill data dependencies. Our algorithm, DLB-MPK, has been shown to be efficient in that it does not increase MPI overhead when compared to the traditional MPK implementation. This is because these collections of vertices $I_k$ grow inwards, keeping the number of halo elements constant while slightly reducing the efficiency of cache blocking. Furthermore, DLB-MPK has the advantage that it uses the same computation and halo communication routines as a traditional distributed MPK. Therefore, it can be easily integrated into existing libraries and can be used as a drop-in replacement for traditional distributed matrix power kernels.

We used the roof{}line model to explain expected performance behavior using key metrics extracted from our selection of test hardware platforms. After that, we gave an example of how one may tune DLB-MPK for optimal performance. To evaluate the performance of DLB-MPK, we first gave a snapshot summary of the optimally tuned performance as compared to the traditional MPK on modern multicore CPUs. We observed a node-wide average (maximum) speedup
of $1.6\times$ ($2.5\times$), $1.7\times$ ($2.4\times$), and $1.6\times$ ($2.7\times$) for large in-memory datasets on ICL, SPR, and MIL, respectively

Then, strong scaling characteristics of DLB-MPK were studied, where we observed the influence of caches and communication on performance.
Finally, DLB-MPK was integrated into an application using a Chebyshev method for the time evolution of quantum states for the Anderson model of localization. This enabled us to perform weak scaling investigations on up to eight Sapphire Rapids nodes, in which we observed a speed-up of up to $4\times$ when compared to the traditional MPK implementation. Future work will be directed towards the integration of GPGPU support for DLB-MPK.

\begin{funding}
This work was supported by NHR@FAU, which funded by German Federal Ministry of Education and Research and the state governments participating on the basis of the resolutions of the GWK for the national high-performance computing at German universities (NHR) and by the German Federal Ministry of Education and Research (BMBF) through the project "StroemungsRaum," 16ME0707, which is part of the initiative "Neue Methoden und Technologien f\"ur das Exascale-H\"ochstleistungsrechnen" (SCALEXA).
\end{funding}

\theendnotes

\bibliographystyle{SageH}
\bibliography{bib}

\begin{thebibliography}{29}
\providecommand{\natexlab}[1]{#1}
\providecommand{\url}[1]{\texttt{#1}}
\providecommand{\urlprefix}{URL }
\expandafter\ifx\csname urlstyle\endcsname\relax
  \providecommand{\doi}[1]{DOI:\discretionary{}{}{}#1}\else
  \providecommand{\doi}{DOI:\discretionary{}{}{}\begingroup
  \urlstyle{rm}\Url}\fi

\bibitem[{Alappat et~al.(2020{\natexlab{a}})Alappat, Basermann, Bishop, Fehske,
  Hager, Schenk, Thies and Wellein}]{RACE}
Alappat C, Basermann A, Bishop AR, Fehske H, Hager G, Schenk O, Thies J and
  Wellein G (2020{\natexlab{a}}) A recursive algebraic coloring technique for
  hardware-efficient symmetric sparse matrix-vector multiplication.
\newblock \emph{ACM Trans. Parallel Comput.} 7(3).
\newblock \doi{10.1145/3399732}.

\bibitem[{Alappat et~al.(2022)Alappat, Hager, Schenk and Wellein}]{RACEMPK}
Alappat C, Hager G, Schenk O and Wellein G (2022) Level-based blocking for
  sparse matrices: Sparse matrix-power-vector multiplication.
\newblock \emph{IEEE Transactions on Parallel and Distributed Systems} 34(2):
  1--18.
\newblock \doi{10.1109/TPDS.2022.3223512}.

\bibitem[{Alappat et~al.(2023)Alappat, Thies, Hager, Fehske and
  Wellein}]{alappat2023algebraic}
Alappat C, Thies J, Hager G, Fehske H and Wellein G (2023) Algebraic temporal
  blocking for sparse iterative solvers on multi-core {CPUs}.
\newblock \emph{arXiv:2309.02228,} Submitted.

\bibitem[{Alappat et~al.(2020{\natexlab{b}})Alappat, Hofmann, Hager, Fehske,
  Bishop and Wellein}]{10.1007/978-3-030-50743-5_21}
Alappat CL, Hofmann J, Hager G, Fehske H, Bishop AR and Wellein G
  (2020{\natexlab{b}}) Understanding {HPC} benchmark performance on {I}ntel
  {B}roadwell and {C}ascade {L}ake processors.
\newblock In: Sadayappan P, Chamberlain BL, Juckeland G and Ltaief H (eds.)
  \emph{High Performance Computing}. Cham: Springer International Publishing.
\newblock ISBN 978-3-030-50743-5, pp. 412--433.

\bibitem[{Anderson(1958)}]{AndersonLocalization}
Anderson PW (1958) Absence of diffusion in certain random lattices.
\newblock \emph{Phys. Rev.} 109: 1492--1505.
\newblock \doi{10.1103/PhysRev.109.1492}.

\bibitem[{Davis and Hu(2011)}]{10.1145/2049662.2049663}
Davis TA and Hu Y (2011) The {University} of {Florida} sparse matrix
  collection.
\newblock \emph{ACM Trans. Math. Softw.} 38(1).
\newblock \doi{10.1145/2049662.2049663}.

\bibitem[{Demmel et~al.(2008)Demmel, Hoemmen, Mohiyuddin and Yelick}]{4536305}
Demmel J, Hoemmen M, Mohiyuddin M and Yelick K (2008) Avoiding communication in
  sparse matrix computations.
\newblock In: \emph{2008 IEEE International Symposium on Parallel and
  Distributed Processing}. pp. 1--12.
\newblock \doi{10.1109/IPDPS.2008.4536305}.

\bibitem[{Fehske et~al.(2009)Fehske, Schleede, Schubert, Wellein, Filinov and
  Bishop}]{FEHSKE20092182}
Fehske H, Schleede J, Schubert G, Wellein G, Filinov VS and Bishop AR (2009)
  Numerical approaches to time evolution of complex quantum systems.
\newblock \emph{Physics Letters A} 373(25): 2182--2188.
\newblock \doi{10.1016/j.physleta.2009.04.022}.

\bibitem[{Janarek et~al.(2020)Janarek, Delande, Cherroret and
  Zakrzewski}]{QuantumBoomerangInteraction}
Janarek J, Delande D, Cherroret N and Zakrzewski J (2020) Quantum boomerang
  effect for interacting particles.
\newblock \emph{Phys. Rev. A} 102: 013303.
\newblock \doi{10.1103/PhysRevA.102.013303}.

\bibitem[{Janarek et~al.(2022)Janarek, Gr\'emaud, Zakrzewski and
  Delande}]{QuantumBoomerangNoTRSym}
Janarek J, Gr\'emaud B, Zakrzewski J and Delande D (2022) Quantum boomerang
  effect in systems without time-reversal symmetry.
\newblock \emph{Phys. Rev. B} 105: L180202.
\newblock \doi{10.1103/PhysRevB.105.L180202}.

\bibitem[{Karypis and Kumar(1998)}]{Karypis_1998}
Karypis G and Kumar V (1998) \emph{METIS: A Software Package for Partitioning
  Unstructured Graphs, Partitioning Meshes, and Computing Fill-Reducing
  Orderings of Sparse Matrices}.

\bibitem[{Kreutzer et~al.(2014)Kreutzer, Hager, Wellein, Fehske and
  Bishop}]{doi:10.1137/130930352}
Kreutzer M, Hager G, Wellein G, Fehske H and Bishop AR (2014) A unified sparse
  matrix data format for efficient general sparse matrix-vector multiplication
  on modern processors with wide {SIMD} units.
\newblock \emph{SIAM Journal on Scientific Computing} 36(5): C401--C423.
\newblock \doi{10.1137/130930352}.

\bibitem[{Langguth et~al.(2019)Langguth, Arevalo, Hustad and Cai}]{9005849}
Langguth J, Arevalo H, Hustad KG and Cai X (2019) Towards detailed real-time
  simulations of cardiac arrhythmia.
\newblock In: \emph{2019 Computing in Cardiology (CinC)}. pp. Page 1--Page 4.
\newblock \doi{10.22489/CinC.2019.301}.

\bibitem[{Langguth et~al.(2015)Langguth, Sourouri, Lines, Baden and
  Cai}]{7155461}
Langguth J, Sourouri M, Lines GT, Baden SB and Cai X (2015) Scalable
  heterogeneous {CPU}-{GPU} computations for unstructured tetrahedral meshes.
\newblock \emph{IEEE Micro} 35(4): 6--15.
\newblock \doi{10.1109/MM.2015.70}.

\bibitem[{Loe et~al.(2020)Loe, Thornquist and
  Boman}]{doi:10.1137/1.9781611976137.4}
Loe JA, Thornquist HK and Boman EG (2020) Polynomial preconditioned gmres in
  trilinos: Practical considerations for high-performance computing.
\newblock In: \emph{Proceedings of the 2020 SIAM Conference on Parallel
  Processing for Scientific Computing (PP)}. SIAM, pp. 35--45.
\newblock \doi{10.1137/1.9781611976137.4}.

\bibitem[{Luxburg(2004)}]{spectral_clustering}
Luxburg U (2004) A tutorial on spectral clustering.
\newblock \emph{Statistics and Computing} 17: 395--416.
\newblock \doi{10.1007/s11222-007-9033-z}.

\bibitem[{McQueen et~al.(2016)McQueen, Meil{\u{a}}, VanderPlas and
  Zhang}]{JMLR:v17:16-109}
McQueen J, Meil{\u{a}} M, VanderPlas J and Zhang Z (2016) Megaman: Scalable
  manifold learning in {P}ython.
\newblock \emph{Journal of Machine Learning Research} 17(148): 1--5.
\newblock \urlprefix\url{http://jmlr.org/papers/v17/16-109.html}.

\bibitem[{Moghimi(2023)}]{moghimi2023downfall}
Moghimi D (2023) {Downfall}: Exploiting speculative data gathering.
\newblock In: \emph{32th USENIX Security Symposium (USENIX Security 2023)}.

\bibitem[{Mohiyuddin et~al.(2009)Mohiyuddin, Hoemmen, Demmel and
  Yelick}]{10.1145/1654059.1654096}
Mohiyuddin M, Hoemmen M, Demmel J and Yelick K (2009) Minimizing communication
  in sparse matrix solvers.
\newblock In: \emph{Proceedings of the Conference on High Performance Computing
  Networking, Storage and Analysis}, SC '09. New York, NY, USA: Association for
  Computing Machinery.
\newblock ISBN 9781605587448, pp. 1--12.
\newblock \doi{10.1145/1654059.1654096}.

\bibitem[{Prat et~al.(2019)Prat, Delande and Cherroret}]{QuantumBoomerang}
Prat T, Delande D and Cherroret N (2019) Quantum boomeranglike effect of wave
  packets in random media.
\newblock \emph{Phys. Rev. A} 99: 023629.
\newblock \doi{10.1103/PhysRevA.99.023629}.

\bibitem[{Simpson et~al.(2018)Simpson, Pasadakis, Kourounis, Fujita, Yamaguchi,
  Ichimura and Schenk}]{10.1145/3218176.3218232}
Simpson T, Pasadakis D, Kourounis D, Fujita K, Yamaguchi T, Ichimura T and
  Schenk O (2018) Balanced graph partition refinement using the graph
  p-laplacian.
\newblock In: \emph{Proceedings of the Platform for Advanced Scientific
  Computing Conference}, PASC '18. New York, NY, USA: Association for Computing
  Machinery.
\newblock ISBN 9781450358910, pp. 1--11.
\newblock \doi{10.1145/3218176.3218232}.

\bibitem[{Tal‐Ezer and Kosloff(1984)}]{ChebyshevTimeEvolution}
Tal‐Ezer H and Kosloff R (1984) An accurate and efficient scheme for
  propagating the time dependent {S}chr{\"o}dinger equation.
\newblock \emph{The Journal of Chemical Physics} 81(9): 3967--3971.
\newblock \doi{10.1063/1.448136}.

\bibitem[{Treibig et~al.(2010)Treibig, Hager and Wellein}]{5599200}
Treibig J, Hager G and Wellein G (2010) {LIKWID}: A lightweight
  performance-oriented tool suite for x86 multicore environments.
\newblock In: \emph{2010 39th International Conference on Parallel Processing
  Workshops}. pp. 207--216.
\newblock \doi{10.1109/ICPPW.2010.38}.

\bibitem[{Vatai et~al.(2020)Vatai, Singhal and Suda}]{10.1145/3368474.3368494}
Vatai E, Singhal U and Suda R (2020) Diamond matrix powers kernels.
\newblock In: \emph{Proceedings of the International Conference on High
  Performance Computing in Asia-Pacific Region}, HPCAsia2020. New York, NY,
  USA: Association for Computing Machinery.
\newblock ISBN 9781450372367, p. 102–113.
\newblock \doi{10.1145/3368474.3368494}.

\bibitem[{Vuduc and Demmel(2003)}]{10.5555/1023242}
Vuduc RW and Demmel JW (2003) \emph{Automatic performance tuning of sparse
  matrix kernels}.
\newblock PhD Thesis.
\newblock AAI3121741.

\bibitem[{Williams et~al.(2009)Williams, Waterman and
  Patterson}]{10.1145/1498765.1498785}
Williams S, Waterman A and Patterson D (2009) Roofline: an insightful visual
  performance model for multicore architectures.
\newblock \emph{Commun. ACM} 52(4): 65–76.
\newblock \doi{10.1145/1498765.1498785}.

\bibitem[{Yamazaki et~al.(2014{\natexlab{a}})Yamazaki, Anzt, Tomov, Hoemmen and
  Dongarra}]{6877272}
Yamazaki I, Anzt H, Tomov S, Hoemmen M and Dongarra J (2014{\natexlab{a}})
  Improving the performance of {CA}-{GMRES} on multicores with multiple {GPUs}.
\newblock In: \emph{2014 IEEE 28th International Parallel and Distributed
  Processing Symposium}. pp. 382--391.
\newblock \doi{10.1109/IPDPS.2014.48}.

\bibitem[{Yamazaki et~al.(2014{\natexlab{b}})Yamazaki, Rajamanickam, Boman,
  Hoemmen, Heroux and Tomov}]{7013063}
Yamazaki I, Rajamanickam S, Boman EG, Hoemmen M, Heroux MA and Tomov S
  (2014{\natexlab{b}}) Domain decomposition preconditioners for
  communication-avoiding {Krylov} methods on a hybrid {CPU}/{GPU} cluster.
\newblock In: \emph{SC '14: Proceedings of the International Conference for
  High Performance Computing, Networking, Storage and Analysis}. pp. 933--944.
\newblock \doi{10.1109/SC.2014.81}.

\bibitem[{Zambetaki et~al.(1997)Zambetaki, Li, Economou and
  Soukoulis}]{PhysRevB.56.12221}
Zambetaki I, Li Q, Economou EN and Soukoulis CM (1997) Localization in weakly
  coupled planes and weakly coupled wires.
\newblock \emph{Phys. Rev. B} 56: 12221--12231.
\newblock \doi{10.1103/PhysRevB.56.12221}.

\end{thebibliography}

\begin{acks}
The authors gratefully acknowledge the computing time provided to them on the high-performance computer Noctua 2 at the NHR Center PC2. These are funded by the German Federal Ministry of Education and Research and the state governments participating on the basis of the resolutions of the GWK for the national high-performance computing at universities (www.nhr-verein.de/unsere-partner). We would also like to thank Johannes Langguth at Simula Research Laboratory for the thought provoking discussions, as well as his assistance in generating the \texttt{Lynx} matrices for our performance investigations. H.F.\ and G.W.\ acknowledge the hospitality at Los Alamos National Laboratory and the paper-related discussions with A.~Saxena and A.~R.~Bishop. 
\end{acks}

\end{document}